\documentclass[11pt]{article}

\usepackage[pdftex]{graphicx}
\usepackage{epstopdf}

\usepackage{natbib,geometry,html,rotate,lscape,epsfig,pslatex,
	colortbl, xspace,sectsty, amsthm,float,xy,xr,
	amssymb,amsfonts,amsbsy,amsmath,hyperref,delarray,
	bm,color,multirow,latexsym,fancyhdr,fontenc, anysize, url, subfigure,
	rotating,ifthen,fancybox,fullpage,amscd,pifont, enumerate, mathtools}
\usepackage[section]{placeins}

\newtheorem{thm}{Theorem}
\newtheorem*{thm*}{Theorem}

\newtheorem*{lma*}{Lemma}
\newtheorem{defi}{Definition}
\newtheorem{cor}{Corollary}
\newtheorem{prop}{Proposition}

\definecolor{DarkBlue}{rgb}{0.05,.0,.7}
\definecolor{DarkRed}{rgb}{.7,0,.4}

\def\bco{\iffalse}

\newcommand{\bea}{\begin{eqnarray*}}
	\newcommand{\eea}{\end{eqnarray*}}
\newcommand{\be}{\begin{eqnarray}}
\newcommand{\ee}{\end{eqnarray}}
\newcommand{\ed}{\end{document}}
\newcommand{\no}{\noindent}

\newcommand{\btab}{\begin{tabular}}
\newcommand{\etab}{\end{tabular}}
\newcommand{\bc}{\begin{center}}
\newcommand{\ec}{\end{center}}

\newcommand{\bi}{\begin{itemize}}
\newcommand{\ei}{\end{itemize}}
\newcommand{\bfi}{\begin{figure}}
\newcommand{\efi}{\end{figure}}
\newcommand{\ben}{\begin{enumerate}}
\newcommand{\een}{\end{enumerate}}
\newcommand{\bdes}{\begin{description}}
\newcommand{\edes}{\end{description}}
\newcommand{\bay}{\begin{array}}
\newcommand{\eay}{\end{array}}

\newcommand{\nn}{\nonumber}

\def\Var{{\rm Var}}

\def\diag{{\rm diag}}

\DeclareMathOperator*{\argmin}{arg\,min}

\DeclareMathOperator*{\arginf}{arg\,inf}
\def\R{\mbb{R}}
\def\Om{\Omega}
\def\ci{\cite}
\def\cp{\citep}
\def\d2{d_2}

\def\mbb{\mathbb}

\def\s1n{\sum_{i=1}^n}
\def\p1n{\prod_{i=1}^n}
\def\i01{\int_0^1}
\def\1d{{T_\delta}}
\def\1d{{1_\delta}}

\def\F{Fr\'echet }
\graphicspath{{figures/}}

\begin{document}
	
	\thispagestyle{empty}
	
	\begin{center}
		{\Large {\bf Conditional Wasserstein Barycenters and Interpolation/Extrapolation
				of Distributions}}\footnote{Research supported by NSF Grants DMS-1712864 and DMS-2014626}
		\vspace{.2in}
		
		May 2021

		\vspace{.3in}

		Jianing Fan and Hans-Georg M\"uller\\
		
		\vspace{.05in}
		
		Department of Statistics, University of California, Davis, CA 95616,
		USA\\
	\end{center}

\vspace{.25in}

\begin{abstract}
	Increasingly complex data analysis tasks motivate the study of the dependency of  distributions of multivariate continuous random variables on scalar or vector predictors.   Statistical regression models for distributional responses so far have primarily been investigated for the case of one-dimensional response distributions. We investigate  here the  case  of multivariate response distributions while adopting the 2-Wasserstein metric in the distribution space. The challenge is that  unlike the situation in the univariate case, the optimal transports that correspond to geodesics in the space of distributions with the 2-Wasserstein metric do not  have an explicit representation for multivariate distributions.  We show that under some regularity assumptions the conditional Wasserstein barycenters constructed for a geodesic in the Euclidean predictor space form a corresponding geodesic in the Wasserstein distribution space and demonstrate how the notion of conditional barycenters can be harnessed to interpolate as well as extrapolate multivariate distributions. The utility of distributional inter- and extrapolation is explored in simulations and examples.  We study both  global parametric-like and  local smoothing-like models to implement  conditional Wasserstein barycenters and establish asymptotic convergence properties for the corresponding estimates. For algorithmic implementation  we make use of a Sinkhorn entropy-penalized algorithm.   Conditional Wasserstein barycenters and distribution extrapolation are illustrated  with applications in climate science and studies of aging.\\
	
	\no KEY WORDS: Optimal Transport, Fr\'echet Mean, Wasserstein Metric, Geodesics,  Sinkhorn Penalty,  Climate Change, Baltimore Longitudinal Study of Aging.
\end{abstract}

\thispagestyle{empty}

\section{Introduction}
Optimal transport and the associated Wasserstein distance, also known as earth mover's distance \cp{rubner:2000}, have been widely used in analyzing probability distributions and their relationships  \cp{villani:2008}. Especially the Wasserstein barycenter problem \cp{agueh:2011} has met with growing interest in many fields including machine learning \cp{genevay:2018,frogner:2015,rabin:2011}, physics \cp{buttazzo:2012} and economics \cp{carlier:2010}. 
The $p$-Wasserstein  metric in the space of distributions is motivated by the Monge–-Kantorovich transportation problem, where the Kantorovich version of the  problem \cp{villani:2008} is 
\bea
W_p^p(\mu,\nu) = \inf_{\mathbf{\mu} \in \Pi(\mu,\nu)}E_{(X,Y)\sim \mathbf{\mu}}\|X-Y\|^p.
\eea
Here $X,Y$ are random variables in $\R^d$, $\mu,\nu$ are probability measure supported on a set $M \subset \R^d$, and $\Pi(\mu,\nu)$ is the space of joint probability measures on $M\times M$ with marginals $\mu$ and $\nu$. There is a close connection to  Monge's transport problem \cp{monge:1781}, where  optimal transport is characterized by  
\bea
OT^p(\mu,\nu)= \inf_{T: T_\# \mu = \nu}E_{X\sim \mu}\|X-T(X)\|^p
\eea
and the  optimization is taken over all push-forward maps $T$ that map $\mu$ to $\nu$. The push-forward maps $T$ are Borel maps  $M\rightarrow M$ and with $\nu_1$ denoting any probability measure on $M$,  $\nu_2= T_{\#}\nu_1$ stands for the push-forward measure of $\nu_1$, defined as the measure satisfying  $\nu_2(M_1) = \nu_1(T^{-1}(M_1))$ for any measurable set $M_1\subset M$. If it exists, the minimizer $\arginf_{T: M\rightarrow M}E_{X\sim \mu}\|X-T(X)\|^p$ is the  optimal transport plan. In general,  the Kanorovich and Monge problems  do not admit the same solution. But when Monge's problem has a minimizer $T_0$, it is also a solution of Kantorovich's problem. If $\mu$ is absolutely continuous, the two problems are equivalent \cp{villani:2008} .

\begin{figure}
	\centering
	\includegraphics[width=5.5cm]{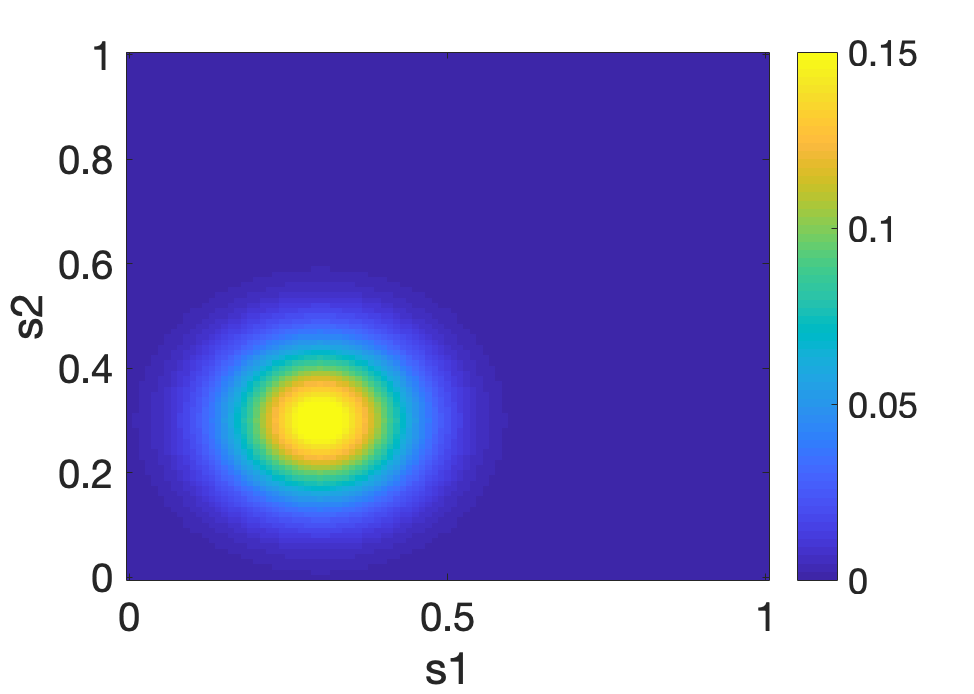}
	\includegraphics[width=5.5cm]{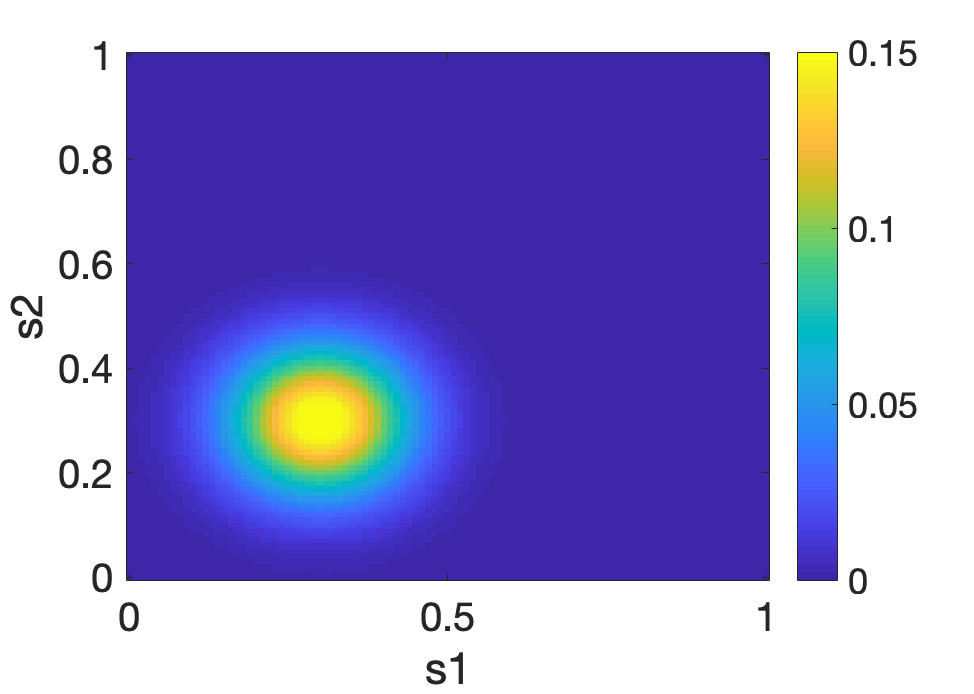}\\
	\includegraphics[width=5.5cm]{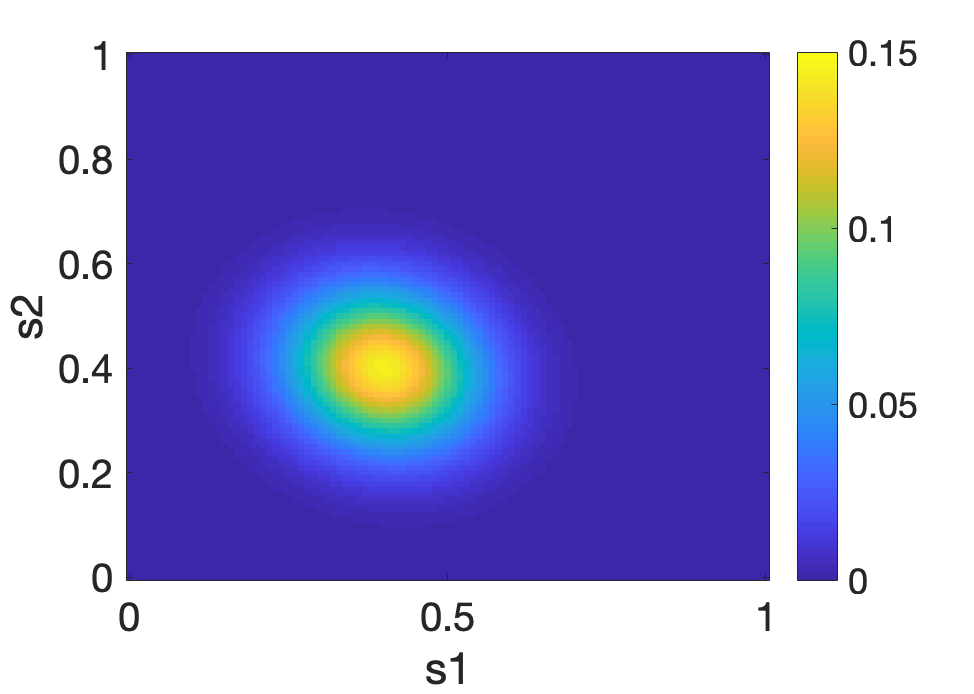}
	\includegraphics[width=5.5cm]{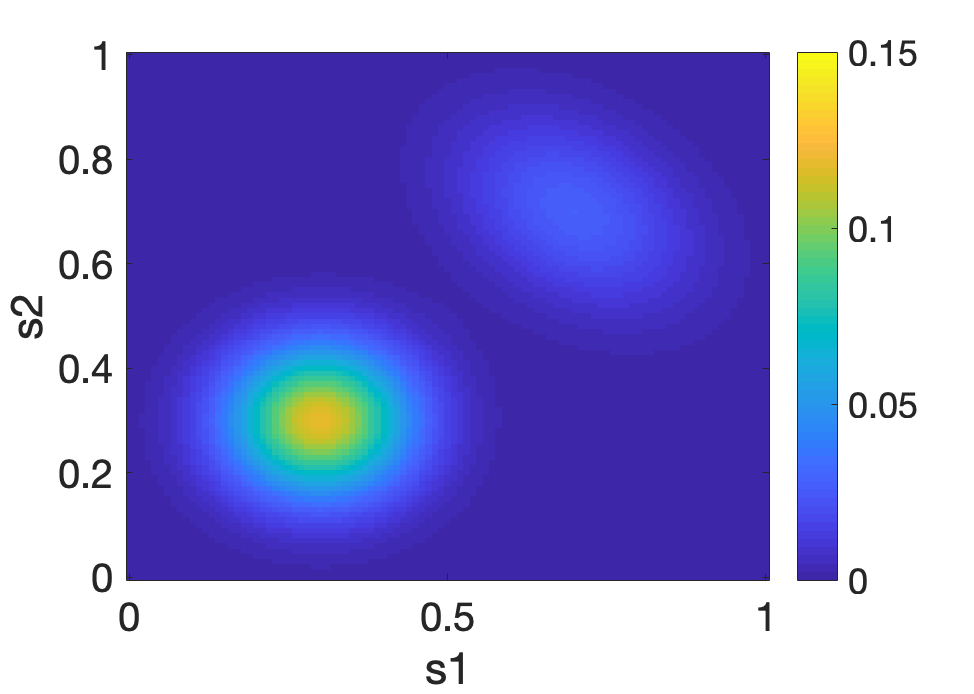}\\
	\includegraphics[width=5.5cm]{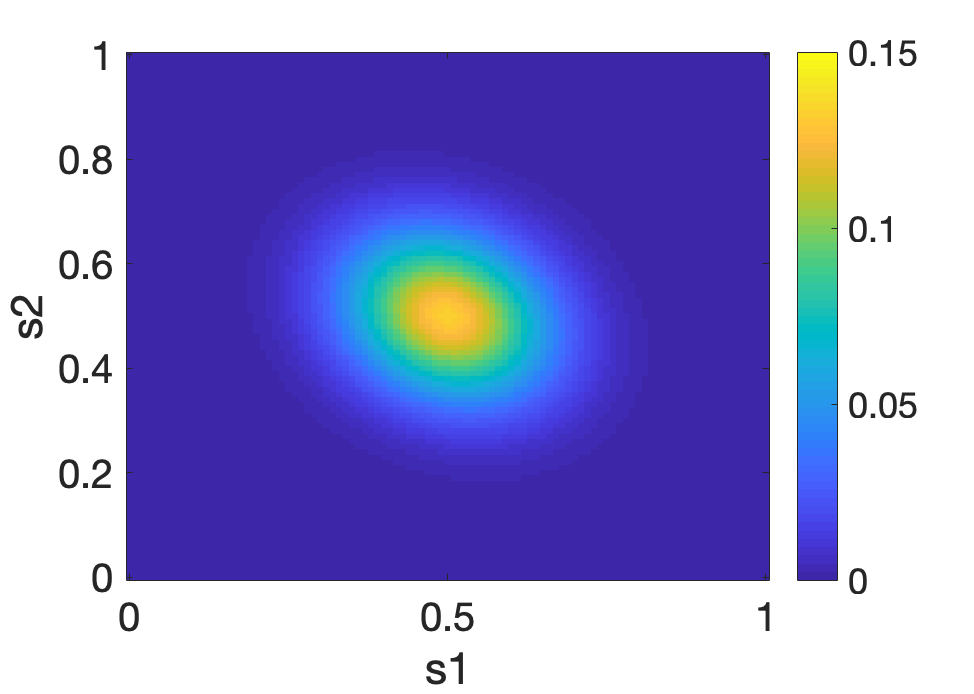}
	\includegraphics[width=5.5cm]{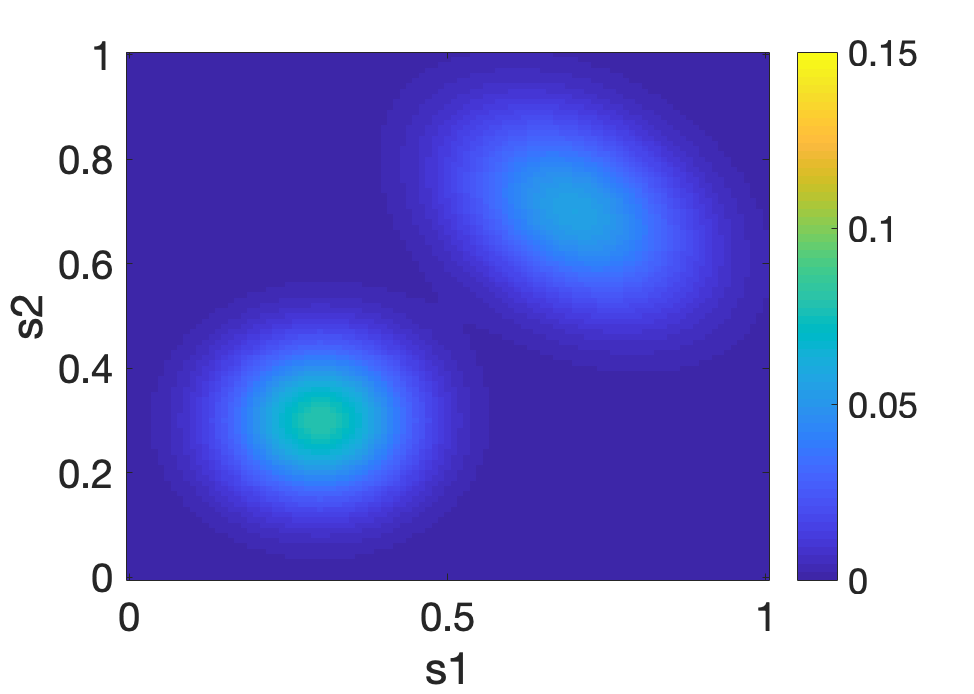}\\
	\includegraphics[width=5.5cm]{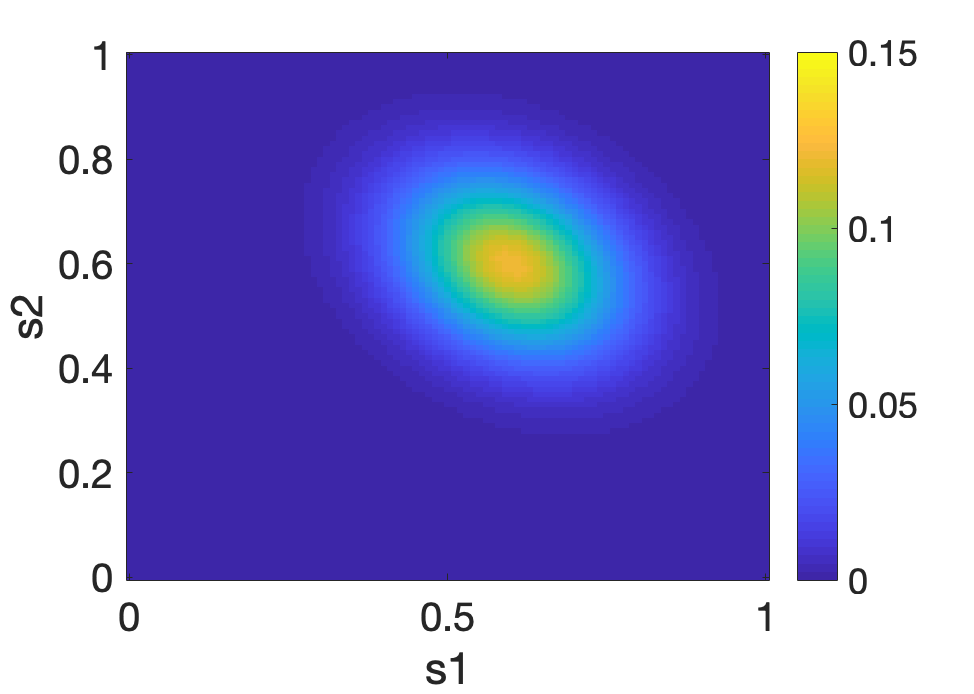}
	\includegraphics[width=5.5cm]{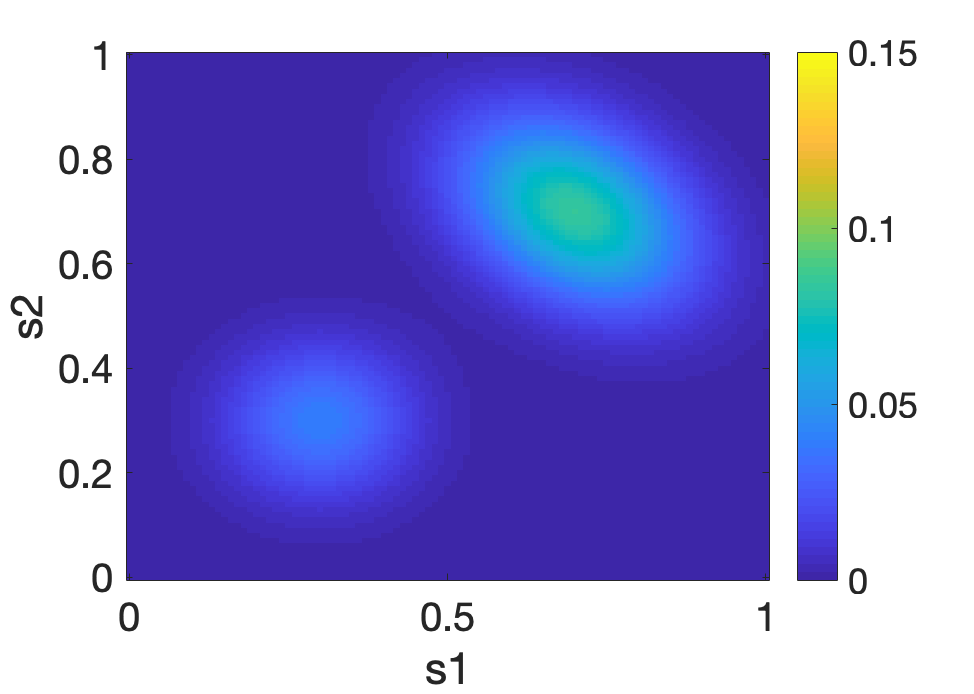}\\
	\includegraphics[width=5.5cm]{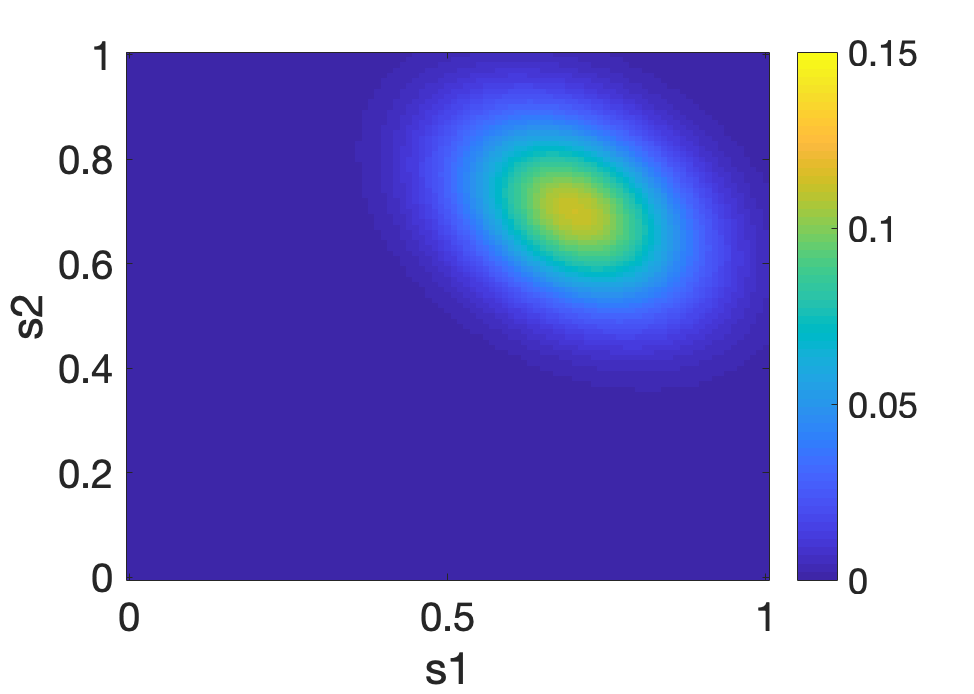}
	\includegraphics[width=5.5cm]{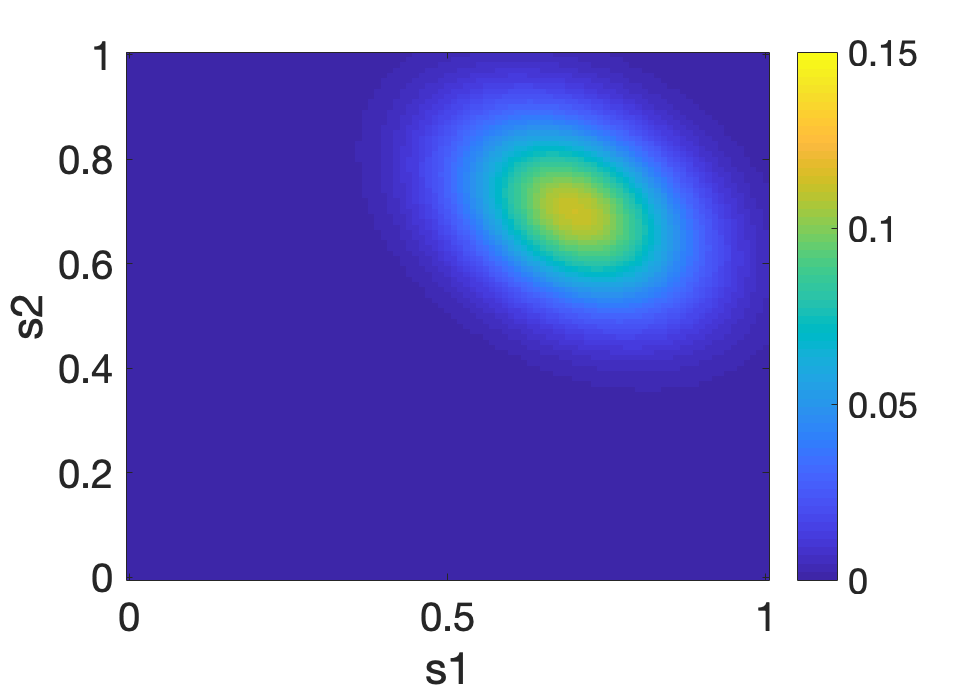}\\
	\caption{Wasserstein (left panel) and $L^2$ (right panel) geodesics at levels $x=0,0.25,0.5,0.75,1$,  from top to bottom, %where $x=0$ corresponds to $N((4,4),[1\, 0;0\, 1])$, that 
		interpolating  the Gaussian distributions $N((4,4),[1\, 0;0\, 1])$ (for $x=0$) and $N((6,6),[1.25\, -0.25;-0.25\, 1.25])$ (for $x=1$) and illustrated as heat maps of the respective distributions.} %.  Wasserstein interpolation with the proposed global method is shown in the left panels and $L^2$ interpolation in the right panels. }
	\label{f0}
\end{figure}

For the special case of  distributions  on the real line $R^1$, the $p$-Wasserstein distance between probability distributions is well known to correspond to  the $L^p$ distance between their quantile  functions. If $F_{\mu_1}, F_{\mu_2}$ denote the cumulative distribution functions of measures $\mu_1$ and $\mu_2$,  the $p$-Wasserstein distance can be written as
\be
\label{wassR1}
W_p^p(\mu_1,\mu_2) = \int_0^1 (F_{\mu_1}^{-1}(s)-F_{\mu_2}^{-1}(s))^p ds,
\ee
where $F_1^{-1}, F_2^{-1}$ are left inverses of $F_1, F_2$.
The most commonly  used Wasserstein distances are the 1-Wasserstein and 2-Wasserstein distances, not least since their  geodesics are easily  interpretable.   A comparative example of geodesics in  the 2-Wasserstein space and in the space of distributions with the $L^2$ distance between densities is provided in Figure \ref{f0} for the case of two-dimensional Gaussian densities , where it is seen that the 2-Wasserstein geodesics are more intuitive and better interpretable than   the $L^2$  geodesics.  Accordingly, we  focus here on 2-Wasserstein barycenters for distributions in the finite-dimensional  Euclidean space $R^d$ and consider their extension to conditional barycenters.

For  a random measure $\nu$ taking values in 2-Wasserstein space $\mathcal{W}_2$, the  2-Wasserstein barycenter of $\nu$ is defined as \cp{le:2017} 
\be
\label{wb}
B(\nu) = \arginf_{\mu\in \mathcal{W}_2} E W^2_2(\mu,\nu).
\ee
Similarly, when a sample of measures $\{\nu_i\}$, $i=1,2,\ldots,n$ is observed, the sample 2-Wasserstein barycenter \cp{agueh:2011} is
\bea
B(\nu_1,\ldots,\nu_n) = \arginf_{\mu\in \mathcal{W}_2} \sum_{i=1}^{n}W_2^2(\nu_i,\mu).
\eea
For  $n$ nonnegative weights  with $w_i>0, \, \sum_{i=1}^{n}w_i=1$, a weighted version of the 2-Wasserstein sample barycenter  is  analogoulsy defined as 
\be
\label{wwb}
B_{(w_1,\ldots,w_n)}(\nu_1,\ldots,\nu_n) = \arginf_{\mu\in \mathcal{W}_2} \sum_{i=1}^{n}w_iW_2^2(\nu_i,\mu).
\ee

Sample or empirical barycenters can also be viewed as sample \F means \citep{frechet:1948} of $\{\nu_i\}$ in the Wasserstein space. Existence and uniqueness of 2-Wasserstein barycenters has been shown  under the assumption of  absolute continuity of the measures $\nu_i$, where  \cite{le:2017} also studied the convergence  of sample barycenters towards the population barycenter. 
While  the properties of empirical barycenters and their  computation have been well studied, the situation is quite different for conditional barycenters. From a statistical perspective, the conditional barycenter problem was addressed  in \cite{petersen:2019} for the special setting of probability measures on $R^1$ as responses.  In the one-dimensional case one can alternatively pursue global transformations that map the Wasserstein space of one-dimensional distributions to a Hilbert space $L^2$  \cp{mull:16:1}  or alternatively of log maps to tangent bundles in the Wasserstein manifold \cp{bigo:17,bigo:18,mull:20:7}, although both come at the price of a metric distortion. After applying such a transformation, functional regression models designed for linear  spaces naturally become applicable in the ensuing Hilbert spaces \cp{horv:12,mull:16:3}. 

In this paper we aim to address the problem of estimating conditional barycenters in general $R^d$ spaces, where such transformations are not known or face major difficulties in theory and implementation so that the barycenter problem  cannot be linearized in this way.  Then the notion of conditional barycenters  replaces the traditional conditional expectation and forms the backbone  for the proposed regression models with responses that are distributions in $\R^p$ and are coupled with Euclidean predictors.  
The paper is organized as follows. Section 2 introduces  the problem of conditional barycenters and the statistical models that are proposed  for their 
implementation.  Algorithmic aspects for obtaining the conditional Wasserstein barycenter in $R$ and $R^d$, $d\geq 2$ respectively, are discussed in section 3. Section 4 contains a study of the asymptotic behavior of corresponding estimates of  the  conditional barycenters. In Sections 5 and 6, we provide simulation results and data analysis   to  illustrate the methodology and provide additional motivation. All  proofs are in the Supplement.

\section{Global and Local Models for  Conditional Barycenters}
In standard regression modeling in Euclidean spaces, one postulates an explicit model for the conditional expectation $E(Y|X)$ for responses $Y$, given  predictors $X$, to quantify their relationship. A classical model is  $E(Y|X)=\beta X$ for global linear models with a parameter vector $\beta$. The  
utility of such  a model is often not more than  providing a useful approximation \cp{buja:19,buja:19:1}. Simplified and often approximative regression models  are  necessary as the direct modeling of the joint distribution of $(X,Y)$, from which the actual regression relation derives, usually proves too complex for statistical modeling.  It is possible in some cases to relax the assumption of global linearity and to adopt less restrictive local linear models that adapt better to nonlinear shapes. The situation becomes more complex when non-Euclidean data such as elements of the Wasserstein space are
involved in the joint distribution. In the situation we consider here, responses $Y$ are distributions and one then needs an appropriate 
definition of the conditional barycenter, and, importantly, of the specific regression model through which it can be implemented. 

Consider random objects $(X,\nu)$ with $\nu$ taking values in the 2-Wasserstein space $\mathcal{W}_2(M)$ of probability measures supported on a compact domain $M\in R^d$, where  $X\in R^q, \, \text{ with } d,q \ge 1.$ More specifically, to exclude complications that arise when dealing with discrete measures and are of no interest for our purposes, we restrict the  probability measures  $\nu$ to the subset of  absolutely continuous measures so that each measure is associated with a unique density $f_{\nu}$ and denote this subset of probability measures as $\Omega = \mathcal{W}_2(M) \cap \mathcal{C}(M)$,  $\mathcal{C}(M)$ denoting  the set of absolutely continuous measures on $M$. Since the density space equipped with the Wasserstein metric is not a Hilbert space, the usual notion of expectation is not sensible and is replaced by the  \F mean \citep{frechet:1948}, which is the barycenter given in (\ref{wb}),  with $\mathcal{W}_2$ replaced by $\Omega$. The concept of conditional barycenters  was introduced in \ci{petersen:2019} for metric space-valued responses and shown to apply to one-dimensional distributions. Its general definition for probability distributions on spaces $R^d$ or on more general metric spaces is 
\begin{defi}(Conditional Wasserstein Barycenter).
	Assume $(X,\nu)$ has a joint distribution. The conditional Wasserstein barycenter is defined as the \F regression, or conditional \F mean with respect to the  2-Wasserstein distance, i.e., 
	\begin{equation}\label{cwb}
	\mu_0(x) = \argmin_{\mu\in \Omega} E(W_2^2(\mu,\nu)|X=x).
	\end{equation}
\end{defi}

To implement this general concept, one needs specific models for these conditional barycenters, in the same way as one needs 
specific regression models to implement conditional expectations for Euclidean data,  such as $E(Y|X)=\beta X$ for a global linear model 
or $E(Y|X)=m(X)$ with a twice differential function $m$ for a local linear model. 
Specifically, we aim to obtain conditional Wasserstein barycenters from samples $\{X_i,f_i\},\, i = 1,2,\ldots,n$, where the $f_i$ are densities supported on $M$. Here we use densities to represent  probability measures in $\Omega$, as they are assumed to be well defined  and are easier to  interpret
statistically  compared to other equivalent representations. 

It has remained an open question whether  the    \F regression approach can be applied to multivariate distributions.  
\F regression  provides versions of a  global (linear) model and a  local (linear) model, designed  for metric space valued random objects satisfying 
certain regularity  conditions that were shown to be met for one-dimensional 
distributions   \cp{petersen:2019}.  The technical difficulties one faces for the case of multivariate distributions differ fundamentally from the situation for  one-dimensional distributions, as it is not possible anymore to  take advantage of $L^2$ vector space operations in the space of quantile functions as per (\ref{wassR1}), which is a crucial element of the previous analysis.

In this paper we show that  under certain assumptions the extension of global and local linear models for  conditional barycenters to the much larger class of multivariate distributions is feasible, and study the convergence of corresponding estimates   to the population targets. 
Relevant applications include  extrapolating multivariate distributions beyond the range of the predictor $X$ similar to linear extrapolation in the 
Euclidean case, in addition to interpolation.  Consistency of  estimates for the global model is attained if the conditional barycenters have a specific form, while the local model more broadly  approximates the true conditional barycenters under mild assumptions and works well for low-dimensional predictors $X$.  The local model is especially useful to interpolate conditional barycenters for  continuous predictors that may be observed on a discrete grid while one desires conditional barycenters for all predictor values without postulating a global linear model.   

Let $\nu_i$ be random probability measures with density $f_i, \, i=1,\dots,n,$ where we write $\nu$ and $f$ for generic measures and densities.
The global model for conditional Wasserstein barycenter is defined as
\begin{equation}\label{gm}
\mu_{G}(x) = \argmin_{\mu\in \Omega} E(s_G(X,x)W_2^2(\mu,\nu)),
\end{equation}
where $s_G(X,x) = 1+(X-E(X))^{t}\Var(X)^{-1}(x-E(X))$ is a weight function that is linear in $x$. With $\mathcal{\lambda}$ denoting  the Lebesgue measure on $M$, the density corresponding  to $\mu_{G}(x)$ is  $f_{G}(x,\cdot) = \frac{d\mu_{G}(X)}{d\lambda}$.
The corresponding sample version for the  global conditional Wasserstein barycenter estimate is 
\begin{equation}\label{ge}
\hat{\mu}_{G}(x) = \argmin_{\mu\in \Omega} \frac{1}{n}\sum_{i=1}^{n}s_{iG}(x)W_2^2(\mu,\nu_i),
\end{equation}
where $s_{iG}(x)= 1+(X_i-\bar{X})^{t}\Sigma^{-1}(x-\bar{X})$ is the empirical version of $s_G(X,x)$. The corresponding density is $\hat{f}_{G}(x,\cdot) = \frac{d\hat{\mu}_{G}(X)}{d\lambda}$. These notions are analogous to the extension of the usual mean to the \F mean, namely, they are extensions of the characterization of a linear regression with scalar response, which is found to minimize (\ref{gm}) and (\ref{ge}) for the Euclidean metric,  to the case of general metric space valued responses; additional motivation and connections to M estimation can be found in 
\ci{petersen:2019}. 

For  a kernel function $K$ on $M$ which is a probability density and a vector of positive bandwidths $h \in R^{d}$, the local model for conditional Wasserstein barycenters is 
\begin{equation}\label{lm}
\mu_{L,h}(x) = \argmin_{\mu\in \Omega} E(s_{L,h}(X,x)W_2^2(\mu,\nu)),
\end{equation}
where $s_{L,h}(X,x)= \frac{1}{\sigma_{0}^2} K_{h}(X-x)[1 -\mu_{1}^t\mu_{2}^{-1}(X-x)]$, $\sigma_{0}^{2} = \mu_{0}-\mu_{1}^t\mu_2^{-1}\mu_{1}$, $\mu_0 = E(K_h(X - x)),\, \mu_1 = E(K_h(X - x)(X-x)),\,\mu_2=E(K_h(X - x)(X-x)(X-x)^t)$, and the corresponding density is $f_{L,h}(x,\cdot) = \frac{d\mu_{L,h}(x)}{d\lambda}$. 

The motivation again is the extension of means to \F means and rewriting the local linear estimates as M-estimators. 
The corresponding sample version is 
\begin{equation}\label{le}
\hat{\mu}_{L,h}(x) = \argmin_{\mu\in \Omega} \frac{1}{n}\sum_{i=1}^{n}s_{iL,h}(x)W_2^2(\mu,\nu),
\end{equation}
where $s_{iL,h}(x)= \frac{1}{\hat{\sigma}_0^2}K_{h}(X_i-x)(1 -\hat{\mu}_{1}^t\hat{\mu}_{2}^{-1}(X_i-x))$, $\hat{\sigma}_{0}^{2} = \hat{\mu}_{0}-\hat{\mu}_{1}^t\hat{\mu}_{2}^{-1}\hat{\mu}_{1}$ and $\hat{\mu}_j = \frac{1}{n}\sum_{i=1}^{n}K_h(X_i - x)(X_i-x)^j$, with corresponding density $\hat{f}_{L,h}(x,\cdot) = \frac{d\hat{\mu}_{L,h}(x)}{d\lambda}$. We note that the weights used in (\ref{lm}) and (\ref{le}) are not constrained to be positive and indeed some of the weights are usually negative, especially when the predictors are near a boundary of the predictor domain.  It is straightforward to see, for example in the Euclidean case,  that negative weights are an inherent necessity and are indeed required to represent global and local linear regression relations. 

While these global and local models to implement conditional barycenters can be viewed as instantiations of the general Fr\'echet regression paradigm, 
our focus here is to study the features of the population versions and the properties of the resulting estimates in the multivariate distribution case.  For this important case the properties of these models have remained unknown so far and  implementations also involve non-trivial computational aspects that will be explored in the following section. 

\section{Computational Considerations}
%\subsection{One dimensional probability measures}
When the dimension $d$ of the random probability measures that we study here is more than 1, i.e., $d \geq 2$, one does not have an analytic form for the barycenter  and the optimization algorithms to obtain it are complex, in contrast to  the case $d=1$, where  the quantile representation of Wasserstein distance as per (\ref{wassR1}) leads to  an explicit solution via  the $L^2$ mean of the quantile functions. 
The  computation of Wasserstein barycenters in multidimensional Euclidean space has been intensively studied  \cp[e.g.,][]{rabin:2011,alvarez:2016,dvurechenskii:2018,peyr:19}, and one of the
most popular  methods utilizes the Sinkhorn divergence  \citep{cuturi:2013}, which is an entropy-regularized version of the Wasserstein distance that allows for computationally efficient  solutions of the barycenter problem, however at the cost of introducing a bias, as is common for regularized estimation.  Due to the gain in efficiency, we adopt this approach in our implementations.  %We first review the straightforward case  $d=1$ and the consider the multivariate case $d\geq 2$ in the next subsection.

For the straightforward case   $d=1$, where  the Wasserstein distance is the $L^2$ distance of the quantile functions  
$W_2^2(\mu_1,\mu_2) = \int_0^1 (F_{\mu_1}^{-1}(t)-F_{\mu_2}^{-1}(t))^2  dt,$ the following result has been established previously, which greatly facilitates the study of conditional barycenters for this case.
%which facilitates to fit models for conditional barycenters and to obtain their convergence properties, using the following known result. 
\begin{prop}[Global and local estimates for $d=1$ \cp{petersen:2019}]\label{wb1d}
	For one-dimensional probability measures  $\mu_{i},\, i=1,2,\ldots,n$, denoting by $\mathcal{Q}$ the space of quantile functions, the  solutions of the global and local Wasserstein barycenter problems  (\ref{ge}) and (\ref{le}) are obtained as  the distributions with the following quantile functions,
	\bea
	F_{\hat{\mu}_{G}(x)}^{-1}(\cdot) = \argmin_{Q\in \mathcal{Q}} \|Q-\frac{1}{n}\sum_{i=1}^{n}s_{iG}(x)F_{\mu_i}^{-1}(\cdot)\|^2_{2}  \\
	F_{\hat{\mu}_{L,h}(x)}^{-1}(\cdot) = \argmin_{Q\in \mathcal{Q}} \|Q-\frac{1}{n}\sum_{i=1}^{n}s_{iL,h}(x)F_{\mu_i}^{-1}(\cdot)\|^2_{2}.
	\eea
\end{prop}

%\subsection{Multidimensional measures}
%We now turn to the case of multivariate probability measures. In practice, 

For the case $d>1$, the distribution is often only observed on a discrete grid, an assumption that is almost universally made for computational implementations.  Accordingly, we  consider an equidistant rectangular grid $(d_1,\ldots,d_m)$ on the domain $M \subset R^d$. Since $M$ is assumed to be compact, there exists a rectangle $\tilde{M}$ such that $M\subset \tilde{M}$. The grid $(d_1,\ldots,d_m)$ is constructed by first creating an equidistant rectangular grid $(d_1,\ldots,d_{\tilde{m}})$ on $\tilde{M}$ and then $\{d_1,\ldots,d_m\}= \{d_1,\ldots,d_{\tilde{m}}\}\cap M$, then representing the observed densities $f_i(x)$ as length $m$ vectors $(f_i(d_1),f_i(d_2),\ldots,f_i(d_m))$. Prior to computing  the Wasserstein distance, the densities are approximated by the discrete measure vector $\mathbf{r}_{i} = (\frac{f_i(d_1)}{\sum_{j=1}^{m}f_i(d_i)},\frac{f_i(d_2)}{\sum_{j=1}^{m}f_i(d_i)},\ldots,\frac{f_i(d_m)}{\sum_{j=1}^{m}f_i(d_i)})$.  

The following result quantifies the approximation error in terms of the Wasserstein distance between a continuous measure $\nu$ and its discrete approximation $\nu_r$. 
\begin{prop}[Approximation of a continuous measure with a discrete measure]
	Consider an absolutely continuous  probability measure  $\nu$ on $M$ with density $f_{\nu}$ and an  equidistant rectangular grid  $(d_1,...,d_m)$ 
	with probability mass $\mathbf{r}$ as described above. 
	If the density $f_{\nu}$ is Lipschitz continuous, then
	\bea
	W^2_2(\nu,\nu_{\mathbf{r}}) = O(\min_{k,l, k \ne l}\|d_k-d_l\|^2)
	\eea
	as $\min_{k,l, k \ne l}\|d_k-d_l\|^2 \rightarrow 0$.
\end{prop}

Denoting the pairwise distance matrix of the grid by  $D = (\|d_k-d_l\|^2)_{k,l}$,   the 2-Wasserstein distance can be written as 
\be
W_2^2(\mathbf{r}_i,\mathbf{r}_j) = \min_{S\in U(\mathbf{r}_i,\mathbf{r}_j)}\langle S,D \rangle, \label{disc}
\ee
where $U(\mathbf{r}_i,\mathbf{r}_j) = \{S \in  R_{\oplus}^{m\times m}|S\mathbf{1}_{m}=\mathbf{r}_i,\, S'\mathbf{1}_{m}=\mathbf{r}_j\}$. Then the conditional Wasserstein barycenter estimates become 
\begin{align*}
\hat{\mathbf{r}}_{G}(x) = \argmin_{\mathbf{r}\in \mathcal{D}}  \sum_{i=1}^{n}s_{iG}(x)\min_{S_i\in U(\mathbf{r},\mathbf{r}_i)}\langle S_i,D \rangle \\
\hat{\mathbf{r}}_{L,h}(x) = \argmin_{\mathbf{r}\in \mathcal{D}}  \sum_{i=1}^{n}s_{iL,h}(x)\min_{S_i\in U(\mathbf{r},\mathbf{r}_i)}\langle S_i,D \rangle,
\end{align*}
where $\mathcal{D} = \{(r_1,r_2,\cdots,r_m)|r_k \geq 0,\, \sum_{j=1}^{m}r_k = 1\}$. This problem can typically be solved by linear programming \citep{anderes:2016}. However, with a grid of large size, e.g. $100\times 100$ in $R^2$ this may not be satisfactory, as  linear programming does not scale well  and has a computational complexity of the order  $O(nm^3\log m)$. 

To address this problem, we adopt Sinkhorn divergence, a regularized version of Wasserstein distance,  \cp{cuturi:2013}. Various approaches have been developed to compute Sinkhorn barycenters  \cp[e.g.,][]{cuturi:2014,peyre:2015, cuturi:2016}. We use the algorithm of \cite{peyre:2015} as 
we found it to work  well with negative weights. This algorithm  reduces the computational complexity to roughly $O(m^2)$ and it appeared to be even faster than $\sim m^2$ in our implementations. 
The Sinkhorn divergence between discrete measures $\mathbf{r}_i$ and $\mathbf{r}_j$ is defined as \citep{cuturi:2013}
\begin{equation*}
W_{2,\alpha}^2(\mathbf{r}_i,\mathbf{r}_j) = \min_{S\in U_{\alpha}(\mathbf{r}_i,\mathbf{r}_j)}\langle S,D \rangle,
\end{equation*}
where $U_{\alpha}(\mathbf{r}_i,\mathbf{r}_j) = \{S\in U(\mathbf{r}_i,\mathbf{r}_j)|\mathbf{KL}(S|\mathbf{r}_i\mathbf{r}_j') \leq \alpha \}$ and $\mathbf{KL}(p|q)$ is the Kullback–-Leibler divergence. An equivalent version is 
\begin{equation*}
W_{2,\lambda}^2(\mathbf{r}_i,\mathbf{r}_j) = \min_{S\in U(\mathbf{r}_i,\mathbf{r}_j)}\langle S,D \rangle + \frac{1}{\lambda} \sum_{k=1}^m\sum_{l=1}^m S_{kl}\log(S_{kl}).
\end{equation*}
Here $\alpha$ respectively $\lambda$ are regularization parameters. The implementation of  global and local linear estimates of conditional barycenters then amounts to 
\begin{align*}
\hat{\mathbf{r}}_{G}(x,\lambda) = \argmin_{\mathbf{r}\in \mathcal{D}}  \sum_{i=1}^{n}s_{iG}(x)W_{2,\lambda}^2(\mathbf{r}_i,\mathbf{r}) \\
\hat{\mathbf{r}}_{L,h}(x,\lambda) = \argmin_{\mathbf{r}\in \mathcal{D}}  \sum_{i=1}^{n}s_{iL,h}(x)W_{2,\lambda}^2(\mathbf{r}_i,\mathbf{r}).
\end{align*}

For further details we refer to  \cite{peyre:2015}.  It is straightforward to see that Sinkhorn divergence converges to the ordinary Wasserstein distance as $\lambda\rightarrow \infty$, i.e. $\lim_{\lambda\rightarrow \infty}W_{2,\lambda}^2(\mathbf{r}_i,\mathbf{r}_j) = W_{2,\lambda}^2(\mathbf{r}_i,\mathbf{r}_j)$. The convergence rate of empirical Sinkhorn barycenters to the population Sinkhorn barycenter has been established in \cite{bigot:2019}. However, there are only very few results available regarding the convergence of  Sinkhorn barycenters to  Wasserstein barycenters. The available results are restricted to the  Gaussian case  \cp{del:2020}, for which  an explicit expression for the  Sinkhorn barycenter is available. The problem becomes even harder for global and local estimates that we consider here,   due to the presence of negative weights when forming weighted barycenters, for which there are no results available to date.  Theorem 4 below implies that under additional assumptions  the global and local linear  estimates of conditional barycenters under  the Sinkhorn divergence converge to the corresponding conditional barycenters  under  the Wasserstein distance.

\section{Conditional Barycenters as Geodesics}
A connection of interest emerges between  global linear conditional barycenters  obtained along lines in the predictor space that are the geodesics in this Euclidean space and geodesics in the 2-Wasserstein space $\Om$ in special settings.  In a metric space with metric $d$, a constant speed geodesic $\nu(t) \in \Om, \, t\in[0,1],$ connecting two points $\nu_1$ and $\nu_2$ is characterized by $\nu(0) = \nu_1$, $\nu(1) = \nu_2$ and  $d(\nu(t_1),\nu(t_2)) = |t_1-t_2|d(\nu_1,\nu_2)$. If for any two points in a metric space there exists a geodesic connecting them, the space is a geodesic space \cp{bura:01}. 

The 2-Wasserstein space supported on a compact domain $M\in R^d$ is known to be  geodesic  \cp{ambrosio:2008}. However, the geodesics are typically not unique for discrete and mixed type measures, which is another motivation to consider only spaces of distributions with density functions.  To consider  extrapolation of probability measures, we start with the notion of an  extension of a geodesic.  Given a geodesic $\nu(t)$ defined on $t \in [0,1]$, if the geodesic property as defined above continues to hold for $\nu(t)$ with $t \in[t_1,t_2]$, where $t_1<0<1<t_2$, we say that the geodesic can be extended from $[0,1]$ to $[t_1,t_2]$ \cp{ahidar:2019}.
Push-forward maps and optimal transport play an important role in understanding and extending geodesics in the  Wasserstein space. For  absolutely  continuous probability measures  $\nu_1$ and $\nu_2$ on $M$,  an optimal transport map $T$ from $\nu_1$ to $\nu_2$ is a push-forward map  $\nu_2= T_{\#}\nu_1$ such that $W_2^2(\nu_1,\nu_2) = \int_{M}\|x-T(x)\|d\nu_1$.

In \cite{agueh:2011}, the weighted 2-Wasserstein barycenter solutions in equation (\ref{wwb}) are established for each of three special settings: $n=2$; $d=1$; and Gaussian measures. When $n=2$, the  weighted barycenters of $\nu_1,\nu_2$ and  with weight $w_1$ associated with $\nu_1$ varying from $0$ to $1$ and weight $w_2=1-w_1$  associated with $\nu_2$, form  McCann's interpolant \citep{mccann:1997}
\bea
((1-w_1)id + w_1T)_{\#}\nu_1, \quad w_1 \in [0,1], 
\eea
which is the geodesic connecting $\nu_1$ and $\nu_2$. The following result establishes a connection between the global model 
for conditional barycenters and predictors and responses that lie on matching geodesics in their respective spaces. 

\begin{thm}[Geodesic interpolation and extrapolation]
	Consider the sample $\{x_i,\nu_i\}$, $x_i\in R^q$, $i=1,\dots,n$,  and assume there exists a  geodesic  $\nu(t)$, $ t\in [0,1]$
	that uniquely connects measures $\nu(0)$ and $\nu(1)$, such that the responses $\nu_i$ are located on this geodesic, i.e., for each $\nu_i$ there exists a $t_i$ so that 
	$\nu_i=\nu(t_i)$. 
	Assume furthermore that the predictors $x_i$ are located on a geodesic in the Euclidean space, which is a line, such that  $x_i = t_i\mathbf{c}+\mathbf{b}$ for any $i,j = 1,2,\ldots,n$,  where $\mathbf{b},\mathbf{c}\in R^p$ are constant vectors. Then the global model (\ref{ge}) of the conditional Wasserstein barycenter recovers the geodesic $\nu(t),\, t\in[0,1]$ when $x$ is on the line from $\mathbf{b}$ to $\mathbf{c}+\mathbf{b}$. If the geodesic is extendable from $[0,1]$ to $[s_1,s_2]$ and the extension is  unique in the sense that it is the only geodesic connecting $\nu(s_1)$ and $\nu(s_2)$, then the global model recovers the extended  geodesic with $x$  on the line from $\min(t_i)\mathbf{c}$ to $\max(t_i)\mathbf{c}$.
\end{thm}

We note that it is well-known that for Euclidean predictors and responses  where the responses lie on a line or linear surface as predictors vary (i.e., the model is correctly 
specified and there is no noise) satisfy assumptions analogous to those in Theorem 1 and then a least squares fit applied to such data recovers this very line or surface. The above result extends  this basic fact to the case of responses in the Wasserstein space. 

\begin{cor}[Geodesic interpolation and extrapolation for two points]
	If $n=2$, consider the sample $(x_1,\nu_1),\,(x_2,\nu_2)$ , $\mathbf{x}$ the vector connecting $x_2-x_1$ and $\nu_1,\nu_2$ absolutely continuous measures. Then the  global estimation (\ref{ge}) of the conditional Wasserstein barycenter with $x= t(x_2-x_1)+ x_1$ corresponds to the geodesic path $\nu(t)$ from $\nu_1$ to $\nu_2$, $t\in [0,1]$ with $\nu(0) = \nu_1$ and $\nu(1) = \nu_2$. If the geodesic is extendable to $[s_1,s_2]$ with $s_1<0,\, s_2 > 1$ and is unique in the sense that it is the only geodesic connecting $\nu(s_1)$ and $\nu(s_2)$, then the global model recovers this geodesic with $x \in [x_1+s_1(x_2-x_1),x_2 + s_2(x_2-x_1)]$.
\end{cor}

While the connection between weighted Wasserstein barycenters and interpolation has been previously studied \cp{agueh:2011}, the above  results also provide  an approach to extrapolate  measures. The implementation of this approach to measure extrapolation is demonstrated  in the simulation section below. 

\section{Convergence of Conditional Barycenter Estimates}

To study  convergence for M-estimators such as the local and global estimators for conditional barycenters, a  curvature condition at the true minimizer is essential. Such a condition is however generally not available for Wasserstein barycenters when $d>1.$ For example, assume $V_1,V_2,V_3,V_4$ are the vertexes of a regular tetrahedron in $R^3$ such that $\|V_i-V_j\|,\, i\neq j$ is a constant and $\nu_1$ and $\nu_2$ are the discrete measures with   mass $(\frac{1}{2},\frac{1}{2})$ at $V_1,V_2$ and  at $V_3,V_4$,  respectively. Then  the Wasserstein barycenter of $\nu_1$ and $\nu_2$ is not unique, where the discrete measures with (a)  mass $(\frac{1}{2},\frac{1}{2})$ at $\frac{V_1+V_3}{2},\frac{V_2+V_4}{2}$ and (b)  with the same mass  at  $\frac{V_1+V_4}{2},\frac{V_2+V_3}{2}$ are both minimizers of $W_2^2(\mu,\nu_1)+W_2^2(\mu,\nu_2)$. Since absolutely continuous measures can be chosen to be arbitrarily close to the discrete measures $\nu_1$ and $\nu_2$, this makes a curvature condition unattainable for  continuous measures. 

To overcome this problem, we make use of an approach of  \cite{boissard:2015}. This involves the  deformation class of a base measure $\nu_0$, defined as $\{T_\#\nu_0\}$, where the transports $T$ belong to a class of transport maps $\mathcal{T}(M)$ that satisfy certain conditions. In the following,  $GCF(M)$ is the  set of all gradients of convex functions, that is to say the set of all maps $T:M \rightarrow M$ such that there exists a proper convex lower semi-continuous function $\phi:M \rightarrow R$ with $T =\triangledown \phi$. Convergence rates for the global and local estimates of conditional barycenters can then be obtained  under the following conditions.

\noindent (LP) The densities $f$ of the absolutely continuous measures $\nu$ are $\alpha$ differentiable (i.e., the derivatives of order $[\alpha]$ exist and are Lipschitz continuous of order $\alpha -[\alpha]$) and have  uniformly bounded partial derivatives, where $\alpha > \frac{d}{2}$.

\noindent (CD) The marginal density $g$ of $X$ and the conditional densities $g_{\nu}$ of $X|\nu$ exist and are twice continuously differentiable, the latter for all $\nu$, and $\sup_{x,\nu}|g^{''}_{\nu}(x)|<\infty$. Additionally, for any open $U\subset \Omega$, $P(\nu \in U |X=x)$ is continuous as a function of $x$.

\noindent (AD) The random measure $\nu$ is generated from $T_\#\nu_0$, where $T$ is a random map 
in $\mathcal{T}(M)$. The class of transport maps  $\mathcal{T}(M)$ is convex and compact and has the following properties:
\begin{enumerate}[i]
	\item The identity map satisfies  $Id \in \mathcal{T}(M)$,
	\item $\mathcal{T}(M) \subset GCF(M)$,
	\item Any $T_i \in \mathcal{T}(M)$ is one-to-one and onto,
	\item For any $T_i,T_j \in \mathcal{T}$, $T_i\circ T_j^{-1} \in \mathcal{T}(M)$.
\end{enumerate}
Then $\mathcal{T}(M)$ is referred to as class of admissible deformations \cp{boissard:2015}. Some examples of admissible deformations include location-scale families with commuting covariance matrices and measures with the same copula function \cp{bigo:19,pana:19}.

For local estimates,  we require the following  additional condition on the kernel.

\noindent (KN) The kernel $K$ used in  local \F regression is a symmetric probability density function, such that with  $K_{jm} = \int_{R}K^j(u)u^m du$, $|K_{14}|$ and $|K_{26}|$ are both finite.

\begin{thm}[Convergence rate of global model fits]
	If $\nu$ satisfies condition (AD) and the corresponding densities satisfy $f_{\nu}$ satisfy (LP)  
	with $\alpha > \frac{d}{2}$, then 
	\bea
	W_2^2(\mu_G(x) ,\hat{\mu}_{G}(x)) = O_p(n^{-1}).
	\eea
\end{thm}
This shows that using the global estimate for conditional barycenters to track a global target leads to a parametric rate of convergence that does not depend on the dimension of the underlying space, if the distributions are sufficiently smooth.  
\begin{thm}[Convergence rate of local model fits]
	Under condition (AD) , (CD), (KN),  if $h\sim n^{-\frac{1}{5}}$, then when $q =1$
	\bea
	W_2^2(\mu_0(x) ,\hat{\mu}_{L,h}(x)) = O_p(n^{-4/5}).
	\eea
\end{thm}
As in usual real-valued nonparametric regression, the rate of convergence is seen to deteriorate with increasing predictor dimension $q$ and this regression approach is thus subject to the curse of dimensionality, 
but otherwise conforms with the known optimal rate of convergence that can be achieved  for real responses, irrespective of the dimension $d$ of the domain of the distribution. 
\begin{thm}[Convergence of Sinkhorn divergence estimates]
	Under condition (AD) and \\$\mathcal{D} = \{(\frac{f(d_1)}{\sum_{j=1}^mf(d_j)},\ldots,\frac{f(d_j)}{\sum_{j=1}^mf(d_j)})|\nu_{f} \in \Omega\}$, if $\mathbf{r}_i$ is the discrete measure obtained by evaluating  $f_{\nu_i}$ on the discrete grid $(d_1,d_2,\ldots,d_m)$ it holds that 
	\bea
	\lim_{\rho\rightarrow \infty}\lim_{\zeta\rightarrow 0}W_2(\nu_{\hat{\mathbf{r}}_{G}(x,\rho)},\hat{\mu}_{G}(x)) = 0,\\
	\lim_{\rho\rightarrow \infty}\lim_{\zeta\rightarrow 0}W_2(\nu_{\hat{\mathbf{r}}_{L,h}(x,\rho)}, \hat{\mu}_{L,h}(x)) = 0,
	\eea
	where $\nu_{\hat{\mathbf{r}}_{G}(x,\rho)}$ and $\nu_{\hat{\mathbf{r}}_{L,h}(x,\rho)}$ are discrete measures on $(d_1,\ldots,d_m)$ with probability mass $\hat{\mathbf{r}}_{G}(x,\rho)$, respectively  $\hat{\mathbf{r}}_{L,h}(x,\rho)$, and $\zeta$ the length of the diagonal of the small rectangles (bins) defined by the discrete  grid.\end{thm}

This theorem guarantees the effectiveness of approximating the global and local conditional barycenter estimates with their Sinkhorn approximations. These approximations  reduce computation time drastically and are seen in simulations to provide quite reasonable approximations for the  true conditional barycenters. There is a basic trade-off between approximation quality and computation time; as $\lambda$ increases,  the algorithm takes longer to converge to a solution. There is also  a practical limit, as for fixed sample size $n$ for large enough $\lambda$ the approximation is empirically found to break down, leading to unacceptable results.  We found in simulations  that $\lambda \in [\frac{1}{4},1]$ tends to lead to  stable performance and reasonably good approximations for global and local estimates (\ref{ge}), (\ref{le}).

\FloatBarrier 
\section{Simulations}
\subsection{The one-dimensional case}
To investigate the finite sample behavior of the proposed Wasserstein interpolation, we conducted various simulations.  It is instructive to start with  a report on the results of a simulation for  the straightforward and well-explored one-dimensional case.  One-dimensional  predictors $X$ were chosen as uniform $[0,1]$ r.v.s and the response measures to take values in a location-scale family with densities $\sigma^{-1}f_0(\sigma^{-1}(w-\alpha))$ with $f_0(w) \propto \max(0,0.05-(w-0.4)^2))$. The random responses were generated from $\alpha|X=x \sim N(x,0.01)$, $\sigma^2|X=x \sim \frac{5}{6}N(0.1+0.1x,0.01)$, with $[0,1]$ as the domain of all random measures. 
According to  Example 1 in section 2.1, for $\{X,\nu\}$ generated as above, the densities of the conditional Wasserstein barycenter (\ref{cwb}) and of  the global model $\mu_G(x)$ (\ref{gm}) coincide, and correspond to  $(1+0.02x)^{-1/2}f_0((1+0.02x)^{-1/2}(w-0.4-0.2x))$.

\begin{figure}[!htb]
	\centering
	\includegraphics[width=8cm]{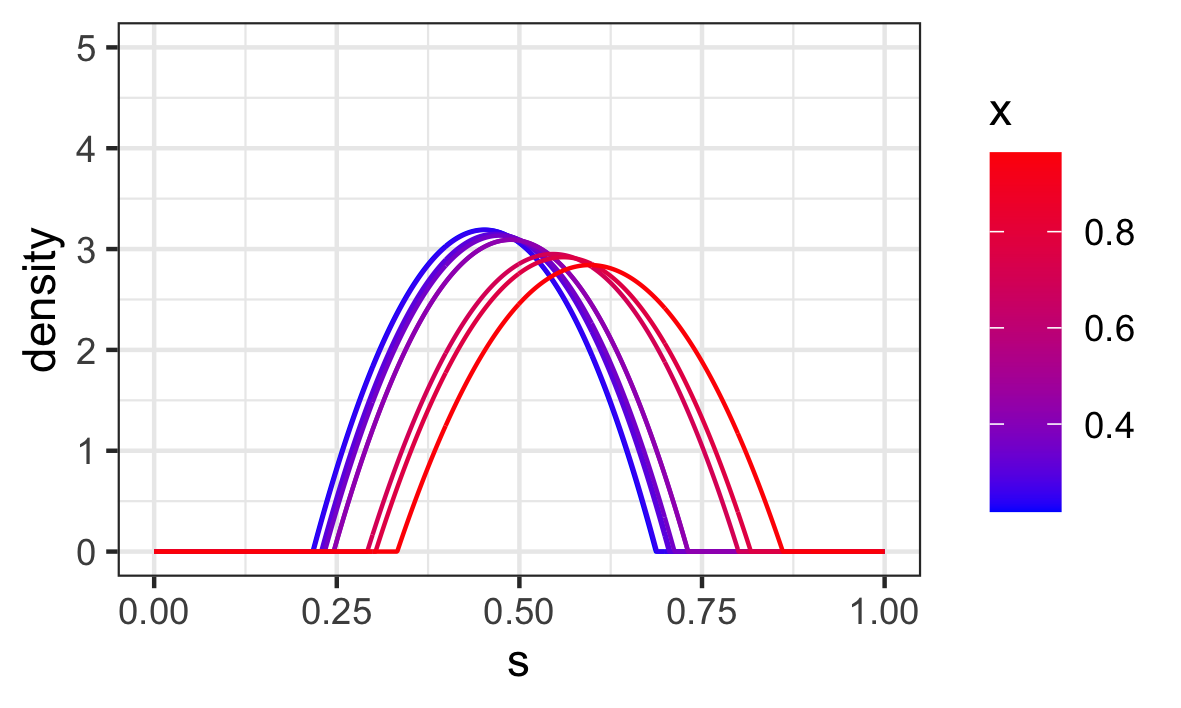}
	\caption{An example of  $10$  response densities with randomly sampled predictor levels.} % for the one-dimensional simulation.}
	\label{f3-1}
\end{figure}

The global and local models were fitted for sample sizes $n= 50,100,150,200$;  the bandwidth  for the local model was chosen as  $0.1$, and  $100$ Monte Carlo runs were performed in each setting;  Figure \ref{f3-1} displays 10 randomly selected responses. The quality of the interpolation was evaluated by mean integrated Wasserstein error (MIWE) over the interpolation domain, defined as 
\be
\text{MIWE}_{i} = E\int_{0}^{1}W_2^2(\hat{\mu}_{G}(x),\mu_G(x))dx,  \label{miwe} 
\ee
where for any $x \in [0,1]$ the fitted model  $\hat{\mu}_{G}(x)$ corresponds to an interpolation and to an extrapolation 
for $x>1$ or $x <0$,  and  estimated by the empirical MIWE (EMIWE) over $M_c$ Monte Carlo runs,
\be
\text{EMIWE}_{i} = \frac{1}{M_c}\sum_{l=1}^{M_c}\int_{0}^{1}W_2^2(\hat{\mu}_{G,l}(x),\mu_G(x))dx.   \label{emiwe}\ee
To evaluate extrapolation on the intervals $[-0.5,0]$ and $[1,1.5]$, we analogously define 
\bea
\text{MIWE}_{e} = E\int_{[-0.5,0] \cup [1,1.5]}W_2^2(\hat{\mu}_{G}(x),\mu_G(x))dx
\eea
and estimates 
\bea
\text{EMIWE}_{e} = \frac{1}{M_c}\sum_{l=1}^{M_c}\int_{[-0.5,0] \cup [1,1.5]}W_2^2(\hat{\mu}_{G,l}(x),\mu_G(x))dx.
\eea

\begin{figure}[!htb]
	\centering
	\includegraphics[width=7.5cm]{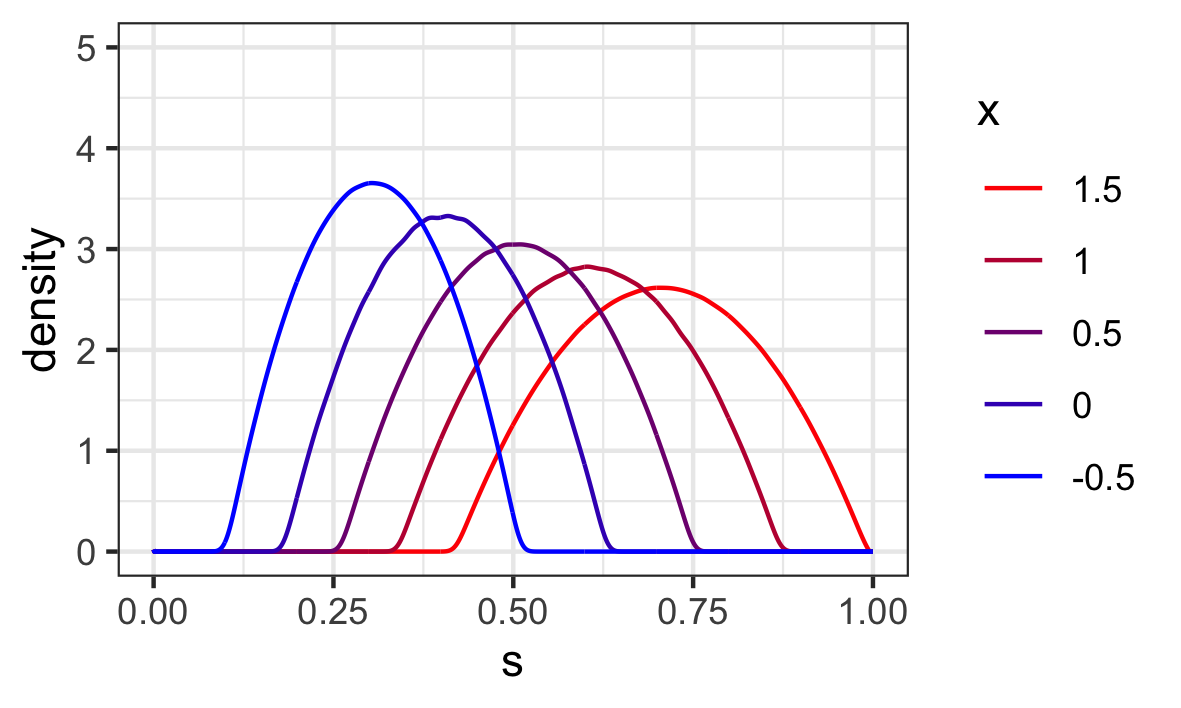}
	\includegraphics[width=7.5cm]{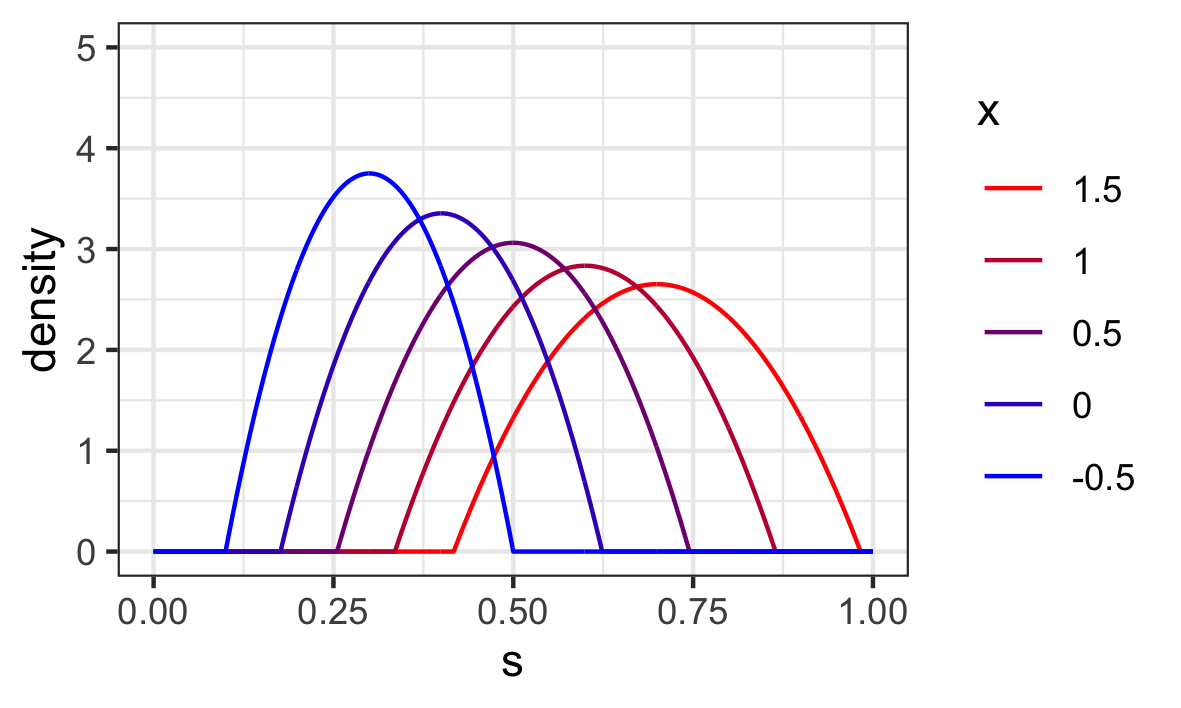}
	\caption{Left panel: Interpolating distributions obtained by fitting the global model. Right panel: True model, with $x = -0.5,0,0.5,1,1.5$. For $x=-0.5$ and $x=1.5$
the fits correspond to 
	extrapolation, as $[0,1]$ is the predictor domain.}
	\label{f3-2}
\end{figure}

When applying the local model,  we only consider interpolation since local fitting is not suited for  extrapolation. The simulation results are shown in Table \ref{tb1} and demonstrate the superior performance of the global model in comparison with the local model for interpolation, which is expected as the true model is the same as global model. Also not unexpectedly,  extrapolation has a much larger error than interpolation. An example of the fitting results at $x = 0,0.25,0.5,0.75,1$ with sample size $100$ is shown in figure \ref{f3-2}, demonstrating very good performance of the proposed interpolation in this setting. 
\vspace{0.1in}
\begin{table}[!htb]
	\centering
	\begin{tabular}{|c|c|c|c|c|}
		\hline 
		Sample size & 50 & 100 & 150 & 200 \\ 
		\hline 
		Global estimate extrapolation ($\text{EMIWE}_{e}$) &  0.00228  & 0.000906 & 0.000713 & 0.000543 \\ 
		\hline 
		Global estimate interpolation ($\text{EMIWE}_{i}$)& 0.000576 &  0.000279 &  0.000217  & 0.000178 \\
		\hline
		Local estimate interpolation ($\text{EMIWE}_{i}$)& 0.00226 &  0.00104 & 0.000599  & 0.000454 \\
		\hline
	\end{tabular}
	\caption{Empirical mean integrated Wasserstein errors (EMIWE) for different sample sizes.}
\label{tb1}
\end{table}

\FloatBarrier 
\subsection{The two-dimensional case}
For this simulation, we  implemented the algorithm in section 3 and constructed distributions  $\alpha|x$ that were obtained by Gaussians that were supplied with changing means and covariances in dependence on a variable $x \in [0,1]$,  truncated on the compact support $[0,1]\times[0,1]$. %  (which covers more than $0.999$ of the probability mass of the standard Gaussian).
% It is well known that when the support is fixed, the uniform distribution achieves the minimum entropy and there is bias that drags the minimizer towards uniform. 
A scalar predictor $X$ was generated from a uniform distribution on $[0,1]$, and the distributional trajectories as  $\alpha|X=x \sim N ((0.4x+0.3\quad 0.4x+0.3)^T,\Sigma(x))$ %  mathbf{I}_2)$
with covariance matrix  $\Sigma(x) = V\Lambda V'$, where $V = \begin{bmatrix}
\frac{\sqrt{2}}{2}& \frac{\sqrt{2}}{2}\\
-\frac{\sqrt{2}}{2}& \frac{\sqrt{2}}{2}
\end{bmatrix}$, $\Lambda =  \diag(\lambda_1,\lambda_2)$ and $(\lambda_1,\lambda_2)|X=x \sim \frac{1}{100}N((1+0.5x\quad 1-0.5x)',\, 0.01\mathbf{I}_2)$. It is easy to check that in this case the global model agrees with the true model. We performed simulation  experiments for sample sizes $n=50,100,150,200$ on a $101$ by $101$ equidistant grid on $[0,1]\times [0,1]$, selecting the Sinkhorn regularization parameter as $\rho = \frac{2}{5}$.

The evaluation of the performance of the global model fit in the two-dimensional case is computationally expensive; due to the  $101 \times 101 = 10201$ grid, the cost  and joint measure matrices have dimension   $10201 \times 10201$ and calculating the exact Wasserstein distance is extremely time consuming. It is therefore expedient to use a shortcut, the mean integrated Sinkhorn error (MISE),
\be
\text{MISE}_{i} = E\int_{0}^{1}W_{2,\rho}^2(\hat{\mu}_{G}(x),\mu_G(x))dx,   \label{mise}\ee
where for $x \in [0,1]$, $ \hat{\mu}_{G}(x)$ is the  interpolation obtained from the global fit and for $x>1$ or $x <0$ these fits are extrapolations. Its  empirical counterpart obtained from  $M_c$ Monte Carlo runs is the empirical MISE, 
\be
\text{EMISE}_{i} = \frac{1}{M_c}\sum_{l=1}^{M_c} \int_{0}^{1}W_{2,\rho}^2(\hat{\mu}_{G,l}(x),\mu_G(x))dx.   \label{emise}\ee
 
\begin{figure}[!htb]
	\centering
	\includegraphics[width=6.5cm]{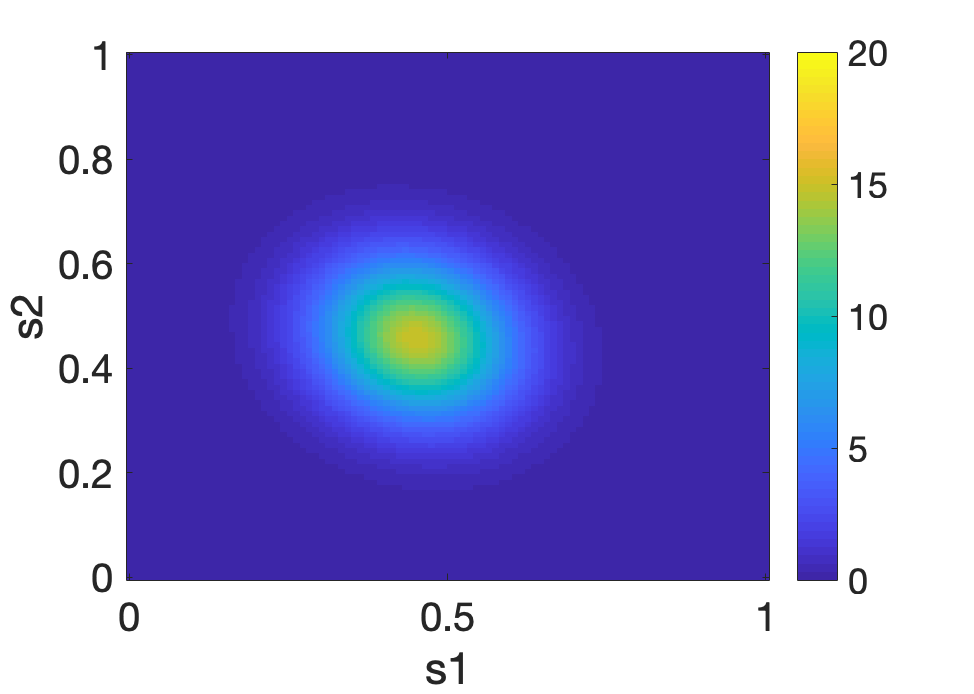}
	\includegraphics[width=6.5cm]{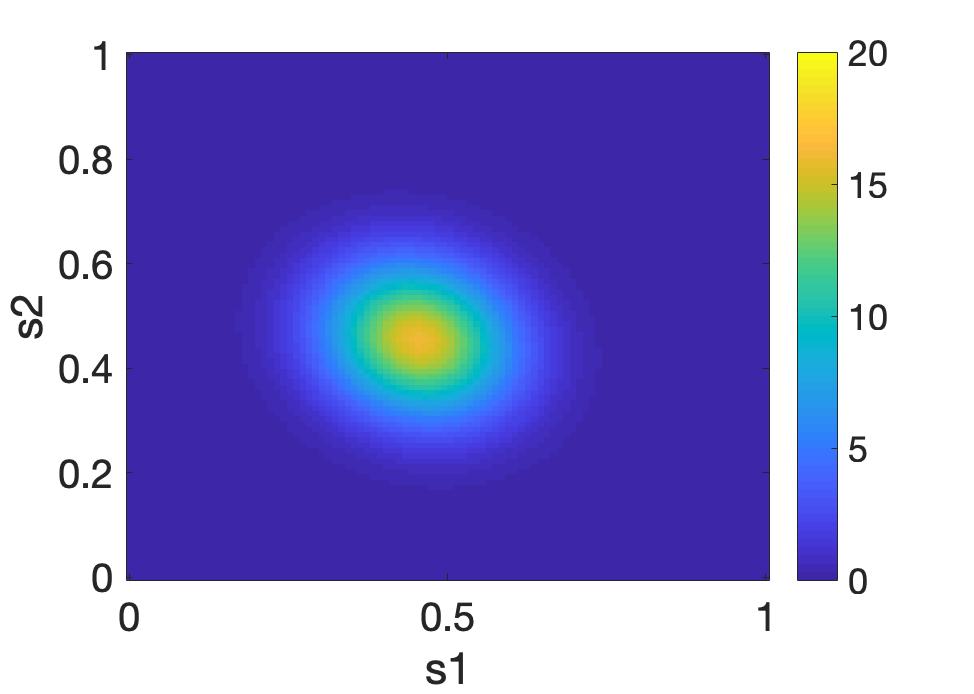}\\
	\includegraphics[width=6.5cm]{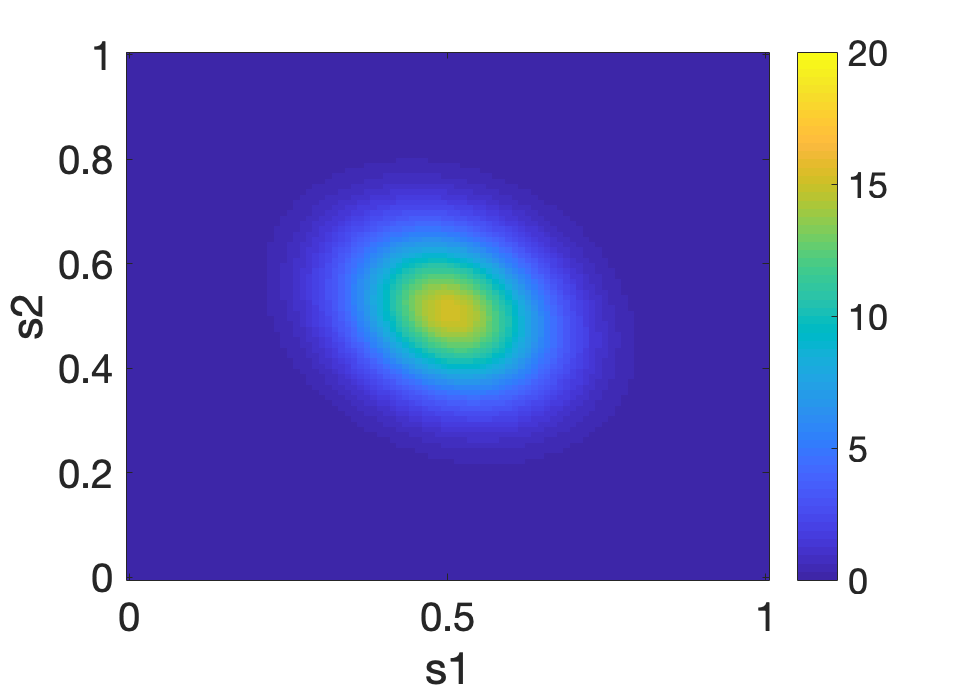}
	\includegraphics[width=6.5cm]{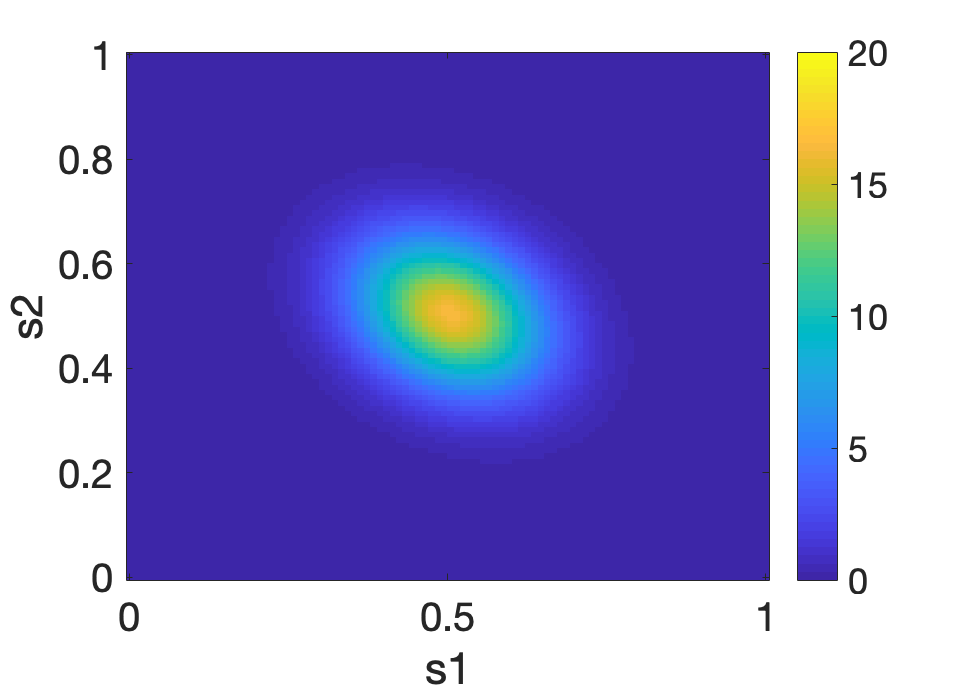}\\
	\includegraphics[width=6.5cm]{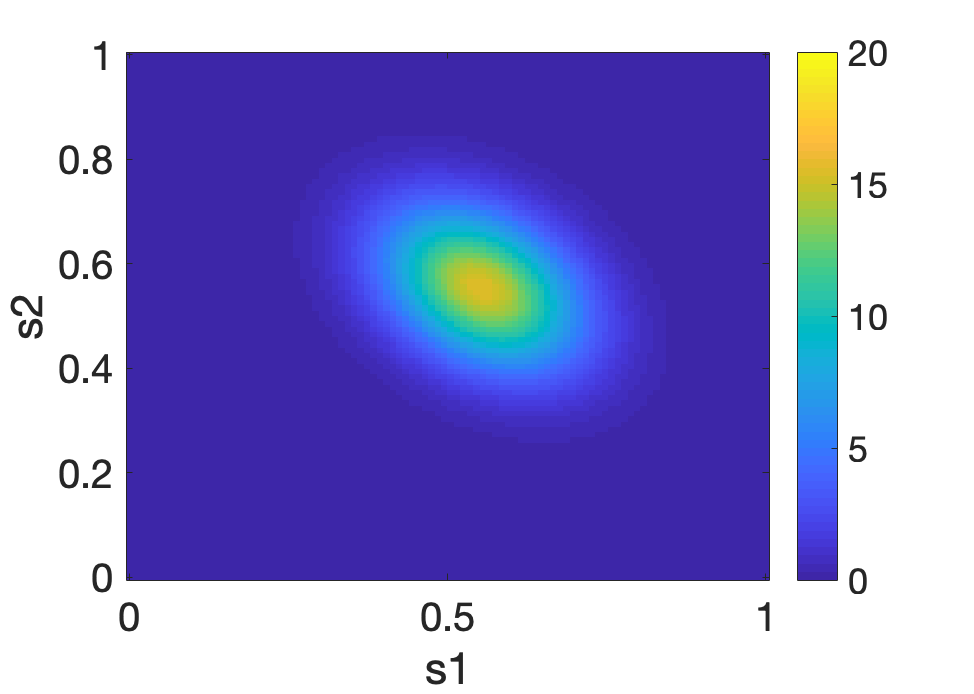}
	\includegraphics[width=6.5cm]{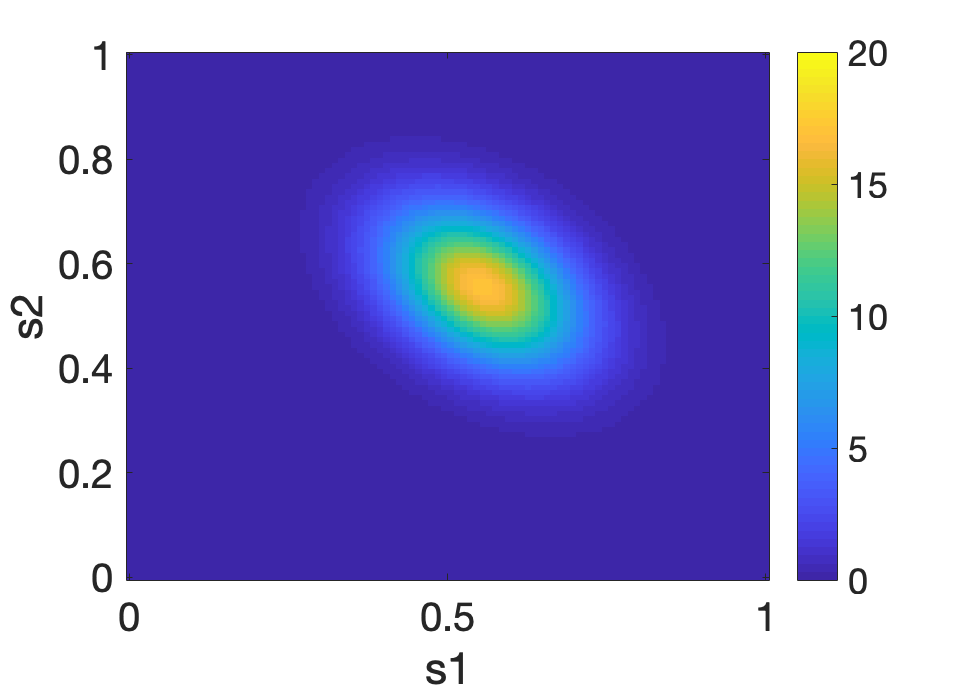}\\
	\caption{Interpolated distributions (left panels) obtained from fitting the global model for 2-dimensional distributions in the simulation setting described in the text with $n=50$ and exact Wasserstein geodesics (right panels), at predictor levels $x=0.25$ (top panels),   $x=0.5$ (middle panels) and $x= 0.75$ (bottom panels).}
	\label{f3-3}
\end{figure}

\begin{figure}[!htb]
	\centering
	\includegraphics[width=6.5cm]{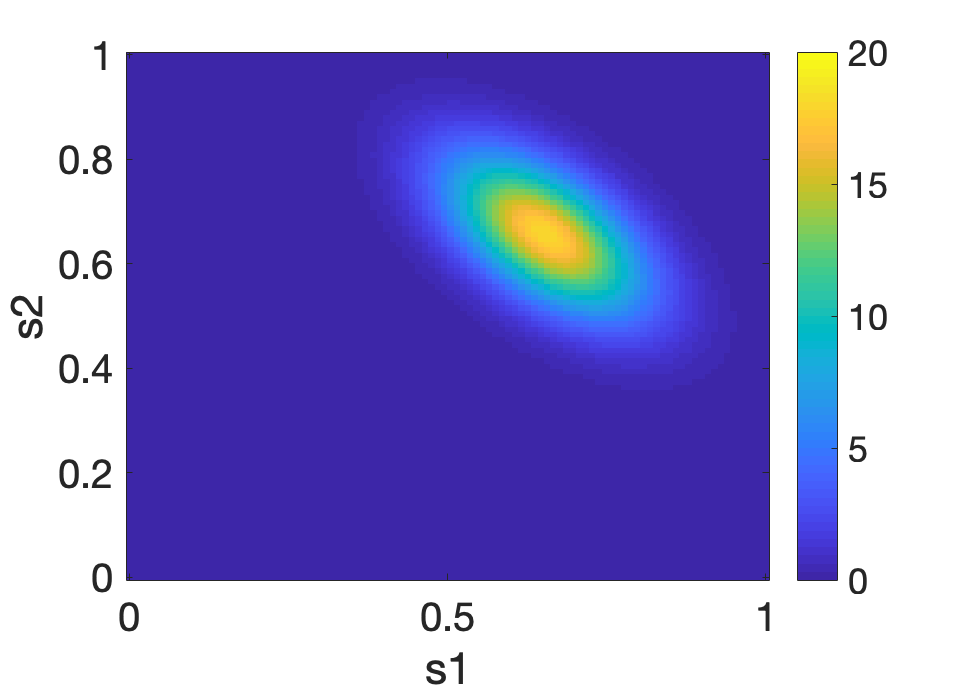}
	\includegraphics[width=6.5cm]{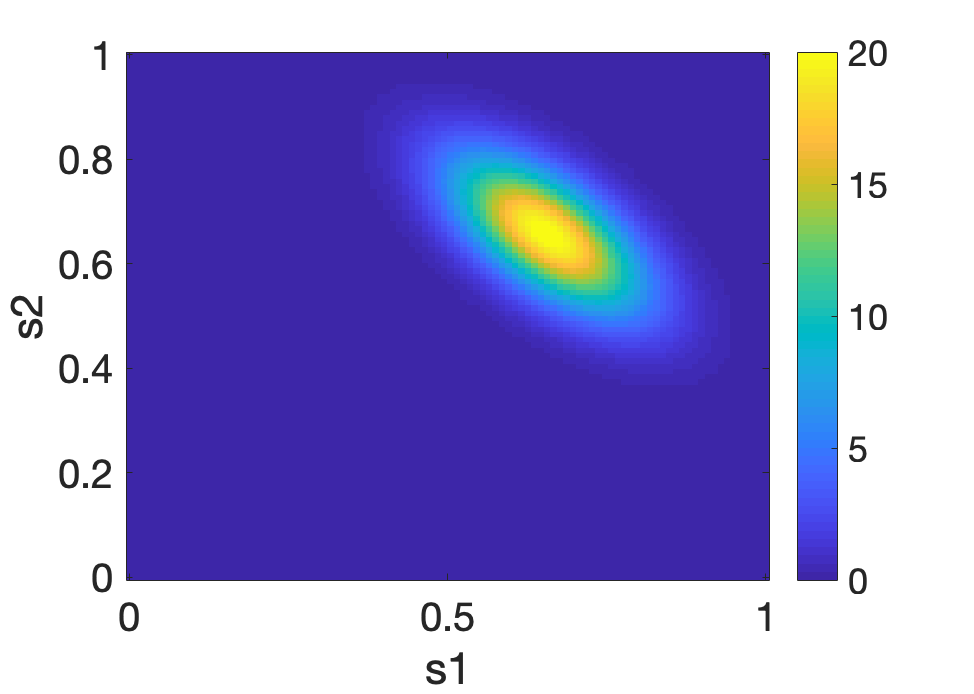}\\
	\includegraphics[width=6.5cm]{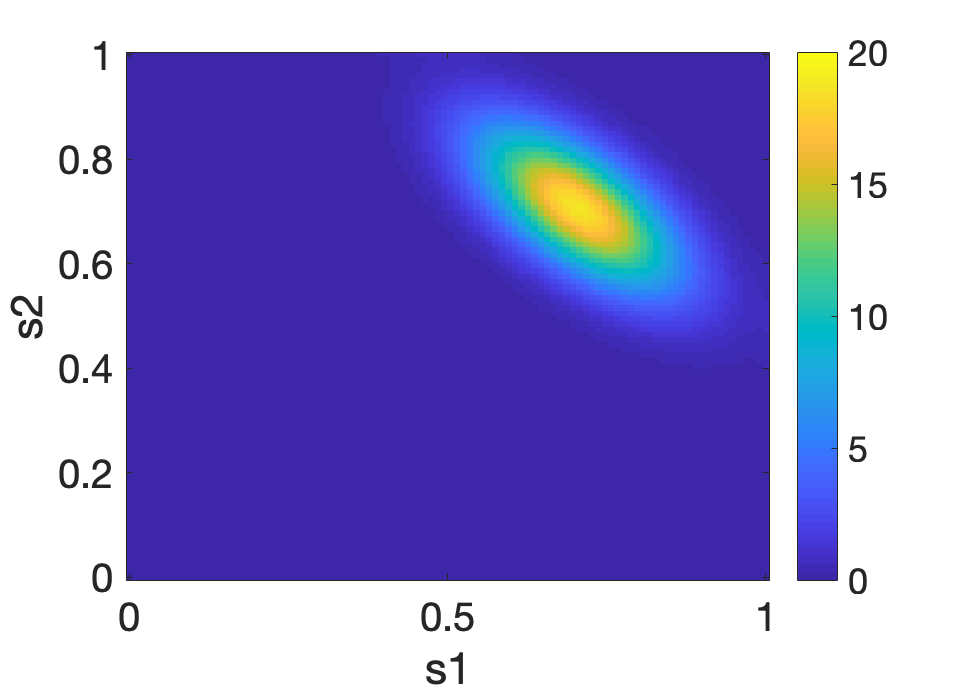}
	\includegraphics[width=6.5cm]{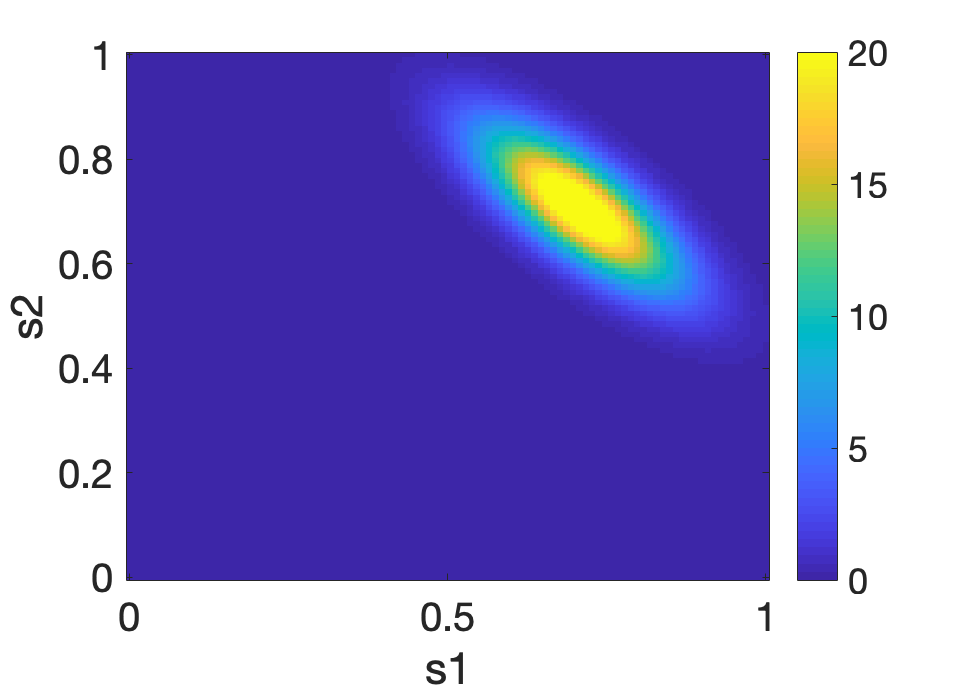}\\
	\caption{Right side extrapolations obtained  by fitting the global model (left panels) for 2-dimensional simulated data with $n=50$, where predictors are randomly sampled on $[0,1]$. available in $[0,1]$, and true extrapolated Wasserstein geodesic (right panels), for mild extrapolation at $x=1.25$ (upper panels) and more extreme extrapolation at $x= 1.5$ (lower panels).}
	\label{f3-4_1}
\end{figure}

\begin{figure}[!htb]
	\centering
	\includegraphics[width=7cm]{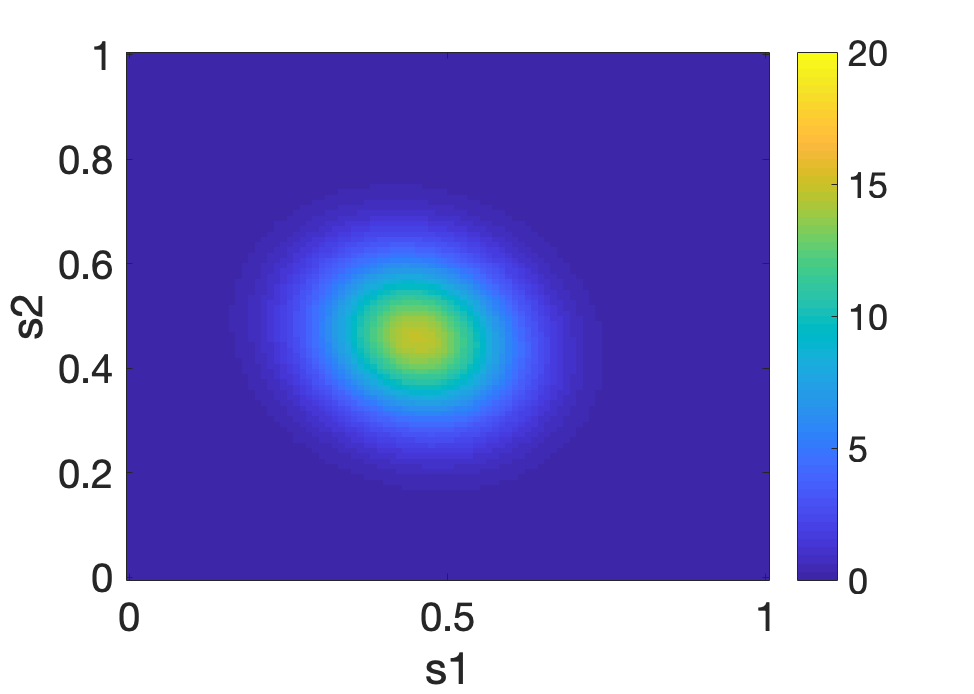}
	\includegraphics[width=7cm]{true0_25}\\
	\includegraphics[width=7cm]{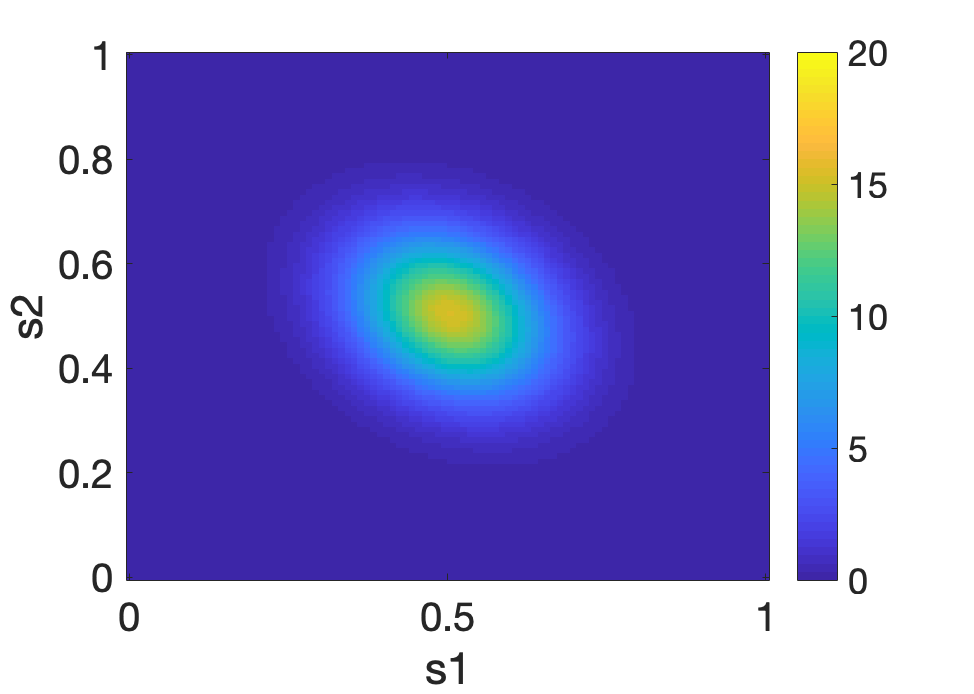}
	\includegraphics[width=7cm]{true0_5}\\
	\includegraphics[width=7cm]{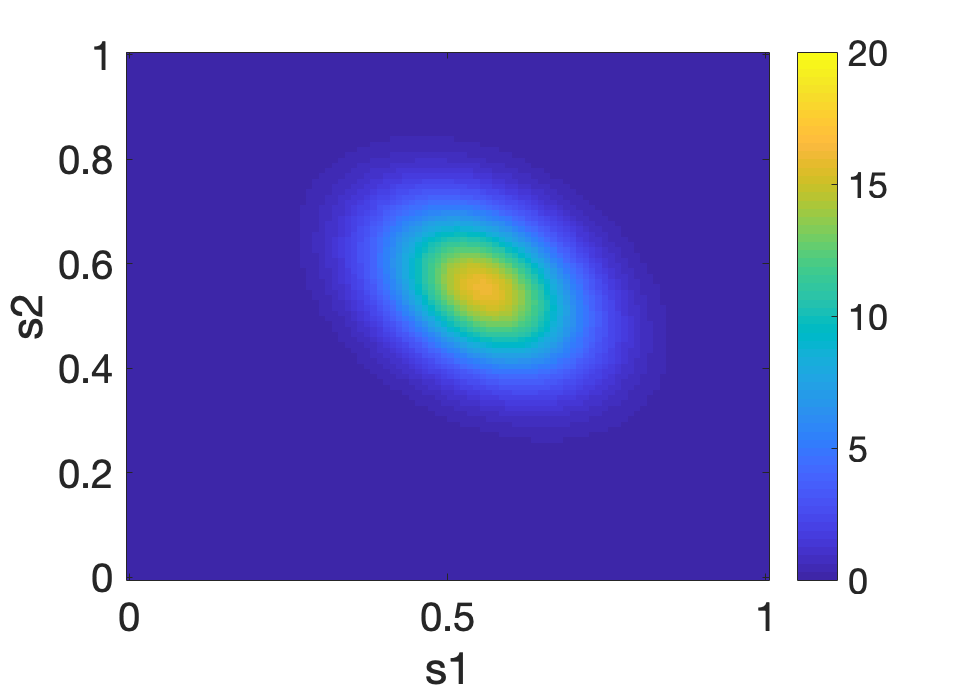}
	\includegraphics[width=7cm]{true0_75}\\
	\caption{Interpolated distributions (left panels) obtained from fitting the local model for 2-dimensional distributions with $n=50$ and exact Wasserstein geodesics (right panels), at predictor levels $x=0.25$ (top panels),   $x=0.5$ (middle panels) and $x= 0.75$ (bottom panels).}
	\label{f3-5}
\end{figure}

The results of the global fits  are visualized in Figures \ref{f3-3} and \ref{f3-4_1}, while those for extrapolation on the left side are in  Figure \ref{f3-4_2}. The local model fits for the same data are shown in Figure  \ref{f3-5}. This indicates that even for small sample sizes such as  $n=50$ these fits work surprisingly well for both interpolation and extrapolation. The  EMISEs of global and local fits  for  $100$ Monte Carlo runs with $\rho = \frac{2}{5}$ in Table \ref{tb2} demonstrate that  EMISE is dominated by  the regularization parameter $\rho$; this conforms with the observation in \ci{janati:2020} that the weighted Sinkhorn barycenters are blurred compared with the actual Wasserstein barycenters.

\vspace{0.2in}
\begin{table}[!htb]
	\centering
	\begin{tabular}{|c|c|c|c|c|}
		\hline 
		Sample size  & $n=50$ & $n=100$ & $n=150$ & $n=200$\\ 
		\hline
		$\rho$ & $\frac{1}{\rho} = 5.5$ & $\frac{1}{\rho} = 4.5$ & $\frac{1}{\rho} = 3.5$ & $\frac{1}{\rho} = 2.5$\\
		\hline 
		Global  extrapolation& 0.0448 & 0.0429 & 0.0413 & 0.0402\\ 
		\hline 
		Global  interpolation& 0.0449  & 0.0432 & 0.0419 & 0.0408  \\
		\hline
		Local  interpolation& 0.0449 & 0.0430 & 0.0414  & 0.0403 \\
		\hline
	\end{tabular}
	\caption{EMISEs (\ref{emise})   for different sample sizes and methods for 2-dimensional distributions.} 
	\label{tb2}
\end{table}

Additional simulations for response distributions with heavier tails can be found in the Supplement. 

\section{Applications}
\FloatBarrier 
\subsection{BLSA data}
The Baltimore Longitudinal Study of Aging (BLSA) data \url{https://www.blsa.nih.gov/} contains various health data collected over a part of the lifespan of included individuals. We extracted  systolic (SBP) and diastolic blood pressure (DBP) measurements, where $8000$ measurements (age of the individual at the time of the measurement, SBP and DBP)  were available for $2801$ individuals.  The number of visiting times for an individual varies from $1$ to $26$ and the range of ages at which measurements were recorded is $[17,75]$. Here we use age as predictor and construct the density responses by kernel density estimators for the joint distribution of SBP and DBP, where in a preprocessing step the data is binned by age at the time of measurements over $20$ equidistant bins between $30$ to $75$; the bandwidth for the kernel smoothing step was chosen using the method of \ci{sheather:1991}, for both this as well as the following data illustration.  %The range of age is chosen such that their are enough observations to construct the densities. The grid on which the 
The joint 2-dimensional  densities were estimated over  $51$ equidistant grid points in each direction over the  domain $[30,130]$ for SBP and $[70,210]$ for DSP.

\begin{figure}[!htb]
 	\centering
	\includegraphics[width=7cm]{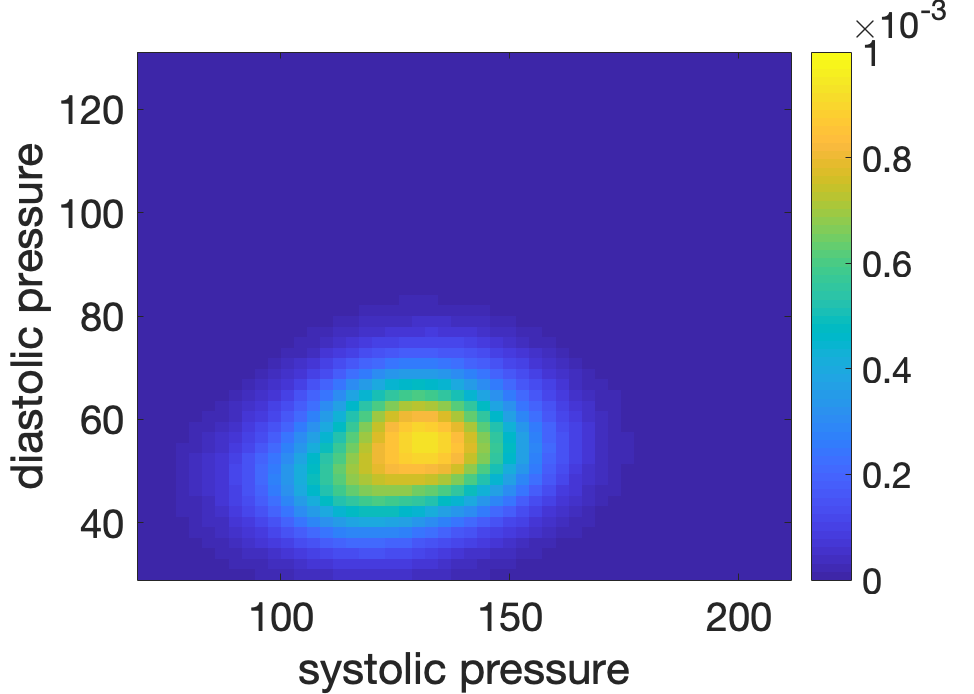}
 	\includegraphics[width=7cm]{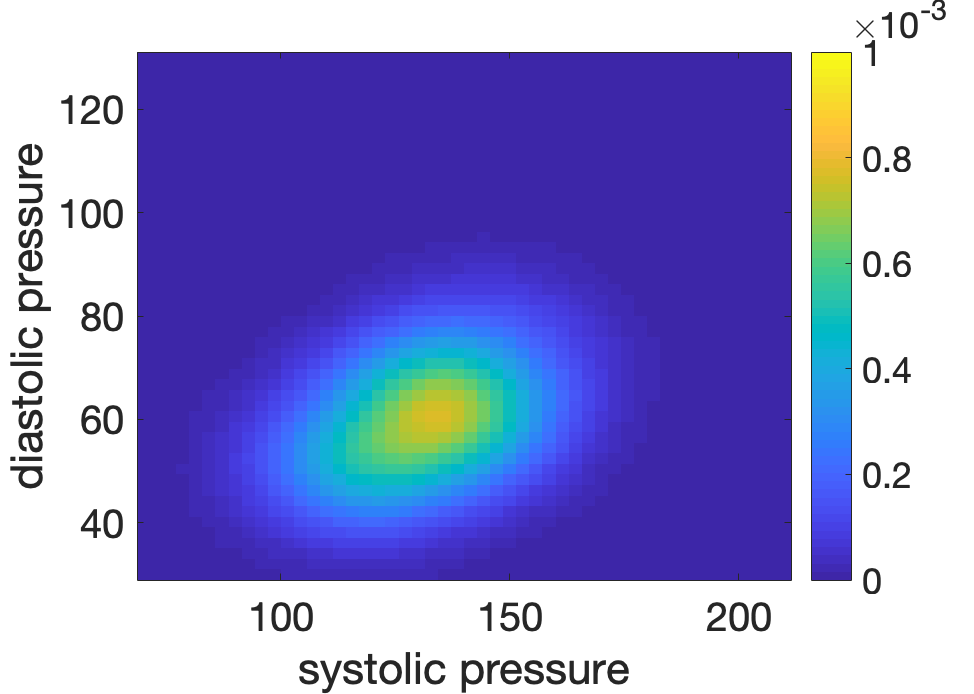}\\
	\includegraphics[width=7cm]{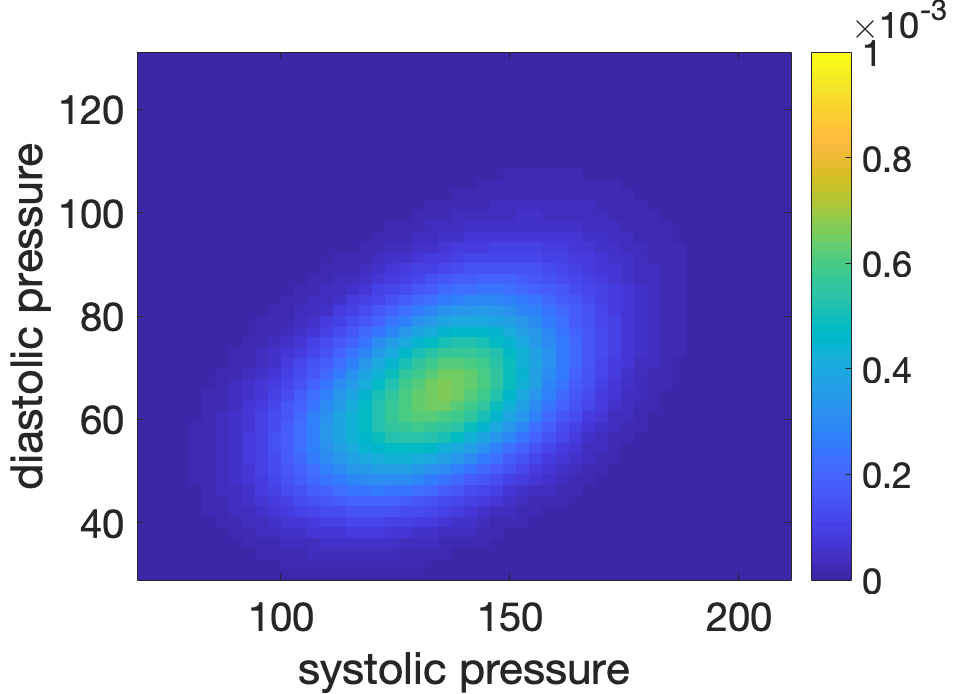}
	\includegraphics[width=7cm]{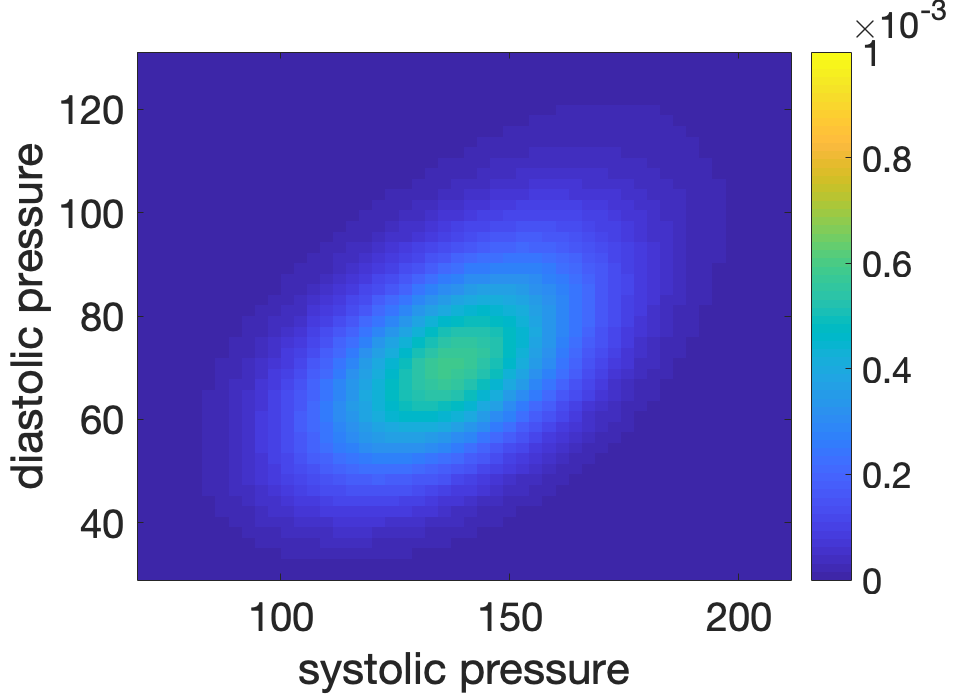}\\
	\includegraphics[width=7cm]{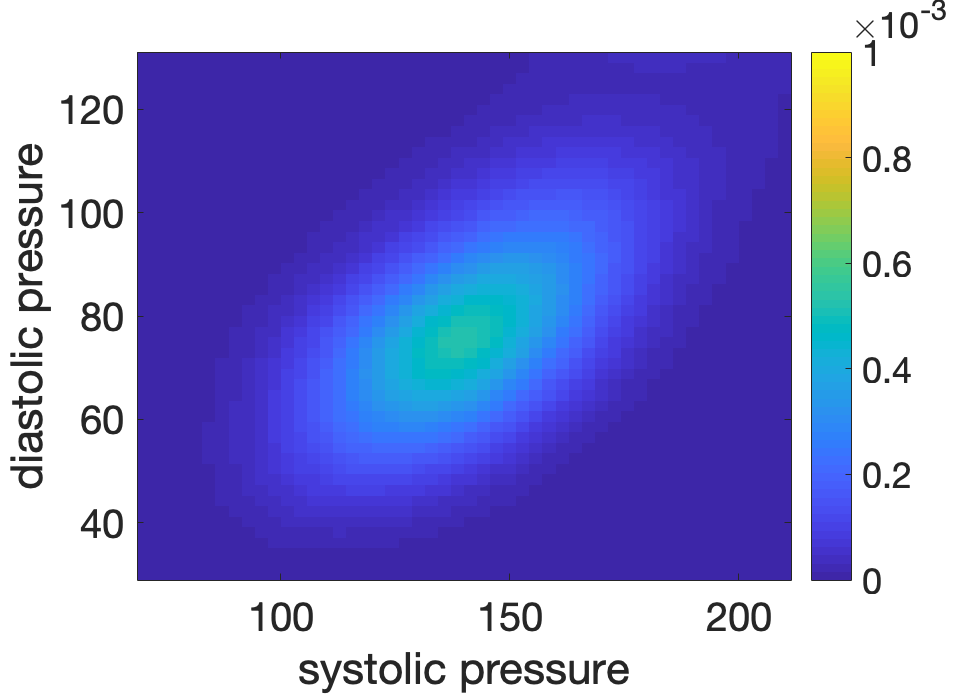}
	\includegraphics[width=7cm]{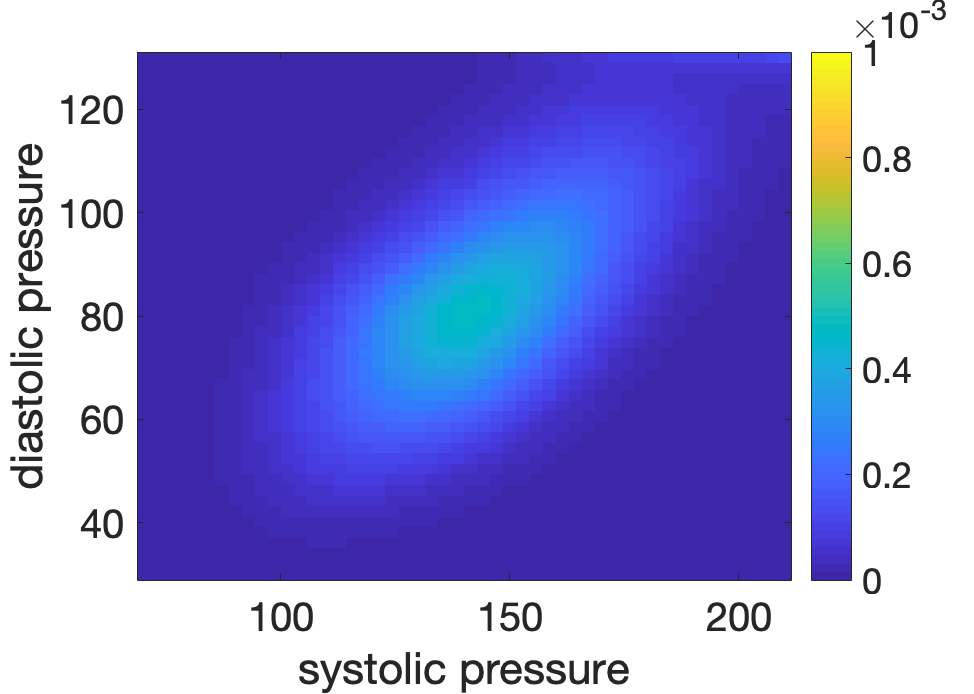}\\
 	\caption{Applying global fitting to the joint (SBP/DBP) distributions for the BLSA data with distributions at $20$ different age bins as inputs. The first row displays the fits for ages $5,25$ (left extrapolation), the second row for ages $45,65$ (interpolation) and the third row for ages $85,105$ (right extrapolation).}
 	\label{f3-7}
 \end{figure}

The distributional fits from the global model in  
Figure \ref{f3-7} for  ages $x= 5,25,85,105$ demonstrate  distributional extrapolation and for ages $x= 45,65$  distributional interpolation. There is a  clear indication of a  distributional trend towards  higher systolic and diastolic blood pressures as age grows, as well as increasing spread. At age $5$ the mode is around $60,120$ and at age $105$ around $90,150$. The interpolated and extrapolated distributions are  concentrated around the diagonal running from $(30,70)$ to $(130,210)$ with the3 exception of age $5$. With increasing age,  the covariance of the distribution increases considerably,  while there are indications that  the correlation of SBP/DSP is quite  stable across age.

\FloatBarrier 
\subsection{Calgary temperature data}
These data, available at \url{https://calgary.weatherstats.ca/}, consist of recordings of  the minimum and maximum temperature for  each day from $1882$ to $2020$ in Calgary (Alberta).  For each year, we constructed the joint distribution  of minimum and maximum temperatures recorded daily for  January, March and June and used  kernel density estimation to obtain estimates for both joint and marginal distributions,  both marginally and jointly. For the marginal distributions both the distributions of minimum and maximum temperature were targeted. For the joint distribution,  a more  expedient way turned out to work with the joint two-dimensional  distribution  of  minimum temperature recorded  for a given day and   the difference between maximum and minimum temperatures recorded for the same day, the latter corresponding to the observed temperature range, as this simple linear  data transformation made it possible to use the same rectangular support for all distributions considered. 

We first fitted the global model for the  marginal one-dimensional density responses and year as predictor to obtain predictions for the future distribution of maximum and minimum temperatures through the proposed Wasserstein extrapolation.  The fitted distributions in Figures \ref{f3-8} (for January), \ref{f3-9_2} (for June)  and \ref{f3-10_1}  (for March, in the Supplement) indicate that the fitted distribution of the minimum  temperature in March and June varies very little over calendar time and while generally  the maximum and minimum temperatures are predicted to move  toward higher values, the maximum temperature in June trends smaller  with increasing extrapolation year. %Another interesting finding is that the distribution is growing sharper except the maximum temperature in March. This indicates that the maximum and minimum temperature tend to have a smaller variation over years in other months but maximum temperature in March tends to have smaller variation.

\begin{figure}[!htb]
	\centering
	\includegraphics[width=6.5cm]{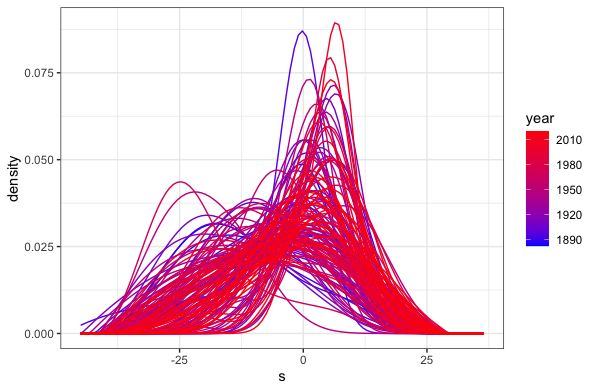}
	\includegraphics[width=6.5cm]{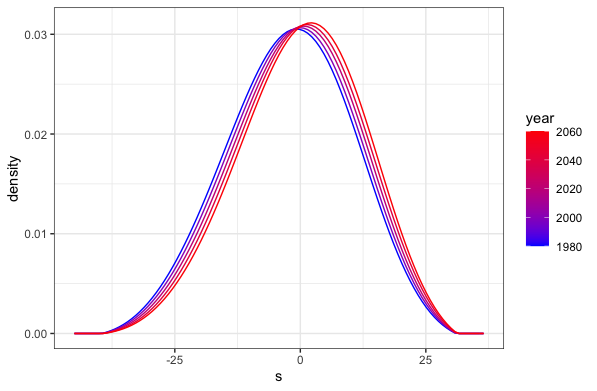}\\
	\includegraphics[width=6.5cm]{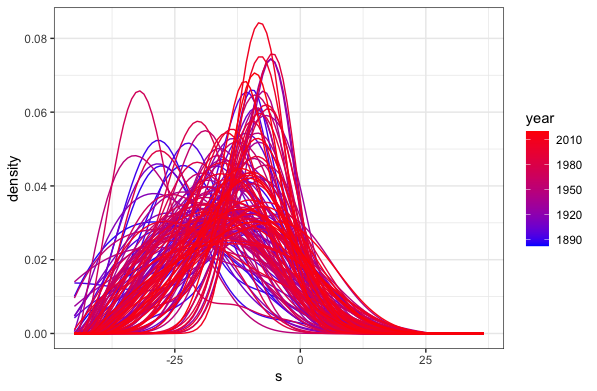}
	\includegraphics[width=6.5cm]{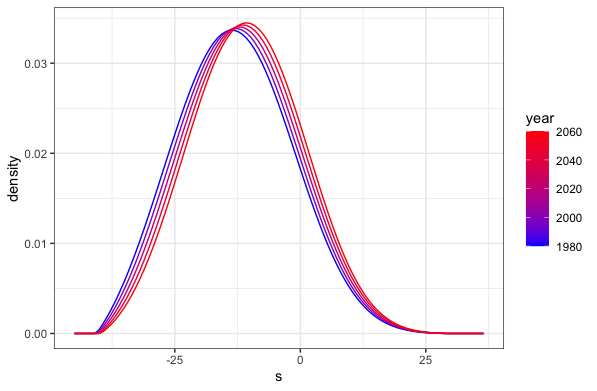}
	\caption{Calgary temperature data. Marginal distributions  of maximum (upper panels) and minimum (lower panels) temperature in January, with observed (kernel smoothed)  densities for these distributions (left panels) and extrapolated global fits (right panel) for $1980, 2000, 2020, 2040, 2060$. The densities are color coded by calendar year.}
\label{f3-8}
	\includegraphics[width=6.5cm]{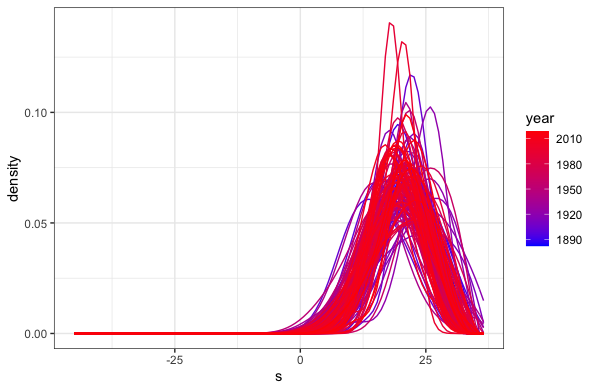}
	\includegraphics[width=6.5cm]{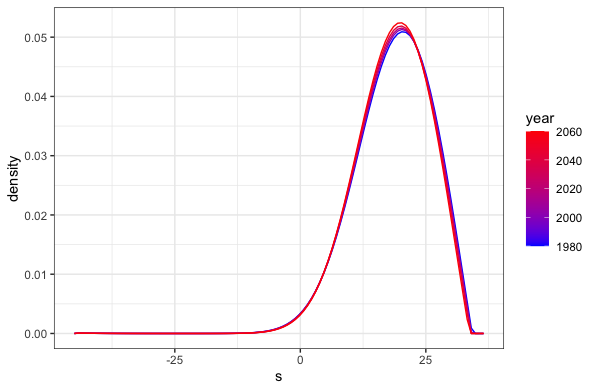}\\
	\includegraphics[width=6.5cm]{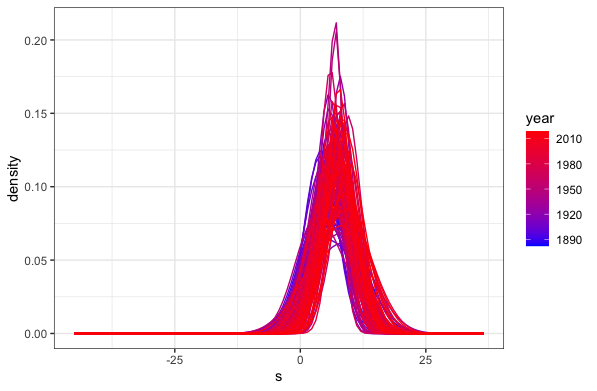}
	\includegraphics[width=6.5cm]{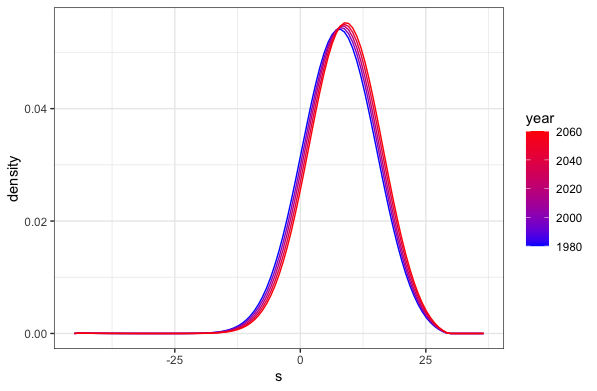}
	\caption{Calgary temperature data. Marginal distributions of maximum and minimum temperature in June, with the same layout as the previous figure.} 
	\label{f3-9_2}
\end{figure}

The fits with the global model of  two-dimensional  densities of maximum-minimum temperature and minimum temperature shown for January in Figure \ref{f3-11} and for March and June in Figures \ref{f3-10_1}, \ref{f3-10} and \ref{f3-12} (in the Supplement) indicate that  for March the primary change over calendar time is in terms of the location of the fitted joint distribution, where the minimum temperature  is increasing,  as shown in Figure \ref{f3-10_1}. As for shape changes,  the fitted joint distribution tends towards a decreased  variance in the $s_2-s_1$ direction, where $s_1$ and $s_2$ are the $x$ and $y$ axes in the two dimensional distribution. For June the changes in the two-dimensional distributions are barely noticeable, which also agrees with the marginal one-dimensional fits displayed in Figure \ref{f3-9_2}. For January, the location of the joint distribution does not have an obvious shift but there is an interesting finding that the joint distribution tends to develop a weak second mode that involves a smaller daily temperature range in the extrapolation towards future calendar years. 

\begin{figure}[!htb]
	\centering
	\includegraphics[width=6.5cm]{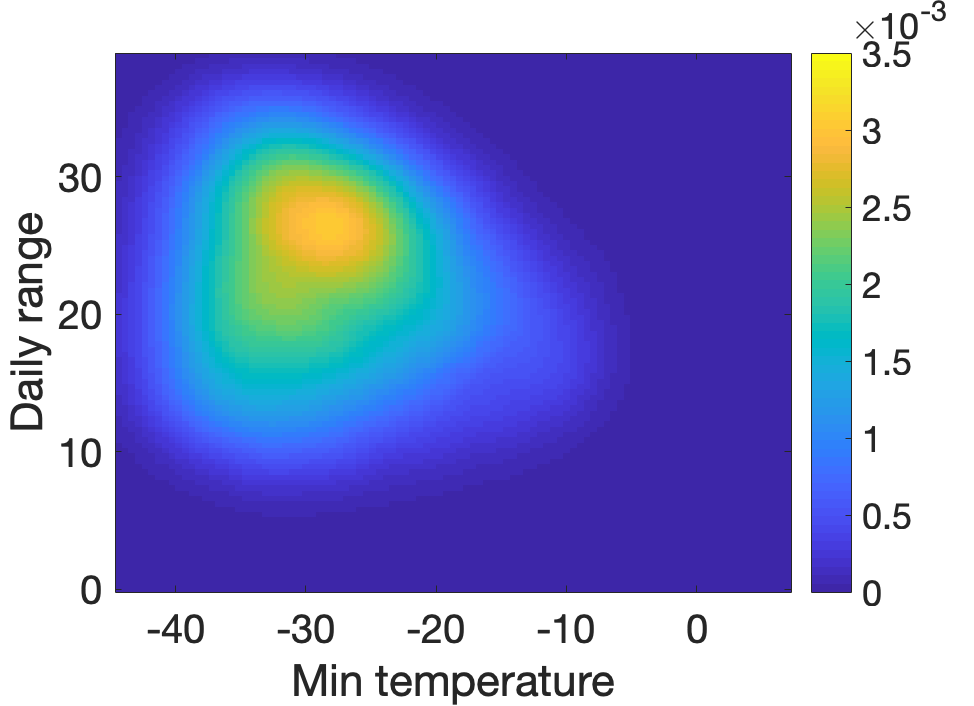}\\
	\includegraphics[width=6.5cm]{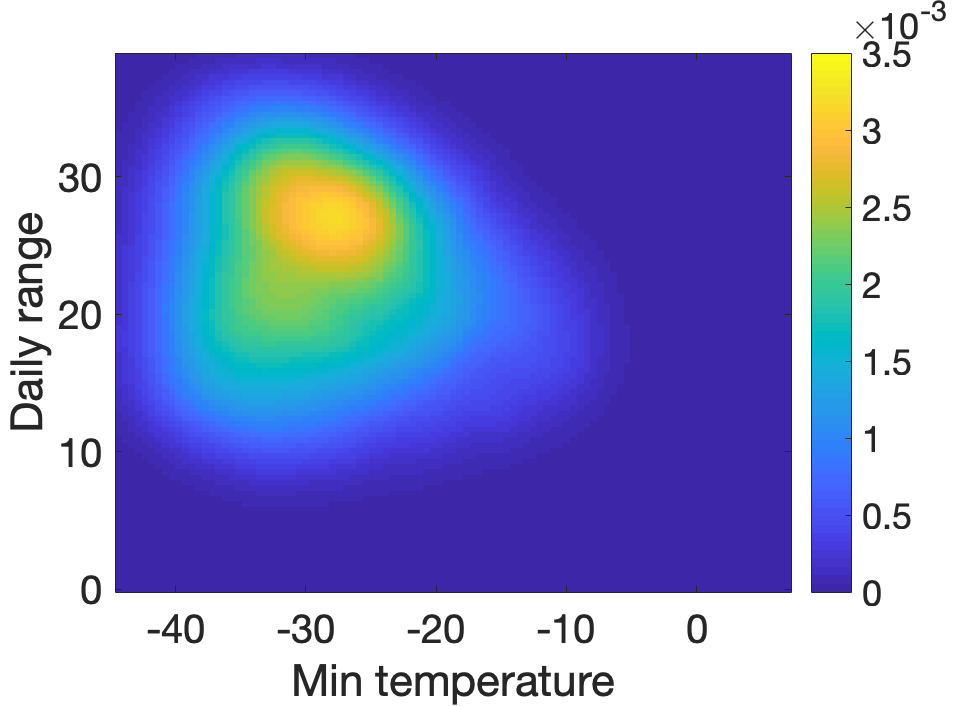}
	\includegraphics[width=6.5cm]{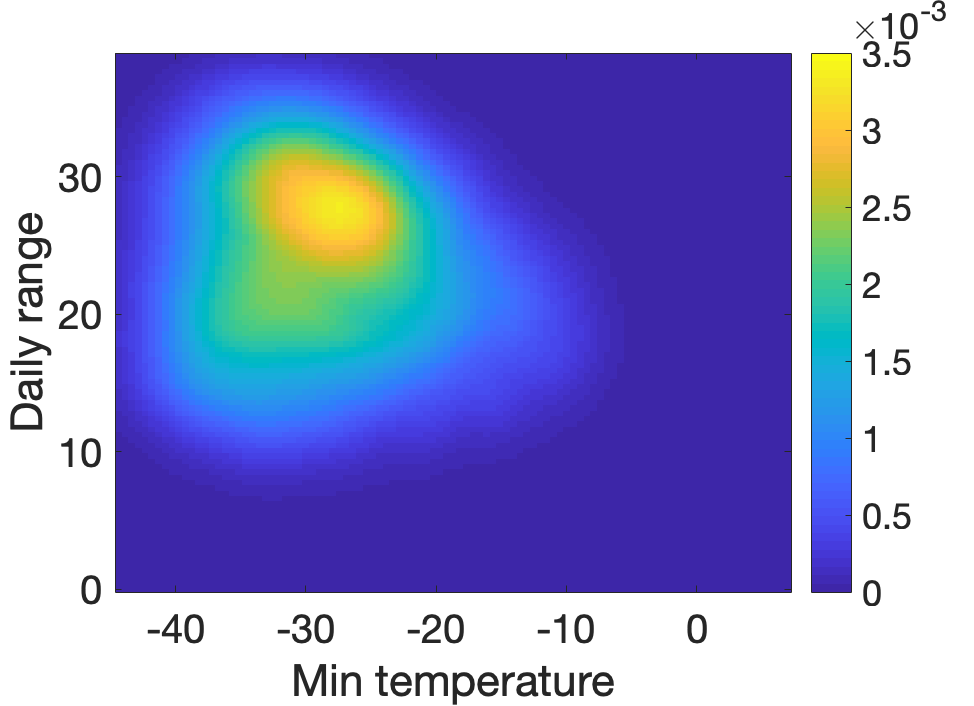}\\
	\includegraphics[width=6.5cm]{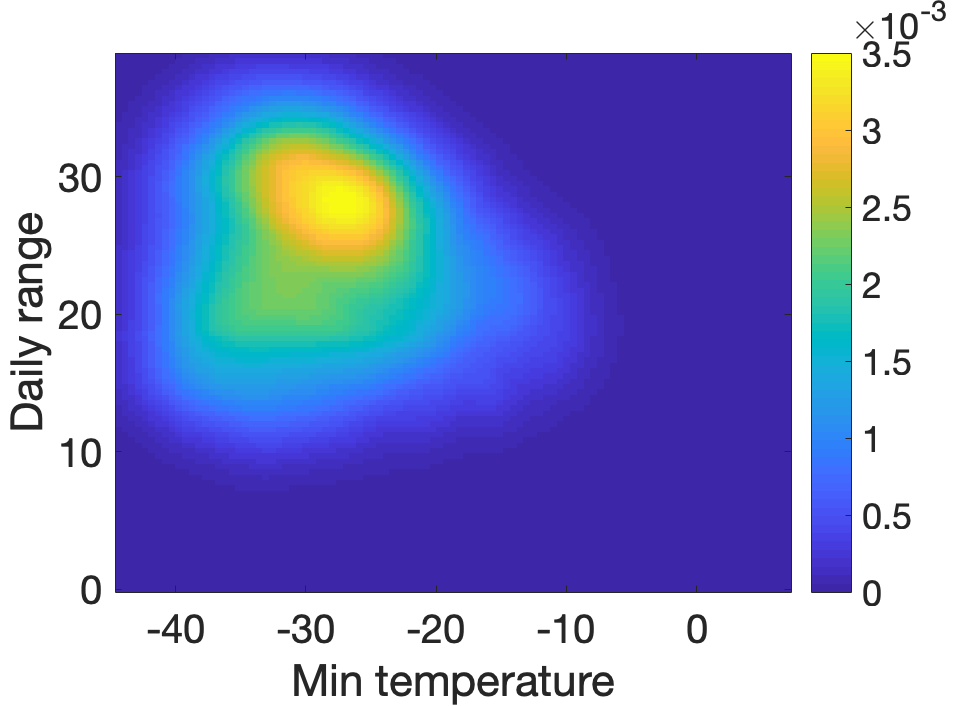}
	\includegraphics[width=6.5cm]{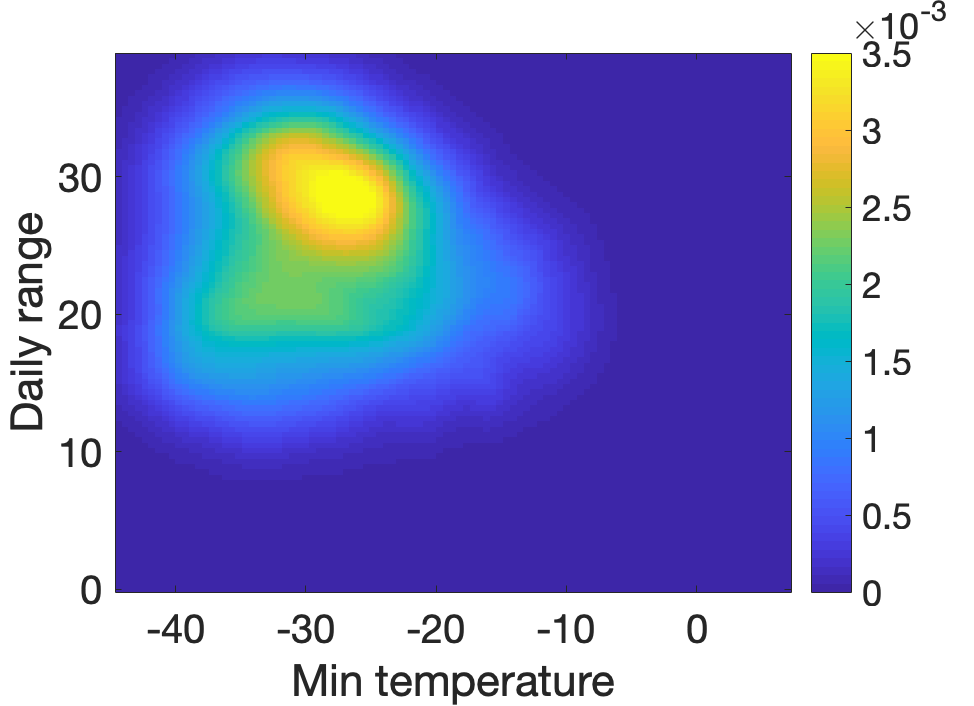}
	\caption{Global fits for the  joint distribution of maximum-minimum (range) and minimum temperature in Calgary for January,  extrapolating to $1980$ (first row), $2000,2020$ (second row) and  $2040,2060$ (third row).}
	\label{f3-11}
\end{figure}

\section{Proofs and Auxiliary Results} 

\subsection{Proof of Proposition 1}
\begin{proof}
	We prove the results for the  global estimates here, the generalization to local estimates is then straightforward.  
	
	By definition $\hat{\mu}_G(x) = \argmin_{\mu \in \Omega} \frac{1}{m}\sum_{i=1}^{n}s_{iG}(x)W_2^2(\mu,\nu_i)$  where in the 1-d case $W_2^2(\mu_1,\mu_2) = \int_{0}^{1}(F^{-1}_{\mu_1}(t)-F^{-1}_{\mu_2}(t))^2dt$, we have
	\bea
	F^{-1}_{\hat{\mu}_G(x)}(t) = \argmin_{Q\in \mathcal{Q}}\frac{1}{n}\sum_{i=1}^{n}s_{iG}(x)\int_{0}^{1}(F^{-1}_{\nu_i}(t)-Q)^2dt.
	\eea
	Then observing $\frac{1}{n}\sum_{i=1}^{n}s_{iG}(x) = 1$, 
	\bea
	\frac{1}{n}\sum_{i=1}^{n}s_{iG}(x)\int_{0}^{1}(F^{-1}_{\nu_i}(t)-F^{-1}_{\mu}(t))^2dt &=&  \int_{0}^{1}(F^{-1}_{\mu}(t)^2 - \frac{2}{n}(\sum_{i=1}^{n}s_{iG}(x)F^{-1}_{\nu_i}(t))F^{-1}_{\mu}(t))dt + C_1\\
	&=& \int_{0}^{1}(F^{-1}_{\mu}(t) - \frac{1}{n}\sum_{i=1}^{n}s_{iG}(x)F^{-1}_{\nu_i}(t))^2dt + C_2,
	\eea
	where $C_1,C_2$ are constants. This implies $F^{-1}_{\hat{\mu}_G(x)}(t) = \argmin_{Q\in \mathcal{Q}}\|Q-\frac{1}{n}\sum_{i=1}^{n}s_{iG}(x)F^{-1}_{\nu_i}(t)\|$ and completes  the proof.
\end{proof}
\subsection{Proof of Proposition 2}
\begin{proof}
	The proof proceeds  by constructing a histogram $H$ such that $W_2^2(\nu_{H},\nu_{f})$ and $W_2^2(\nu_{H},\nu_{\mathbf{r}})$ is of  the desired order.  where $\nu_{H},\nu_{f}$ are continuous measures with density $H$ and $f$ and $\nu_{\mathbf{r}}$ is the discrete measure on $(d_1,\ldots,d_m)$ with probability mass $\mathbf{r}$. In the first step we construct such a histogram. Let $(R_1,\ldots,R_m)$ be the small rectangles created by the grid such that $d_k$ is the vertex of rectangular bin $R_k$ with smallest coordinates in all directions. If $d_k$ reaches the largest value of one coordinate, we can extend the grid so that this rectangle can be included. These rectangles serve as bins of the histogram. Then $H$ is the histogram on these bins with frequencies $(\frac{f_i(d_1)}{\sum_{j=1}^{m}f_i(d_j)},\frac{f_i(d_2)}{\sum_{j=1}^{m}f_i(d_j)},\ldots,\frac{f_i(d_m)}{\sum_{j=1}^{m}f_i(d_j)})$. Let $d_{TV}(\mu_1,\mu_2) = \sup_{B\subset \mathbb{R}^d}|\mu_1(B)-\mu_2(B)|$ be the total variation distance, by Lipschitz continuity we immediately get $d_{TV}(\nu_{H},\nu_{f}) = O(\zeta)$. It has been proved in \ci{gibbs:2002} that $d_P(\mu_1,\mu_2) \leq d_{TV}(\mu_1,\mu_2)$, where $d_P$ is the Prokhorov distance. For any joint distribution $J$ of random variables $X,Y$ with marginals $\mu,\nu$,
	\bea
	E_J\|X-Y\|^2 &\leq& \epsilon^2 P(\|X-Y\|^2\leq \epsilon^2) + \text{diam}(M)P(\|X-Y\|^2 > \epsilon^2)\\
	&=& \epsilon^2 + (\text{diam}(M)-\epsilon^2)P(\|X-Y\|^2 > \epsilon^2).
	\eea
	
	If $d_P(\mu,\nu)\leq \epsilon$, we can choose a coupling $J_1$ so that $P(\|X-Y\|^2 > \epsilon^2) \leq \epsilon$ (\ci{huber:2004},p.27). Thus,
	\bea
	E_{J_1}\|X-Y\|^2 \leq \epsilon^2 + (\text{diam}(M)-\epsilon^2)\epsilon \leq \epsilon^2+ \text{diam}(M)\epsilon
	\eea
	and therefore
	\be\label{ineq}
	W_2^2(\mu,\nu) &\leq& E_{J_1}[\|X-Y\|^2] \leq d_P(\mu,\nu)^2 + \text{diam}(M)d_P(\mu,\nu) \leq d_{TV}^2(\mu,\nu)\\ \nonumber 
	 &\quad& +  \text{diam}(M)d_{TV}(\mu,\nu).
	\ee
	
	We conclude $W^2_2(\nu_{H},\nu_{f}) = O(d_{TV}(\mu,\nu) + d_{TV}^2(\mu,\nu)) =O(\zeta)$. Next we prove $W_2^2(\nu_{H},\nu_{\mathbf{r}}) = o(\zeta^2)$. To do this we construct a push-forward map $\tilde{T}$ from $\nu_H$ to $\nu_{\mathbf{r}}$ such that it maps each bin $R_k$ to the vertex $d_k$. It is obvious that $\tilde{T}_\#\nu_H = \nu_{\mathbf{r}}$ as the probability in each bin $R_k$ equals the probability mass of the corresponding vertex $d_k$. By definition of the 2-Wasserstein distance,
	\be
	W_2^2(\nu_{H},\nu_{f})  \leq \int_{M_0}\|\tilde{T}(w)-w\|^2d\nu_{H} \leq \int_{M_0} \zeta^2d\nu_{H} =\zeta^2= O(\zeta^2), \label{w22}
	\ee
	where $M_0 = \cup_{j=1}^mR_j$. Combining  these two results we have\\ $W_2^2(\nu_{\mathbf{r}},\nu_{f}) \leq 2(W_2^2(\nu_{H},\nu_{f}) + W_2^2(\nu_{H},\nu_{\mathbf{r}})) = O(\zeta)$.
\end{proof}

\subsection{Proof of Theorem 1} 
\begin{proof}
	Since the weights $s_{G}(X,x)=s_{G}(U,u_0)$ if $X =U\mathbf{c}+\mathbf{b}, x= u_0\mathbf{c}+\mathbf{b}$, we only consider the case $X= U$ and $x = u_0$. First we consider the case $u_0 \in [0,1]$;  the extrapolation case can be handled analogously. W.L.O.G. assume $E(U)\leq u_0$. If $u_0 = E(U)$, let $u^*= 0$. Otherwise $s_G(u,u_0)$ is linear in $u$ with positive slope, so there exists $u_1$ be  such that $s_{G}(u_1,u_0) = 0$. In this case define $u^* = \max(u_1,0)$. It is easy to see $u^*<u_0$, as $s_{G}(u_0,u_0) = 1+ (u_0-E(U))^2/\Var(U) > 0$. Because of the positive slope and  linearity of the weights, it holds for any $u$ that $\mathbf{1}_{\{u\in[0,u^*)\}}s_{G}(u,u_0) \leq 0$ and $\mathbf{1}_{\{u\in[u^*,1]\}}s_{G}(u,u_0) \geq 0$, where $\mathbf{1}$ is the indicator function.  Denote $\theta = W_2(\tilde{\nu}(0),\tilde{\nu}(1))$. For $u<u^*$, $W_2(\mu,\tilde{\nu}(u)) \leq W_2(\mu,\tilde{\nu}(u^*)) + W_2(\tilde{\nu}(u^*),\tilde{\nu}(u))$ and for $u\geq u^*$, $|W_2(\mu,\tilde{\nu}(u))| \geq |W_2(\tilde{\nu}(u^*),\mu) - W_2(\tilde{\nu}(u),\tilde{\nu}(u^*))|$. With $W_2(\tilde{\nu}(u),\tilde{\nu}(u^*)) = |u-u^*|\theta$, we have that for any $\mu$
	\bea
	&\quad& E[s_{G}(U,u_0)W_2^2(\mu,\tilde{\nu}(U))\mathbf{1}_{\{U\in [0,u^*)\}}]  \\
	&\geq& E[s_{G}(U,u_0)(W_2(\mu,\tilde{\nu}(u^*)) + W_2(\tilde{\nu}(u^*),\tilde{\nu}(U)))^2\mathbf{1}_{\{U\in [0,u^*)\}}]\\
	&=&  E[s_{G}(U,u_0)(W_2(\mu,\tilde{\nu}(u^*)) +(u^*-U)\theta)^2\mathbf{1}_{\{U\in [0,u^*)\}}] 
	\eea
	and
	\bea
	&\quad& E[s_{G}(U,u_0)W_2^2(\mu,\tilde{\nu}(U))\mathbf{1}_{\{U\in [u^*,1]\}}] \\
	&\geq& E[s_{G}(U,u_0)(W_2(\mu,\tilde{\nu}(u^*)) - W_2(\tilde{\nu}(U),\tilde{\nu}(u^*)))^2\mathbf{1}_{\{U\in [u^*,1]\}}]\\
	&=& E[s_{G}(U,u_0)(W_2(\mu,\tilde{\nu}(u^*)) - (U-u^*)\theta)^2\mathbf{1}_{\{U\in [u^*,1]\}}] ,
	\eea
	where equality is achieved when $\mu = \tilde{\nu}(\tilde{u})$ with $\tilde{u} > u^*$.
	
	Combining the results above with  $E(s_G(U,u_0)) = 1$,  one finds that for any $\mu$
	\bea
	&\quad& E(s_{G}(U,u_0)W_2^2(\mu,\tilde{\nu}(U))) \\
	&=& E[s_{G}(U,u_0)W_2^2(\mu,\tilde{\nu}(U))\mathbf{1}_{\{U\in [0,u^*)\}}] + E[s_{G}(U,u_0)W_2^2(\mu,\tilde{\nu}(U))\mathbf{1}_{\{U\in [u^*,1]\}}]\\
	&\geq& E[s_{G}(U,u_0)(W_2(\mu,\tilde{\nu}(u^*)) +(u^*-U)\theta)^2\mathbf{1}_{\{U\in [0,u^*)\}}]\\
	&\quad& +   E[s_{G}(U,u_0)(W_2(\mu,\tilde{\nu}(u^*)) - (U-u^*)\theta)^2\mathbf{1}_{\{U\in [u^*,1]\}}] \\
	&=& E[s_{G}(U,u_0)(W_2(\mu,\tilde{\nu}(u^*)) +(u^*-U)\theta)^2]\\
	&=& E(s_{G}(U,u_0))EW_2^2(\mu,\tilde{\nu}(u^*)) +  2\theta E[(u^*-U)s_{G}(U,u_0)]E[W_2(\mu,\tilde{\nu}(u^*))] + C_1\\
	&=& EW_2^2(\mu,\tilde{\nu}(u^*)) + 2\theta E[(u^*-U)s_{G}(U,u_0)]EW_2(\mu,\tilde{\nu}(u^*)) + C_1\\
	&=& (W_2(\mu,\tilde{\nu}(u^*))+ \theta E[(u^*-U)s_{G}(U,u_0)]))^2+ C_2,
	\eea
	where $C_1,C_2$ are constants that do not depend on $\mu$. Then observing 
	\bea
	&\quad& E((u^*-U)s_{G}(U,u_0)) \\
	&=& E(u^*-U) + E((U-E(U))(u_0-E(U))/\Var(U))u^* - E(U(U-E(U))(u_0-E(U))/\Var(U))\\
	&=& u^*-E(U)  - (u_0-E(U))\\
	&=& u^*-u_0,
	\eea
	the minimum is achieved when $\mu$ is  on the geodesic and $W_2(\mu,\tilde{\nu}(u^*))  = (u_0-u^*)\theta$. This implies $\hat{\mu}_{G}(u_0) = \tilde{\nu}(u_0)$. For the extrapolation case we simply replace $0$ by $t_1$ and $1$ by $t_2$ in the proof.
\end{proof}

\subsection{Proof of Theorem 2}
\begin{proof}
	This proof uses  Theorem 2 in \ci{petersen:2019}, and the conditions necessary  to apply this result that must be established first. The three conditions are 
	
	(P0) The objects $\mu_{G}(x)$ and $\hat{\mu}_{G}(x)$ exist and are unique, the latter almost surely. Additionally, for any $\epsilon >0$, $\inf_{W_2(\mu,\mu_{G}(x))>\epsilon}E(s_G(X,x)W_2^2(\mu,\nu)) >E(s_G(X,x)W_2^2(\mu_{G}(x),\nu)) $.
	
	(P1) For $\delta > 0$ small enough,
	
	\begin{equation*}
	\int_{0}^{1}\sqrt{1+\log N(\delta\epsilon,B_{\delta}(\mu_{G}(x)),W_2)}d\epsilon <\infty,
	\end{equation*}		
	where $B_{\delta}(\mu_{G}(x),W_2)$ is the $\delta$-ball in 2-Wasserstein space centered at $\mu_{G}(x)$ and $N(\epsilon,\Omega,d)$ is the covering number for $\Omega$ using open balls for radius $\epsilon$.	
	
	(P2) There exists $\eta >0,A >0$ and $\beta>1$, possibly depending on $x$, such that whenever\\ $W_2(\mu_{G}(x),\mu)<\eta$, we have\\ $ E(s_G(X,x)W_2^2(\mu,\nu)) - E(s_G(X,x)W_2^2(\mu_{G}(x),\nu)) - AW_2(\mu,\mu_{G}(x))^{\beta} \geq 0$.
	\vspace{0.1in}
	
	Under conditions (P0)-(P2), it then holds that
	\be\label{gr}
	W_2(\hat{\mu}_G(x),\mu_G(x)) = O_{p}(n^{-1/(2(\beta-1))}).
	\ee

	In the following, we will verify these conditions for $\beta = 2$. We first prove (P1) by establishing an inequality that provides a bound for the covering number. We use that by  (\ref{ineq}) one has the bound $W_2^2(\mu,\nu) = O(d_{TV}(\mu,\nu)) =O( \|f_1-f_2\|_{\infty})$.
	Then we use the fact that by Theorem 2.7.1 of \ci{van:1996},  if $M$ is a bounded, convex subset of $\mathbb{R}^{d}$ with nonempty interior, there exists a constant $A_1$ depending only on $\gamma$ and $d$ such that
	\begin{equation} \label{brk}
	\log N(\epsilon,\mathcal{F}_\gamma,\|\|_{\infty}) \leq A_1(\frac{1}{\epsilon})^{\frac{d}{\gamma}}
	\end{equation}
	for every $\epsilon > 0$ .

	 Applying $W_2^2(\mu,\nu) = O( \|f_1-f_2\|_{\infty})$, we have $B_{A_2\epsilon}(\mu,\|\|_{\infty}) \subset B_{\sqrt{\epsilon}}(\mu,W_2)$ for some constant $A_2$ and  $N(\epsilon,\mathcal{F}_\gamma,W_2) \leq A_3N(\epsilon^2,\mathcal{F}_\gamma,\|\|_{\infty})$ with a constant $A_3$. Then it holds with a constant $A_4$ that\\
	$\sqrt{1+\log N(\delta\epsilon,B_{\delta}(\mu_{G}(x),W_2),W_2)} \leq  \sqrt{1+\log N(\delta\epsilon,\mathcal{F}_\gamma,W_2)} \leq A_4\epsilon^{-d/\gamma}$,\\
	which leads to
	$\int_{0}^{1}\sqrt{1+\log N(\delta\epsilon,B_{\delta}(m_{\oplus}(x)),d)}d\epsilon < \infty,$ i.e., (P1) holds. 
	
	It remains to prove (P0) and (P2) for $\beta =2$. For  $\nu_1 = T_{1\#}\nu_0,\nu_2 = T_{2\#}\nu_0$, $T_2\circ T_1^{-1}$ is a push-forward map that pushes $\nu_1$ to $\nu_2$. Following an argument in \cite{boissard:2015}, using  the  (AD) assumption and Brenier's theorem, $T_2\circ T_1^{-1}$ is the optimal transport map. Thus the 2-Wasserstein distance between $\nu_1,\nu_2$ is
	\be\label{brenier}
	W_2^2(\nu_1,\nu_2) = \int_{M}\|T_2\circ T_1^{-1}(w)-w\|^2d\nu_1 = \int_{M}\|T_1(w)-T_2(w)\|^2d\nu_0.
	\ee
	It is easy to see that  $E(s_G(X,x)) = 1$. For the random response $\nu$ and any fixed $\mu$, assume the optimal transport map from $\mu$ to $\nu_0$ is $T_{\mu}\in \mathcal{T}(M)$ and the one from $\nu$ to $\nu_0$ is $T_{\nu}$. Furthermore, for the global model (\ref{gm}), 
	\bea
	&&E(s_G(X,x)W_2^2(\mu,\nu)) \\
	&=& E(s_G(X,x)\int_{M}\|T_{\mu}(w)-T_\nu(w)\|^2d\nu_0)\\
	&=& E(s_G(X,x)\int_{M}\|T_{\mu}(w) - E(s_G(X,x)T_\nu(w)) + E(s_G(X,x)T(w))-T_\nu(w)\|^2d\nu_0)\\
	&=& E(s_G(X,x)\int_{M}\|T_{\mu}(W)-E(s_G(X,x)T_\nu(w)) \|^2d\nu_0) \\
	&\quad& + E(s_G(X,x)\int_{M}\|E(s_G(X,x)T_\nu(w))-T_\nu(w)\|^2d\nu_0)\\
	&\quad& + 2E(s_G(X,x)\int_{M}(T_{\mu}(w) - E(s_G(X,x)T_\nu(w)))'(E(s_G(X,x)T_\nu(w))-T_\nu(w))d\nu_0)\\
	&=& \int_{M}\|T_{\mu}(w)-E(s_G(X,x)T_\nu(w)) \|^2d\nu_0 + E(s_G(X,x)\int_{M}\|E(s_G(X,x)T_\nu(w))-T_\nu(w)\|^2d\nu_0).
	\eea
	This yields the minimizer \be \label{tt}  \mu_{G}(x) = \tilde{T}_\#\mu_0, \quad\quad \tilde{T} = \arginf_{T_0\in \mathcal{T}(M)}\int_M\|T_0(w)-E(s_G(X,x)T_\nu(w))\|^2d\nu_0,\ee  where the projection map  $\tilde{T}$ is characterized by $\int_M (E(s_G(X,x)T_\nu(w))-\tilde{T}(w))'( T_a(w) -\tilde{T}(w)) d\nu_0\leq 0$ for any $T_a\in\mathcal{T}(M)$. Then by convexity a unique solution exists, so that (P0) is satisfied. 	
	
	Continuing the above argument, 
	\begin{align}
	&\quad\, E(s_G(X,x)W_2^2(\mu,\nu))  \nn \\ \nn
	&=\int_{M}\|T_{\mu}(w)-E(s_G(X,x)T_\nu(w)) \|^2d\nu_0 + E(s_G(X,x)\int_{M}\|E(s_G(X,x)T_\nu(w))-T_\nu(w)\|^2d\nu_0) \\ \nn
	&\geq   \int_{M}\|T_{\mu}(w)-T_{\mu_{G}(x)}(w)\|^2d\nu_0  + \int_{M}\|T_{\mu_{G}(x)}(w)-E(s_G(X,x)T_\nu(w)) \|^2d\nu_0\\ \nn
	&\quad + E(s_G(X,x)\int_{M}\|E(s_G(X,x)T_\nu(w))-T_\nu(w)\|^2d\nu_0)\\ \nn
	&=  W_2^2(\mu,\mu_{G}(x))+ \int_{M}\|T_{\mu_{G}(x)}(w)-E(s_G(X,x)T_\nu(w)) \|^2d\nu_0\\ \nn
	&\quad +  E(s_G(X,x)\int_{M}\|E(s_G(X,x)T(w))-T_\nu(w)\|^2d\nu_0)\\ 
	&=W_2^2(\mu,\mu_{G}(x))+  E(s_G(X,x)W_2^2(\mu_{G}(x),\nu)).\label{eq1}
	\end{align}
	We note that since $E(s_G(X,x)T(w))$ is not necessarily an optimal transport map, one needs to use the projection $\tilde{T}$ in (\ref{tt}). One finds that  (P2) is satisfied with $A =1, \, \beta = 2$, for arbitrary $\eta>0$. Therefore $W_2^2(\mu_G(x) ,\hat{\mu}_{G}(x)) = O(n^{-1})$ by (\ref{gr}). This  completes the proof.
\end{proof}
\subsection*{Proof of Theorem 3} 
\begin{proof}
	We aim to apply  Theorem 3 and Theorem 4 in \ci{petersen:2019}, which hold under the following   conditions. 
	
	(K0) The kernel $K$ is a probability density function, symmetric around zero. Furthermore, defining $K_{jk} = \int_{\mathbb{R}}K^j(x)x^kdx$, $|K_{14}|$ and $|K_{26}|$ are both finite.
	
	(L0) The object $\mu_0(x) $ exists and is unique. For all $n$, $\mu_{L,h}(x)$ and $\hat{\mu}_{L,h}(x)$ exist and are unique, the latter almost surely. Additionally, for any $\epsilon >0$,
	\bea
	\inf_{W_2(\mu,\mu_0(x))>0}\{E(s_G(X,x)W_2^2(\mu,\nu))-E(s_G(X,x)W_2^2(\mu_0(x),\nu))\} &>& 0,\\
	\liminf_{n}\inf_{W_2(\mu,\mu_{L,h}(x))>\epsilon}\{\frac{1}{n}\sum_{i=1}^{n}s_{iL,h}(x)W_2^2(\mu,\nu)-\frac{1}{n}\sum_{i=1}^{n}s_{iL,h}(x)W_2^2(\mu_{L,h}(x),\nu)\} &>& 0.
	\eea
	
	(L1) The marginal density $f$ of $X$ and the conditional densities $g_{\nu}$ of $X|Z=z$ exist and are twice continuously differentiable, the latter for all $\nu \in \Omega$, and $\sup_{x,\nu}|g^{''}_{\nu}(x)|<\infty$. Additionally, for any open $U\subset \Omega$, $P(\nu \in U |X=x)$ is continuous as a function of $x$.
	
	(L2) There exists $\eta_{1}>0$, $C_{1}>0$ and $\beta>1$ such that whenever $d(\omega,\tilde{l}_{\oplus}(x))<\eta_{1}$,
	\bea
	E(s_G(X,x)W_2^2(\mu,\nu))-E(s_G(X,x)W_2^2(\mu_0(x),\nu)) \geq C_{1}W_2(\mu,\mu_0(x))^{\beta_1}
	\eea
	
	(L3) There exists $\eta_2 >0,\,  C_2 >0$ and $\beta_2 >1$, such that  whenever $d(\omega,\tilde{l}_{\oplus}(x))<\eta_{1}$,
	\bea
	\liminf_{n}[\frac{1}{n}\sum_{i=1}^{n}s_{iL,h}(x)W_2^2(\mu,\nu)-\frac{1}{n}\sum_{i=1}^{n}s_{iL,h}(x)W_2^2(\hat{\mu}_{L,h}(x),\nu)] \geq C_2W_2(\mu,\hat{\mu}_{L,h}(x))^{\beta_2}.
	\eea
	
	When the above conditions in addition to  (P1) as stated in the proof of Theorem 2 are satisfied,  
	\begin{equation*}
	W_2(\mu_0(x),\mu_{L,h}(x)) = O_p(h^{2/(\beta_1-1)})
	\end{equation*}
	and if $h \rightarrow 0$ and $nh \rightarrow \infty$, then
	\begin{equation*}
	W_2(\hat{\mu}_{L,h}(x) ,\mu_{L,h}(x) ) = O_p((nh)^{-1/2(\beta_2-1)}).
	\end{equation*}
	
	Under the assumptions (KN) and (CD), we can infer (K0) and (L1), and (P1) was established in the proof of Theorem 2 above. To show that  (L0), (L2) and (L3) hold with $\beta_1 = \beta_2 = 2$, we first transform the distances between measures to distances between optimal transport maps as in (\ref{brenier}). For the random response $\nu$ and any fixed $\mu$, denoting  the optimal transport map from $\mu$ to $\nu_0$ by $T_{\mu}\in \mathcal{T}(M)$ that from $\nu$ to $\nu_0$ by $T_{\nu}$, and those from $\nu_i$ to $\mu_0$ by $T_i$,	and also observing 
	 $\sum_{i=1}^ns_{iL,h}(x)= n$,	
	\bea
	&& E(W_2^2(\mu,\nu)|X = x) = E(\int_{M}\|T_{\mu}(w)-T_{\nu}(w)\|^2d\nu_0|X= x)\\
	&=& E(\int_{M}\|T_{\mu}(w)- E(T_{\nu}(w)|X=x) + E(T_{\nu}(w)|X=x) - T_{\nu}(w)\|^2d\nu_0|X= x)\\
	&=& E(\int_{M}\|T_{\mu}(w)- E(T_{\nu}(w)|X=x)\|^2d\nu_0|X=x) + E(\int_{M}\|E(T_{\nu}(w)|X=x) - T_{\nu}(w)\|^2d\nu_0|X=x) \\
	&\quad& +2E(\int_{M}(E(T_{\nu}(w)|X=x) - T_{\nu}(w))'(T_{\mu}(w)- E(T_{\nu}(w)|X=x))d\nu_0|X=x)\\
	&=& \int_{M}\|T_{\mu}(w)- E(T_{\nu}(w)|X=x)\|^2d\nu_0 + E(\int_{M}\|E(T_{\nu}(w)|X=x) - T_{\nu}(w)\|^2d\nu_0|X=x),
	\eea
	\begin{align} \nn
	&\quad\, \sum_{i=1}^{n}s_{iL,h}(x)W_2^{2}(\mu,\nu_i) = \sum_{i=1}^{n}(s_{iL,h}(x)\int_{M}\|T_{\mu}(w)-T_i(w)\|^2d\nu_0)\\ \nn
	&= \sum_{i=1}^{n}(s_{iL,h}(x)\int_{M}\|T_{\mu}(w) - \frac{1}{n}\sum_{i=1}^{n}s_{iL,h}(x)T_i(w) + \frac{1}{n}\sum_{i=1}^{n}s_{iL,h}(x)T_i(w) -T_i(w)\|^2d\nu_0)\\ \nn
	&= \sum_{i=1}^{n}(s_{iL,h}(x)\int_{M}\|T_{\mu}(w) - \frac{1}{n}\sum_{i=1}^{n}s_{iL,h}(x)T_i(w)\|^2d\nu_0)\\ \nn
	&\quad + \sum_{i=1}^{n}(s_{iL,h}(x)\int_{M}\|\frac{1}{n}\sum_{i=1}^{n}s_{iL,h}(x)T_i(w) -T_i(w)\|^2d\nu_0)\\ \nn
	&\quad+ 2\sum_{i=1}^{n}(s_{iL,h}(x)\int_{M}(\frac{1}{n}\sum_{i=1}^{n}s_{iL,h}(x)T_i(w) -T_i(w))'(T_{\mu}(w) - \frac{1}{n}\sum_{i=1}^{n}s_{iL,h}(x)T_i(w))d\nu_0)\\ \nn
	&= n\int_{M}\|T_{\mu}(w) - \frac{1}{n}\sum_{i=1}^{n}s_{iL,h}(x)T_i(w)\|^2d\nu_0\\
	&\quad+ \sum_{i=1}^{n}s_{iL,h}(x)\int_{M}\|\frac{1}{n}\sum_{i=1}^{n}s_{iL,h}(x)T_i(w) -T_i(w)\|^2d\nu_0.\label{eq2}
	\end{align}
	Then  the minimizer of $E(W_2^2(\mu,\nu)|X = x)$ is seen to be $\tilde{T}_\#\mu_0$ with $\tilde{T} = \arginf_{T_0\in \mathcal{T}(M)}\int_M\|T_0(w)-E(T(w)|X=x)\|^2d\nu_0$ and that of  of $\sum_{i=1}^{n}(s_{iL,h}(x)W_2^{2}(\mu,\nu_i))$ to be $\hat{\mu}_{L,h}(x) = T_{L\#}\nu_0,$ with $$T_L = \arginf_{T_0\in \mathcal{T}(M)}\int_M\|T_0(w)-\frac{1}{n}\sum_{i=1}^{n}s_{iL,h}(x)T_i(w)\|^2d\nu_0.,$$ where $\tilde{T}$ is characterized by $\int_M (E(T(w)|X=x)-\tilde{T}(w))'( T_a(w) -\tilde{T}(w)) d\nu_0\leq 0$ and $T_L$ by \newline $\int_M (\frac{1}{n}\sum_{i=1}^{n}s_{iL,h}(x)T_i(w)-\tilde{T}(w))'( T_a(w) -\tilde{T}(w)) d\nu_0\leq 0$ for any $T_a\in\mathcal{T}(M)$. Then following the same argument as  in (\ref{eq1}) above,  (L0), (L2) and (L3) are satisfied with $\beta_1 = \beta_2= 2$ and thus $W_2^2(\mu_0(x) ,\hat{\mu}_{L,h}(x)) = O(n^{-4/5})$ when $p=1$.
\end{proof}

\subsection{Proof of Theorem 4} 
\begin{proof}
	We prove the convergence for global estimate (\ref{ge}) only, as the extension to the local estimate (\ref{le}) follows analogous arguments.  First we derive that $\sum_{i=1}^{n}(s_{iG}(x)W_2^{2}(\mu,\nu_i))$ is away from minimum outside of a small ball around minimizer, similar with condition (P0) and (L0). We follow the same procedure as (\ref{eq1}) and (\ref{eq2}) to finish the proof. Let $T_i$ be the optimal transport map from $\nu_i$ to $\mu_0$, then
	\bea
	&&\sum_{i=1}^{n}s_{iG}(x)W_2^2(\mu,\nu) =  \sum_{i=1}^{n}(s_{iG}(x)\int_{M}\|T_{\mu}(w)-T_i(w)\|^2d\nu_0)\\
	&=& \sum_{i=1}^{n}(s_{iG}(x)\int_{M}\|T_{\mu}(w) - \frac{1}{n}\sum_{i=1}^{n}s_{iG}(x)T_i(w) + \frac{1}{n}\sum_{i=1}^{n}s_{iG}(x)T_i(w) -T_i(w)\|^2d\nu_0)\\
	&=& \sum_{i=1}^{n}(s_{iG}(x)\int_{M}\|T_{\mu}(w) - \frac{1}{n}\sum_{i=1}^{n}s_{iG}(x)T_i(w)\|^2d\nu_0)\\
	&\quad& + \sum_{i=1}^{n}(s_{iG}(x)\int_{M}\|\frac{1}{n}\sum_{i=1}^{n}s_{iG}(x)T_i(w) -T_i(w)\|^2d\nu_0)\\
	&\quad&+ 2\sum_{i=1}^{n}(s_{iG}(x)\int_{M}(\frac{1}{n}\sum_{i=1}^{n}s_{iG}(x)T_i(w) -T_i(w))'(T_{\mu}(w) - \frac{1}{n}\sum_{i=1}^{n}s_{iG}(x)T_i(w))d\nu_0)\\
	&=& n\int_{M}\|T_{\mu}(w) - \frac{1}{n}\sum_{i=1}^{n}s_{iG}(x)T_i(w)\|^2d\nu_0\\
	&\quad& + \sum_{i=1}^{n}s_{iG}(x)\int_{M}\|\frac{1}{n}\sum_{i=1}^{n}s_{iG}(x)T_i(w) -T_i(w)\|^2d\nu_0\\
	&\geq&   n\int_{M}\|T_{\mu}(w)-T_{\hat{\mu}_{G}(x)}(w) \|^2d\nu_0  + n\int_{M}\|T_{\hat{\mu}_{G}(x)}(w)-\frac{1}{n}\sum_{i=1}^{n}s_{iG}(x)T_i(w) \|^2d\nu_0\\
	&\quad& + \sum_{i=1}^{n}s_{iG}(x)\int_{M}\|\frac{1}{n}\sum_{i=1}^{n}s_{iG}(x)T_i(w)-T_i(w)\|^2d\nu_0\\
	&=&  nW_2^2(\mu,\hat{\mu}_{G}(x))+ n\int_{M}\|T_{\hat{\mu}_{G}(x)}(w)-\frac{1}{n}\sum_{i=1}^{n}s_{iG}(x)T_i(w) \|^2d\nu_0\\
	&\quad& +  \sum_{i=1}^{n}s_{iG}(x)\int_{M}\|\frac{1}{n}\sum_{i=1}^{n}s_{iG}(x)T_i(w)-T_i(w)\|^2d\nu_0\\
	&\geq& nW_2^2(\mu,\hat{\mu}_{G}(x))+ \sum_{i=1}^{n}s_{iG}(x)W_2^2(\hat{\mu}_{G}(x),\nu_i),
	\eea
	where $T_{\mu}\in \mathcal{T}(M)$ is the optimal transport map from $\mu$ to $\nu_0$ and $T_{\hat{\mu}_{G}(x)}(w)$ the optimal transport map from $\hat{\mu}_{G}(x)$ to $\nu_0$. Then 
	for any $\epsilon >0$, 
	\begin{align}
	\inf_{W_2(\mu,\hat{\mu}_{G}(x))>\epsilon}\sum_{i=1}^{n}(s_{iG}(x)W_2^{2}(\mu,\nu_i)) >\sum_{i=1}^{n}(s_{iG}(x)W_2^{2}(\hat{\mu}_{G}(x),\nu_i)).\label{eq3}
	\end{align}
	
	According to (\ref{w22}), $W^2_2(\nu_{\mathbf{r}},\nu_{f}) = O(\zeta^2)$,  and this  convergence is found to be uniform in $\nu_{f}$. Observing that  $\lim_{\rho\rightarrow \infty}W_{2,\rho}^2(\mathbf{r}_i,\mathbf{r}_j) = W_{2}^2(\mathbf{r}_i,\mathbf{r}_j)$ and that by the boundedness of the 
	entropy $\sum_{ij}S_{ij}\log(S_{ij})$ this convergence is uniform in $\mathbf{r}_i,\mathbf{r}_j$ \cp{neumayer:2020}, we conclude 
	\bea
	\sup_{\mu} |W_2^{2}(\nu_{\mathbf{r}_{\mu}},\nu_{\mathbf{r}_i})- W_2^{2}(\mu,\nu_i)| &=&  \sup_{\mu} |W_{2,\rho}^{2}(\nu_{\mathbf{r}_{\mu}},\nu_{\mathbf{r}_i}) - W_{2}^{2}(\nu_{\mathbf{r}_{\mu}},\nu_{\mathbf{r}_i}) + W_{2}^{2}(\nu_{\mathbf{r}_{\mu}},\nu_{\mathbf{r}_i}) - W_2^{2}(\mu,\nu_i)| \\
	&\leq& \sup_{\mu} |W_{2,\rho}^{2}(\nu_{\mathbf{r}_{\mu}},\nu_{\mathbf{r}_i}) - W_{2}^{2}(\nu_{\mathbf{r}_{\mu}},\nu_{\mathbf{r}_i})| + \sup_{\mu}|W_{2}^{2}(\nu_{\mathbf{r}_{\mu}},\nu_{\mathbf{r}_i}) - W_2^{2}(\mu,\nu_i)| 
	\eea
	and therefore
	\bea
	\lim_{\rho\rightarrow \infty}\lim_{\zeta \rightarrow 0}\sup_{\mu} |W_2^{2}(\nu_{\mathbf{r}_{\mu}},\nu_{\mathbf{r}_i})- W_2^{2}(\mu,\nu_i)| = 0,
	\eea
	where $\nu_{\mathbf{r}_{\mu}}, \nu_{\mathbf{r}_i}$ are the discrete measures approximating $\mu, \nu_{\mathbf{r}_i}$, respectively, as in (\ref{disc}). Then 
	\bea
	\lim_{\rho\rightarrow \infty}\lim_{\zeta \rightarrow 0}\sup_{\mu}|\frac{1}{n}\sum_{i=1}^{n}s_{iG}(x)W_2^{2}(\nu_{\mathbf{r}_{\mu}},\nu_{\mathbf{r}_i}) -\frac{1}{n}\sum_{i=1}^{n}s_{iG}(x)W_2^{2}(\mu,\nu_i)|= 0,
	\eea
	\bea
	\lim_{\rho\rightarrow \infty}\lim_{\zeta \rightarrow 0}\inf_{\mu}\frac{1}{n}\sum_{i=1}^{n}s_{iG}(x)W_2^{2}(\nu_{\mathbf{r}_{\mu}},\nu_{\mathbf{r}_i}) = \inf_{\mu}\frac{1}{n}\sum_{i=1}^{n}s_{iG}(x)W_2^{2}(\mu,\nu_i),
	\eea  and for the minimizer 
	$\hat{\mathbf{r}}_{G}(x,\rho)$ satisfying 
	\bea
	\frac{1}{n}\sum_{i=1}^{n}s_{iG}(x)W_2^{2}(\nu_{\hat{\mathbf{r}}_{G}(x,\rho)},\nu_{\mathbf{r}_i}) = \inf_{\mu}\frac{1}{n}\sum_{i=1}^{n}s_{iG}(x)W_2^{2}(\nu_{\mathbf{r}_{\mu}},\nu_{\mathbf{r}_i}),
	\eea
	one has that for  any $\delta>0$, there exist sufficiently large $\rho$ and sufficiently small $\zeta$ such that 
	\bea
	|\sum_{i=1}^{n}s_{iG}(x)W_2^{2}(\nu_{\hat{\mathbf{r}}_{G}(x,\rho)},\nu_{\mathbf{r}_i})- \sum_{i=1}^{n}s_{iG}(x)W_2^{2}(\hat{\mu}_{G}(x),\nu_i)| < \delta.
	\eea
	Combining this with  (\ref{eq3}) completes  the proof of
	\bea
	\lim_{\rho\rightarrow \infty}\lim_{\zeta\rightarrow 0}W_2(\nu_{\hat{\mathbf{r}}_{G}(x,\rho)},\hat{\mu}_{G}(x)) = 0.
	\eea
For the  derivation of 
	\bea
	\lim_{\rho\rightarrow \infty}\lim_{\zeta\rightarrow 0}W_2(\nu_{\hat{\mathbf{r}}_{L,h}(x,\rho)}, \hat{\mu}_{L,h}(x)) = 0
	\eea one proceeds analogously. 
\end{proof}

\section{Concluding Remarks}
In this paper we presented  global and local models for fitting probability measure responses for $d$-dimensional distributions in the 2-Wasserstein space in dependence on scalar or vector predictors.  The global model can be harnessed for  Wasserstein interpolation and extrapolation while the local model is primarily useful  for interpolation. Elucidating  connections between  extrapolation and the extension of geodesics in the 2-Wasserstein space for  both the population level and the sample level leads to a better understanding of  Wasserstein extrapolation.    

For numerical implementations, Sinkhorn divergence, as an approximation of Wasserstein distance, is practically relevant in order to relax the computation complexity an ease the computational burden. The convergence of the resulting Sinkhorn estimates to the targeted Wasserstein estimates is established for the case where the regularization parameter goes to infinity. In the framework of  an admissible family of probability measures, we established the convergence rate for the two estimates. The simulations and data applications indicate that the proposed methodology is useful when data samples consist of multivariate random distributions.

\bco

\appendix
\begin{center}
	{\Large {\bf Supplement}}
\end{center}
\section*{Proofs and Auxiliary Results}

\fi

\bco

\FloatBarrier
\section*{Simulation with Multivariate t Distribution}
In this simulation we generate predictor $X$ from uniform $[0,1]$ and $\nu|X$ from multivariate t distribution with degree of freedom $2$. The conditional mean and covariance given $X=x$ are generated the same way as in section 5.2. $\alpha|X=x \sim N ((0.4x+0.3\quad 0.4x+0.3)^T,\mathbf{I}_2)$ and covariance matrix from $\Sigma = V\Lambda V'$, where $V = \begin{bmatrix}
\frac{\sqrt{2}}{2}& \frac{\sqrt{2}}{2}\\
-\frac{\sqrt{2}}{2}& \frac{\sqrt{2}}{2}
\end{bmatrix}$, $\Lambda =  \diag(\lambda_1,\lambda_2)$ and $(\lambda_1,\lambda_2)|X=x \sim \frac{1}{100}N((1+0.5x\quad 1-0.5x)',0.01\mathbf{I}_2)$.

\begin{figure}[!htb]
	\centering
	\includegraphics[width=6cm]{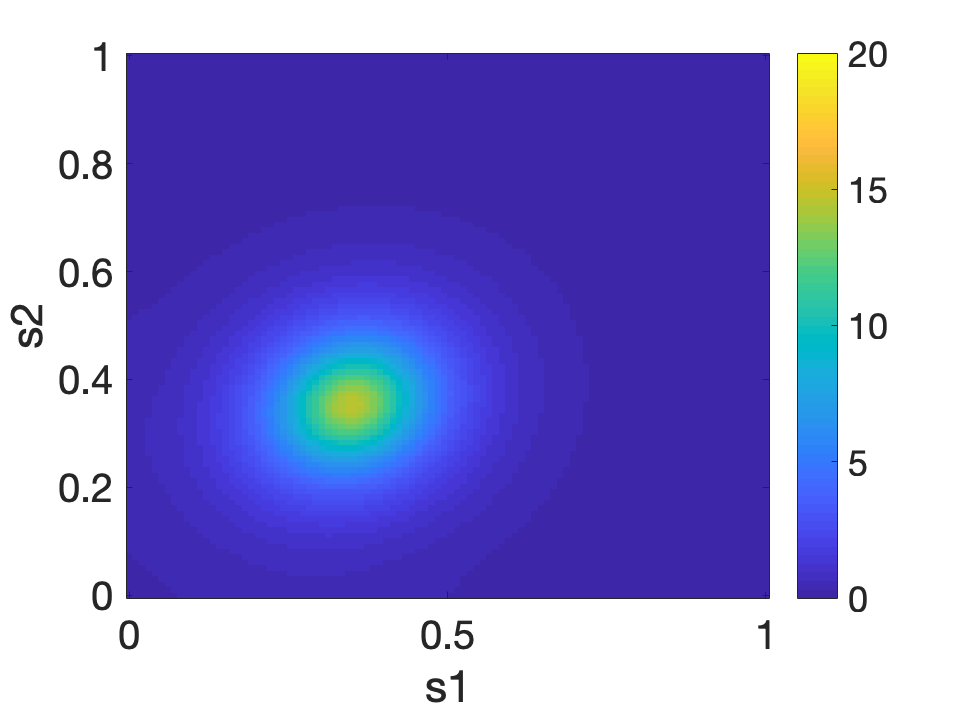}
	\includegraphics[width=6cm]{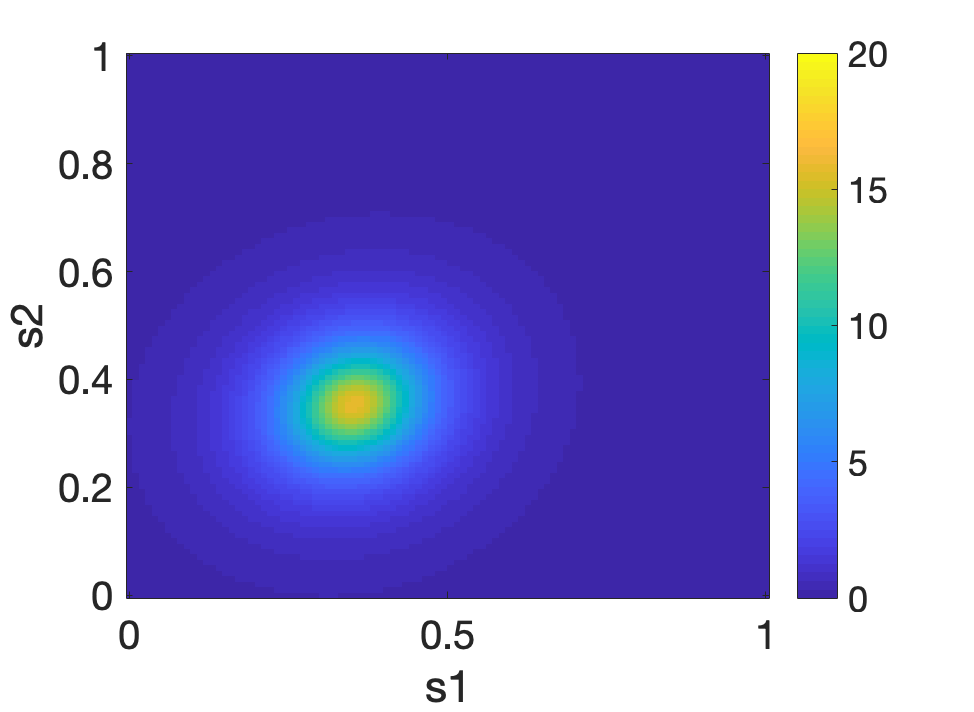}\\
	\includegraphics[width=6cm]{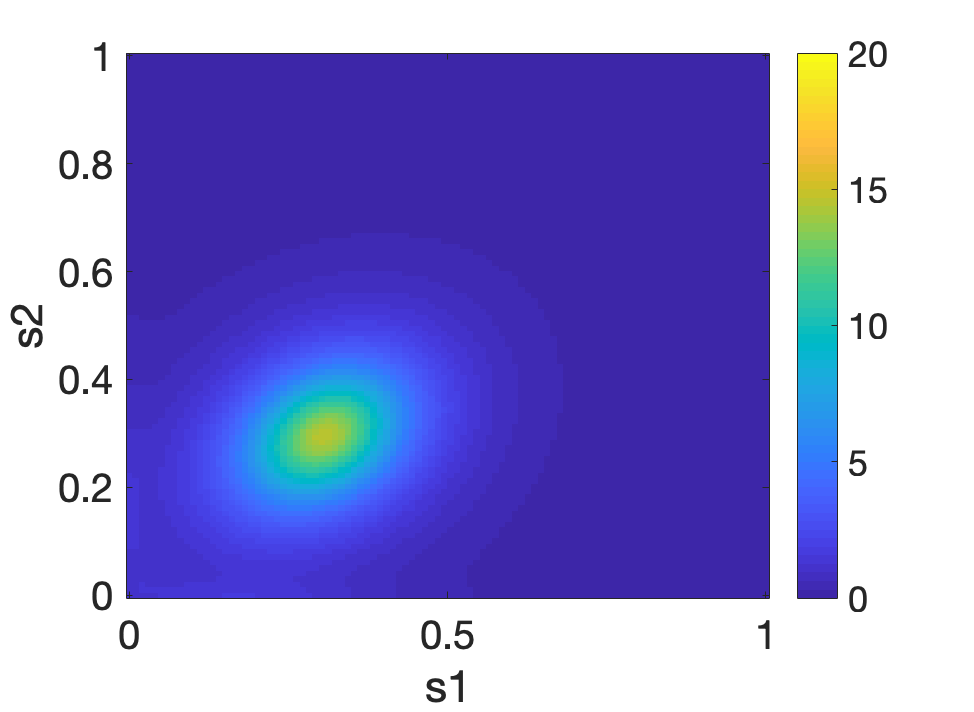}
	\includegraphics[width=6cm]{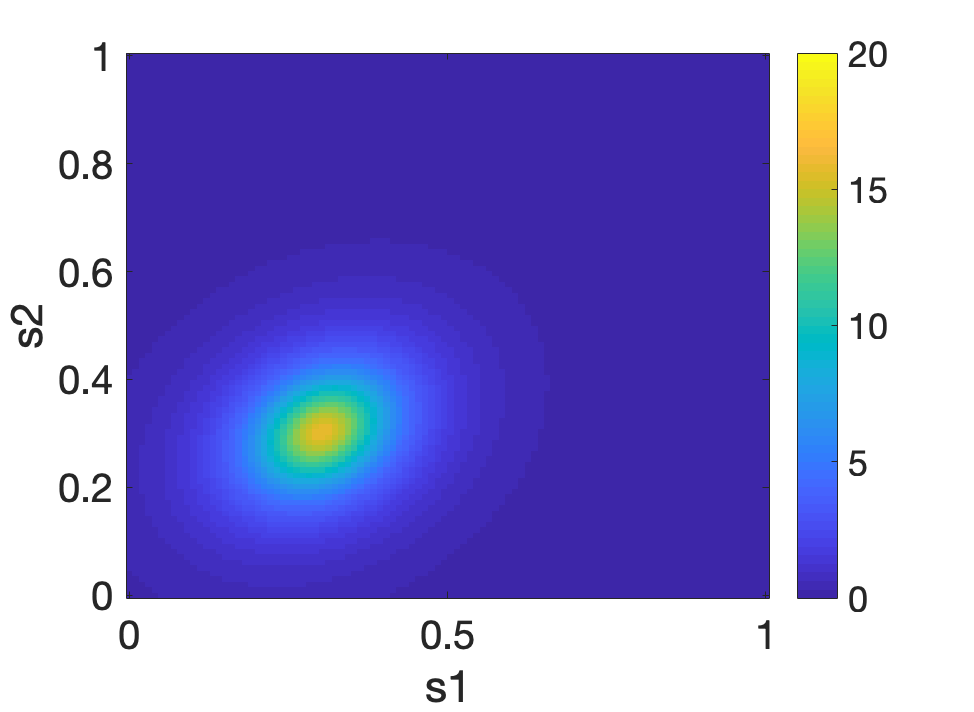}\\
	\caption{Left extrapolations by global model of multivariate t response simulation with $n=50$. The left panels are estimates and right panels are true models. The top two are $x=1.25$ and the bottom two $x= 1.5$.}
	\label{f3-13}
	\includegraphics[width=6cm]{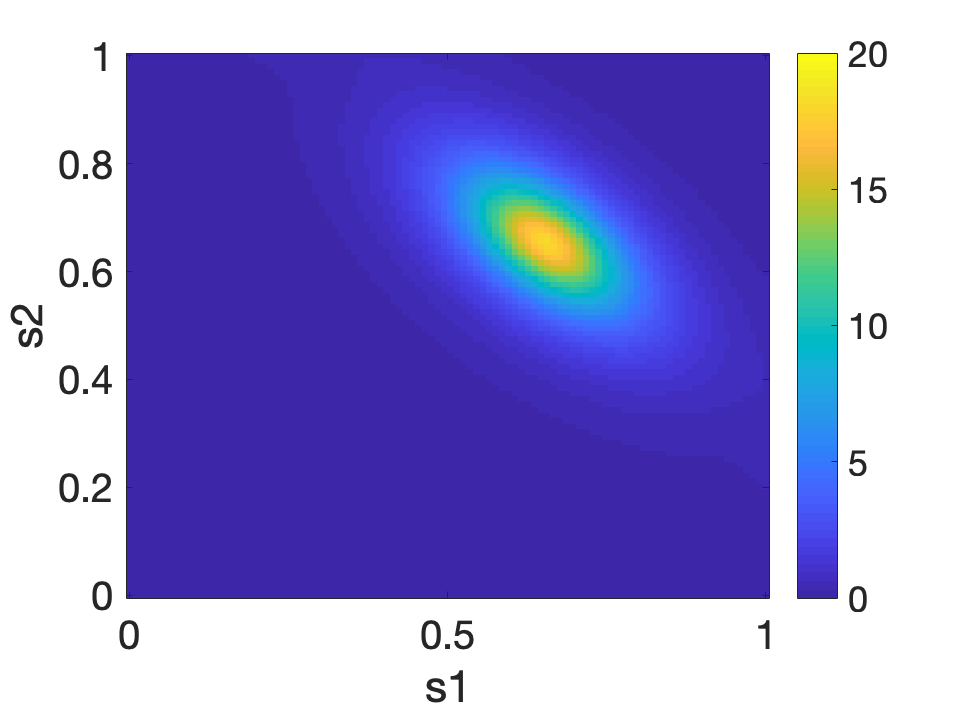}
	\includegraphics[width=6cm]{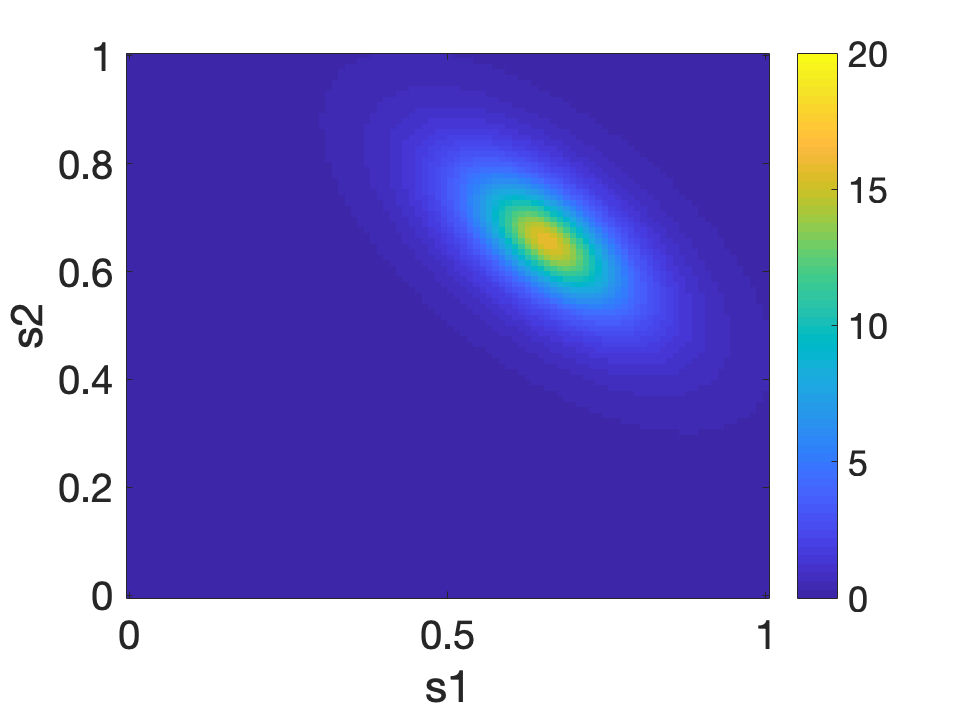}\\
	\includegraphics[width=6cm]{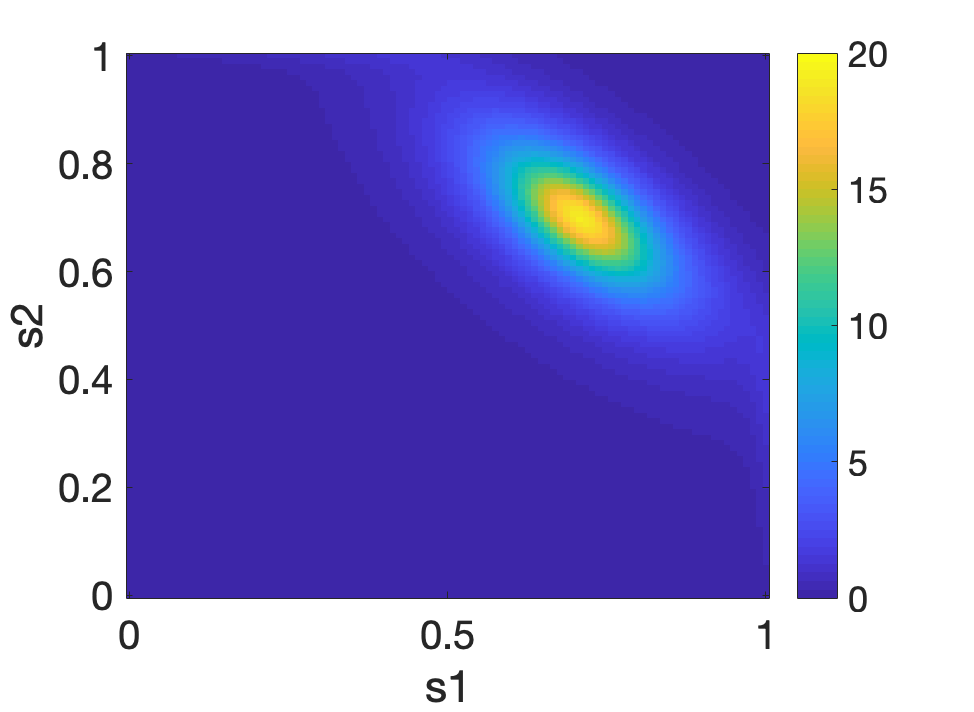}
	\includegraphics[width=6cm]{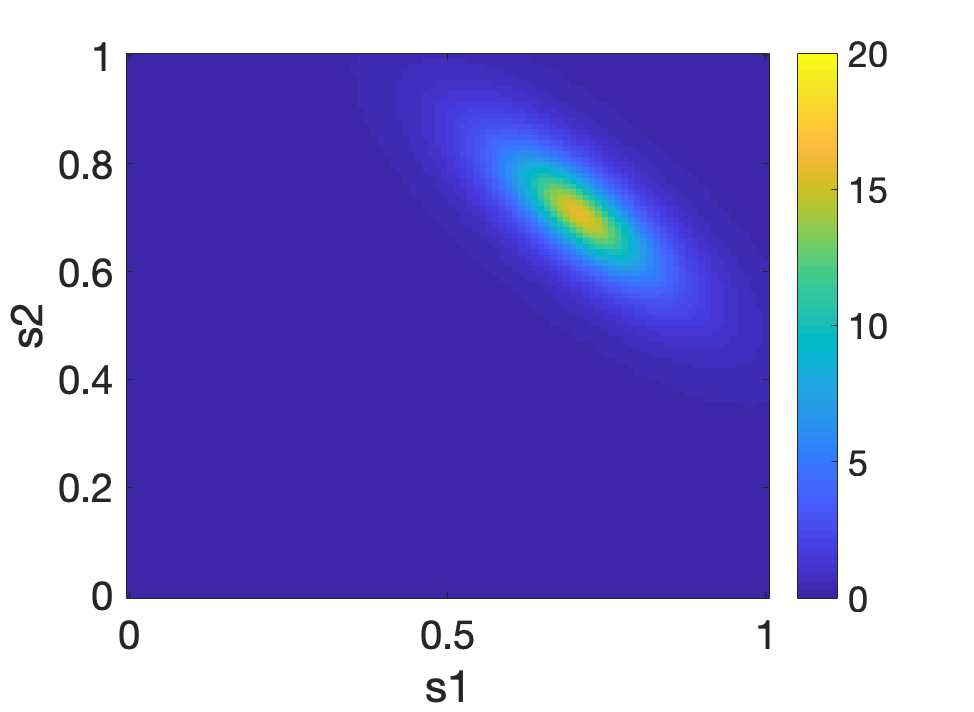}\\
	\caption{Right extrapolations by global model of multivariate t response simulation with $n=50$. The left panels are estimates and right panels are true models. The top two are $x=1.25$ and the bottom two $x= 1.5$.}
	\label{f3-14}
\end{figure}

\begin{figure}[!htb]
	\centering
	\includegraphics[width=6.5cm]{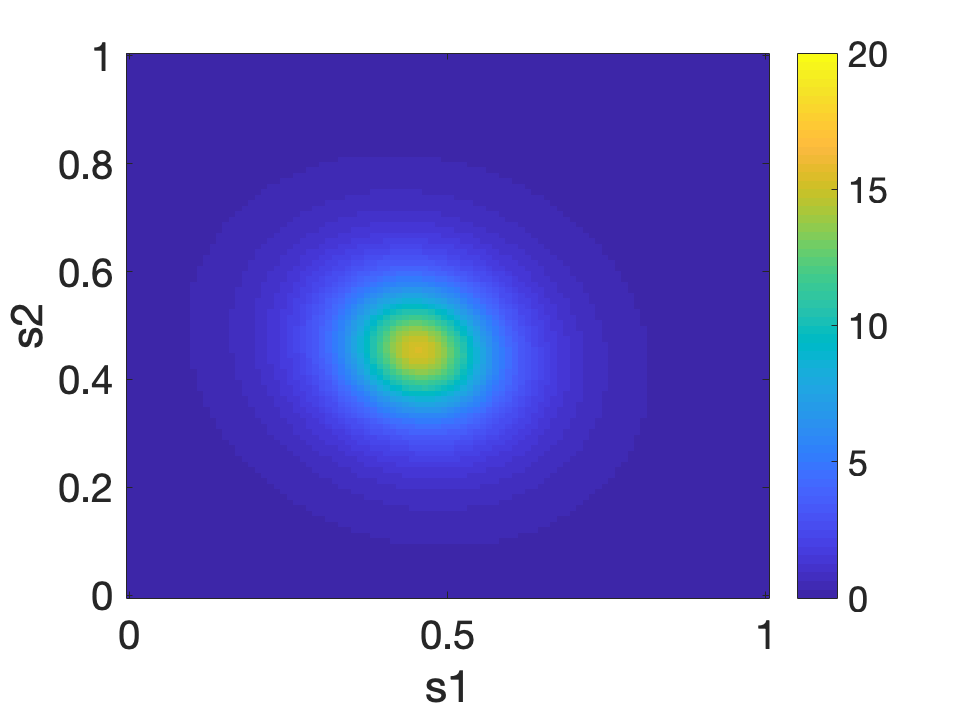}
	\includegraphics[width=6.5cm]{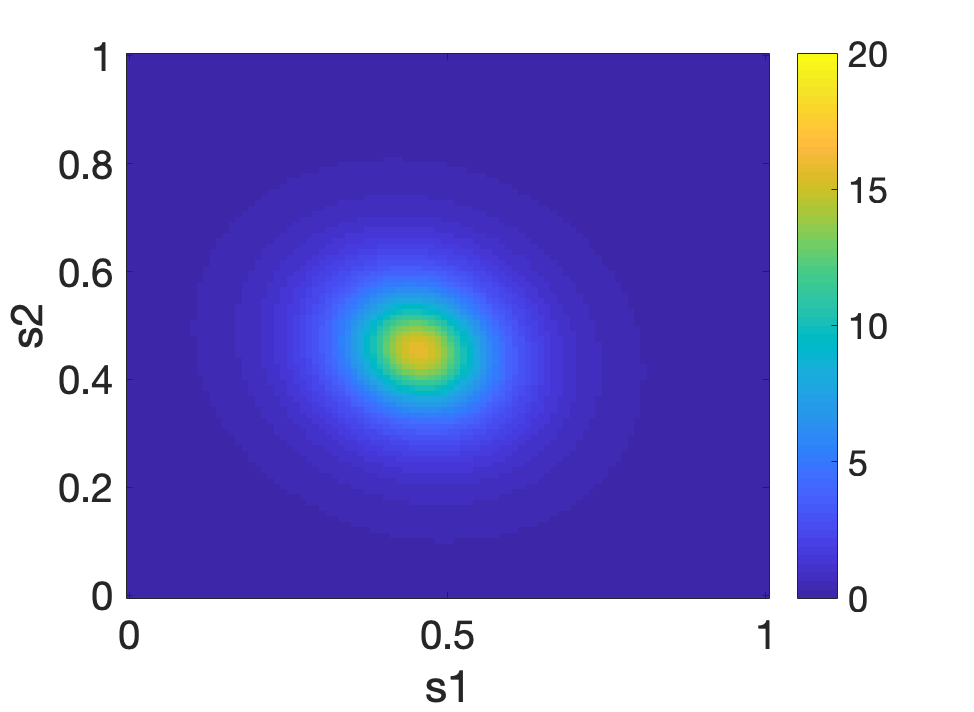}\\
	\includegraphics[width=6.5cm]{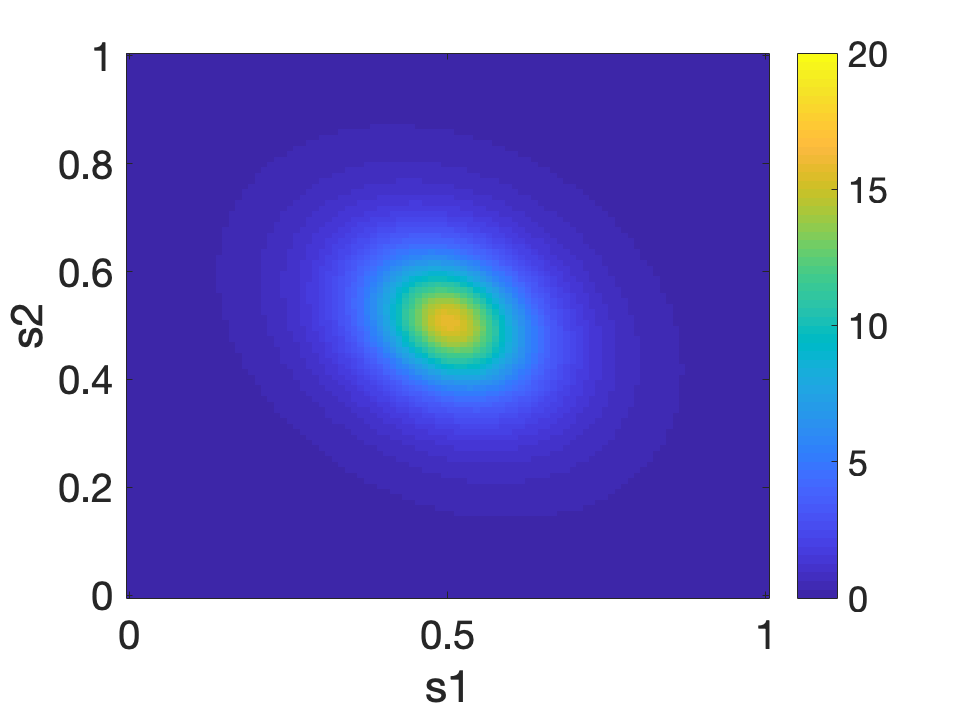}
	\includegraphics[width=6.5cm]{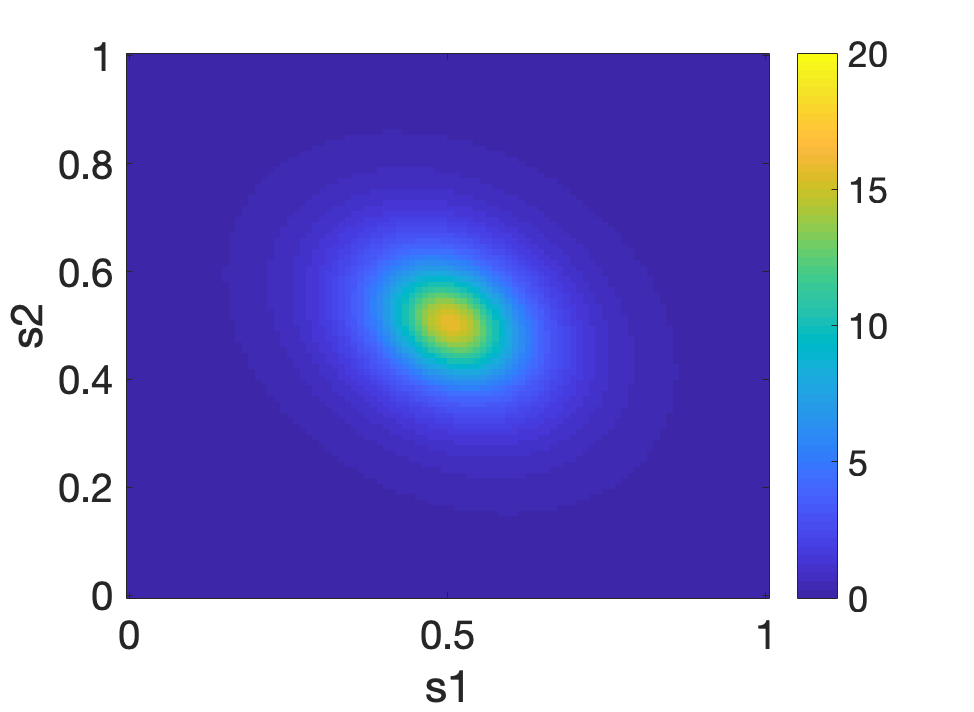}\\
	\includegraphics[width=6.5cm]{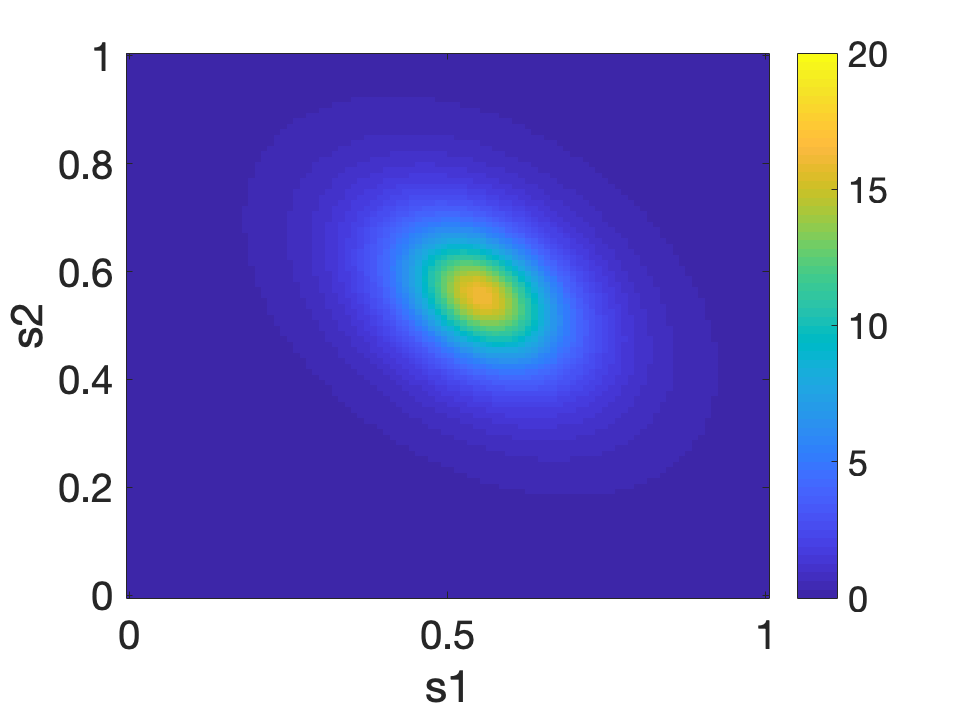}
	\includegraphics[width=6.5cm]{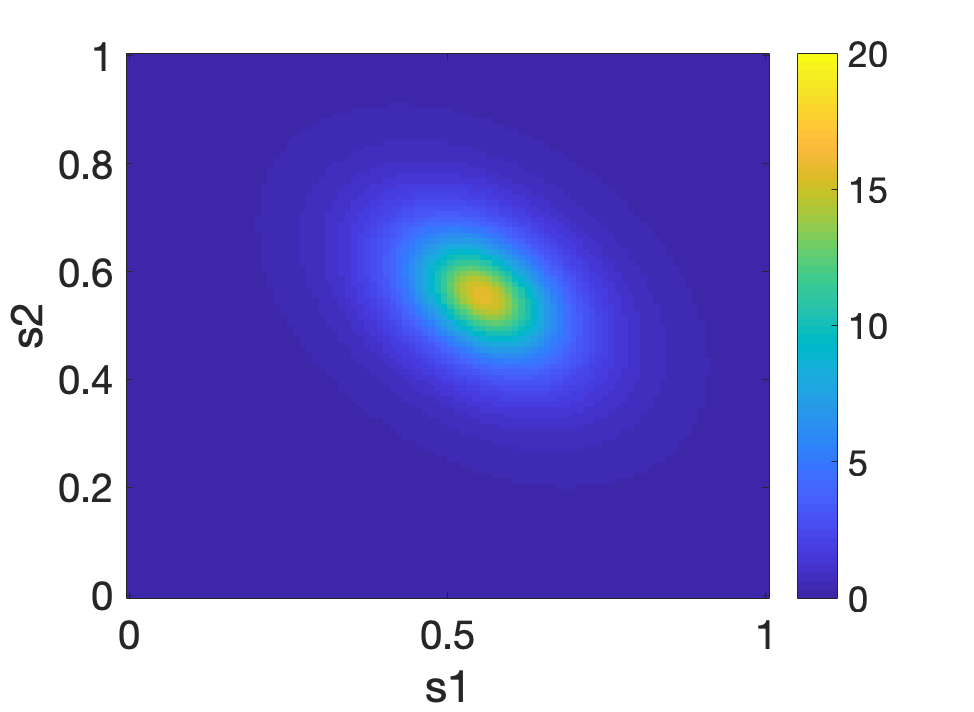}\\
	\caption{Interpolations by global model of multivariate t response simulation with $n=50$.  The left panels are estimates and right panels are true models. The top row is $x=0.25$, middle row $x=0.5$ and the bottom row $x= 0.75$.}
	\label{f3-15}
\end{figure}

\begin{figure}[!htb]
	\centering
	\includegraphics[width=7cm]{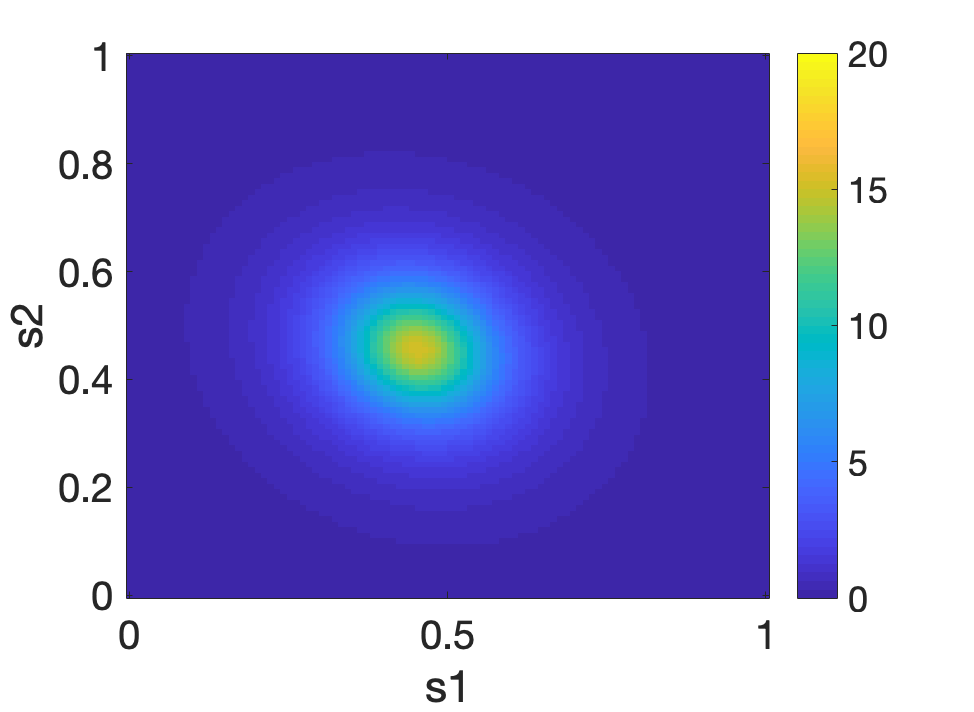}
	\includegraphics[width=7cm]{mvt_g0_25_t}\\
	\includegraphics[width=7cm]{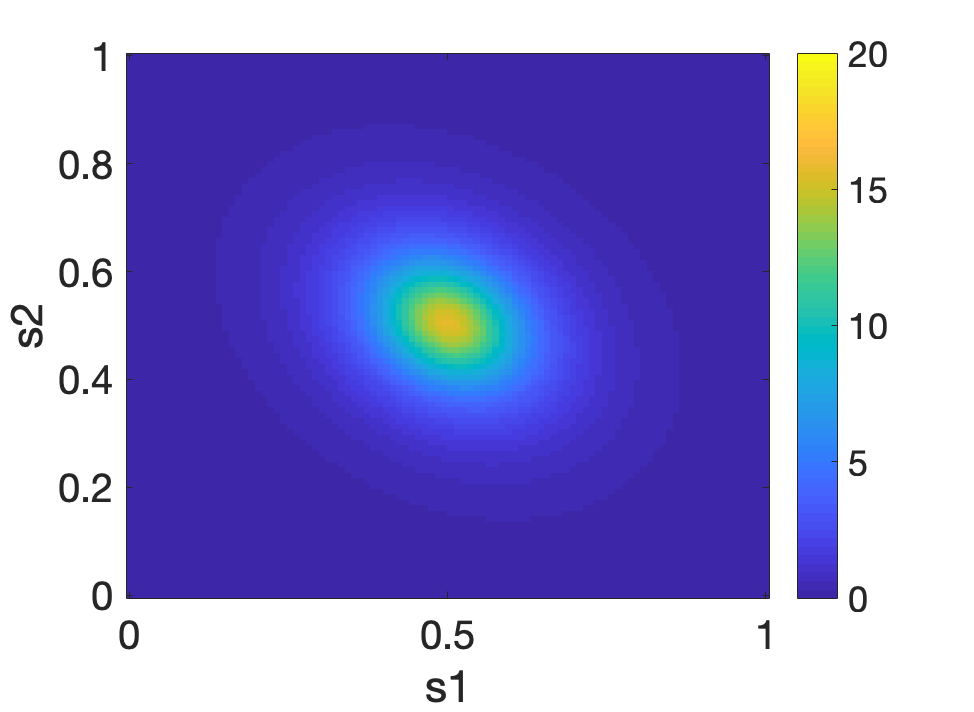}
	\includegraphics[width=7cm]{mvt_g0_5_t}\\
	\includegraphics[width=7cm]{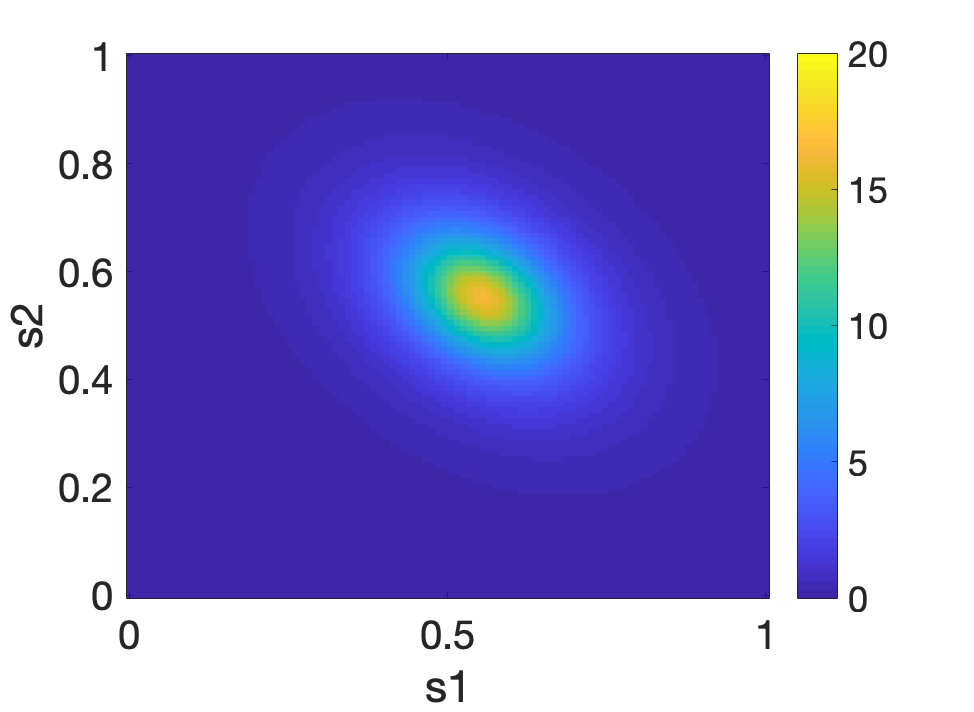}
	\includegraphics[width=7cm]{mvt_g0_75_t}\\
	\caption{Interpolations by local model of multivariate t response simulation with $n=50$. The left panels are estimates and right panels are true models. The top row is $x=0.25$, middle row $x=0.5$ and the bottom row $x= 0.75$.}
	\label{f3-16}
\end{figure}

The fitted results of global and local models are provided in figure \ref{f3-13}, \ref{f3-14},  \ref{f3-15} and \ref{f3-16}, where we can see that the method is also working well for this heavier tail family of distribution. The EMISEs of $100$ runs are provided in \ref{tb3}. The error is a little higher than the Gaussian case probably because of the heavier tail not included in the compact support.

\vspace{1.5in}
\begin{table}[!htb]
	\centering
	\begin{tabular}{|c|c|c|c|c|}
		\hline 
		Sample size  & $n=50$ & $n=100$ & $n=150$ & $n=200$\\ 
		\hline
		$\rho$ & $\frac{1}{\rho} = 5.5$ & $\frac{1}{\rho} = 4.5$ & $\frac{1}{\rho} = 3.5$ & $\frac{1}{\rho} = 2.5$\\
		\hline 
		Global  extrapolation& 0.0813 & 0.0796 & 0.0783 & 0.0774\\ 
		\hline 
		Global  interpolation& 0.0811  & 0.0795 & 0.0781 & 0.0772 \\
		\hline
		Local  interpolation& 0.0755 & 0.0743 & 0.0734  & 0.0723 \\
		\hline
	\end{tabular}
	\caption{Table of EMISE for different sample size in 2d simulation}
	\label{tb3}
\end{table}

\fi

\newpage

\section*{ Supplement: Additional Figures}
\begin{figure}[!htb]
	\centering
	\includegraphics[width=6cm]{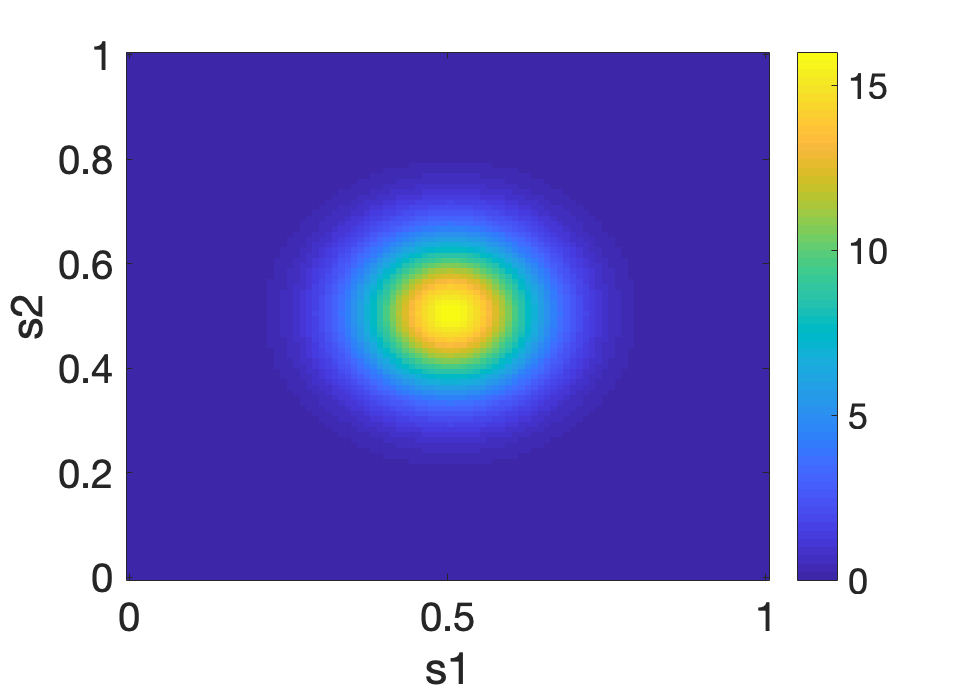}
\includegraphics[width=6cm]{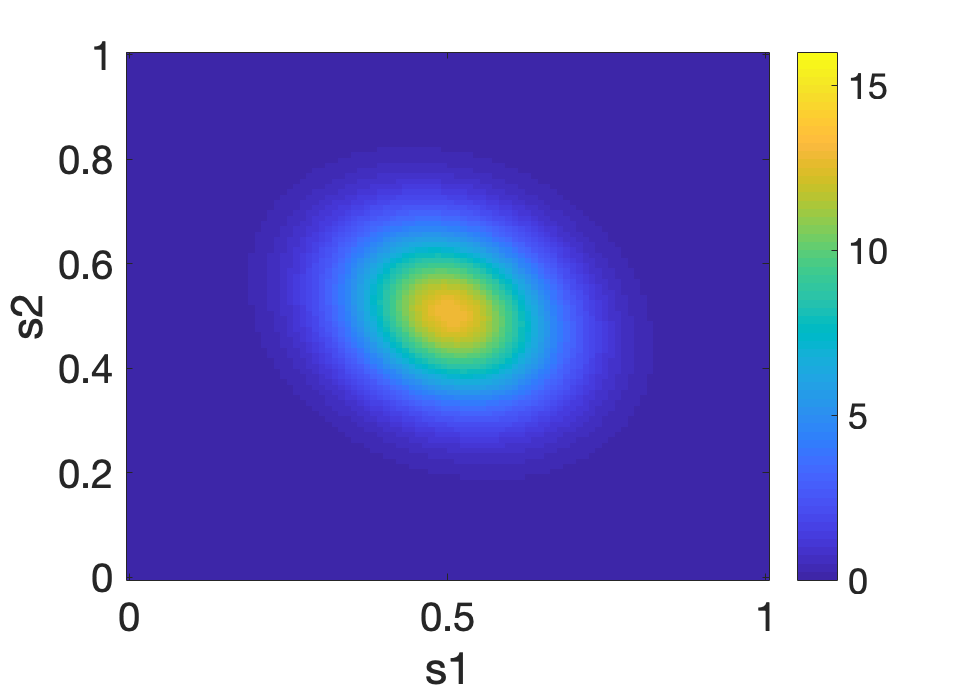}\\
\includegraphics[width=6cm]{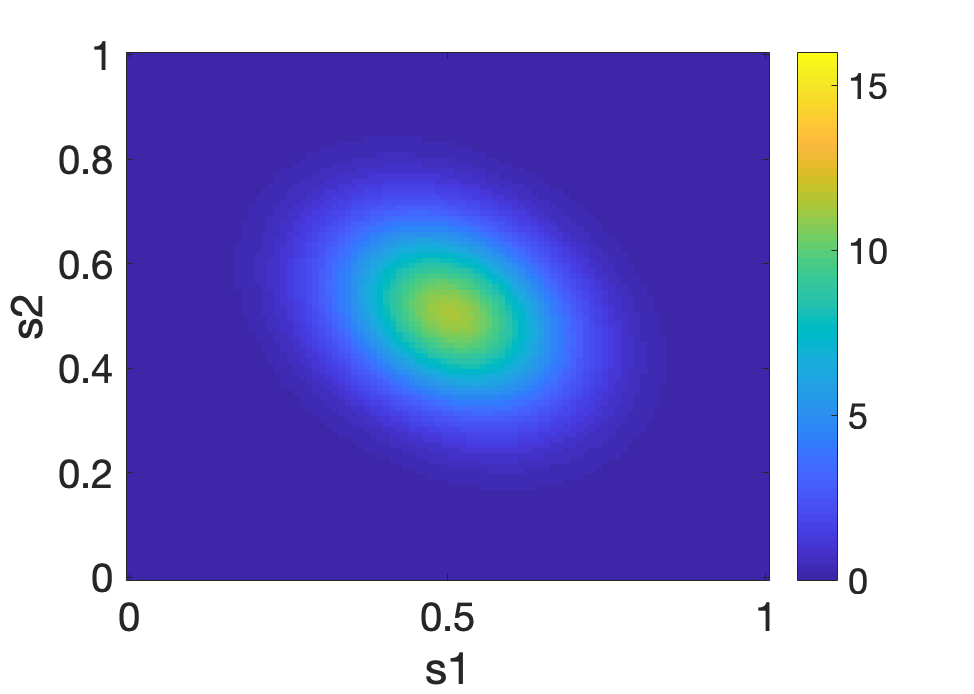}
\includegraphics[width=6cm]{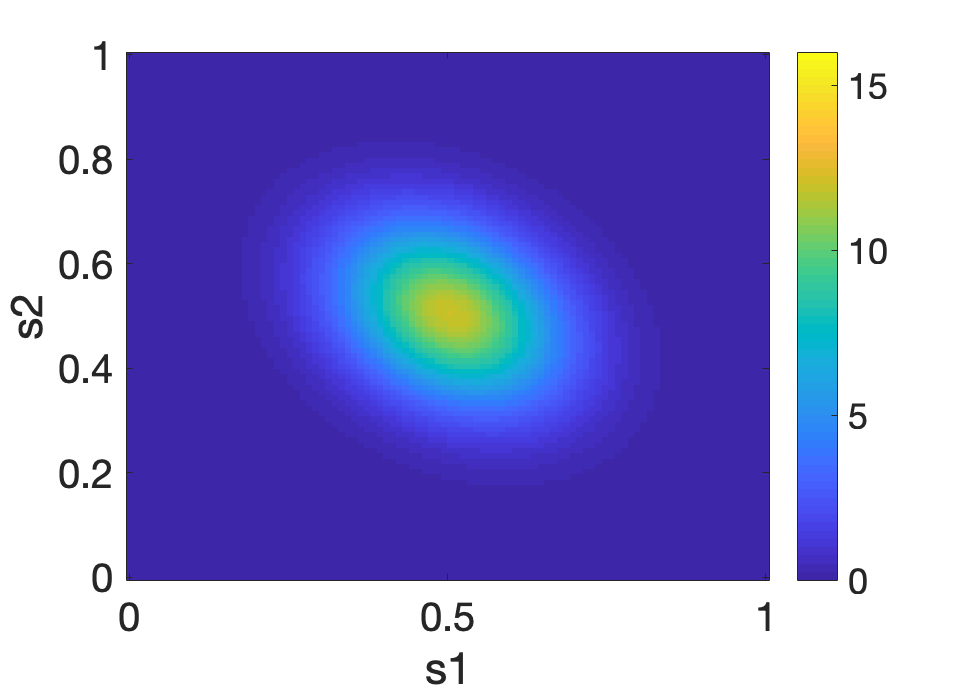}\\
\caption{Extrapolation by fitting the  global model for two-dimensional distributions when $n=2$. The top panels are heat maps for the distributions $(0, N((5,5), [ 1\quad 0 ; 0 \quad 1]))$ and $(1, N((5,5), [ 1.25\quad -0.25 ; -0.25 \quad 1.25]))$, corresponding to the distributions at $x=0$ and $x=1$.  The bottom left panel displays the global extrapolation  at $x= 1.5$ and the bottom right is the true extrapolation.}
	\label{f3-6_3}
\end{figure}

\begin{figure}[!htb]
\centering
	\includegraphics[width=6cm]{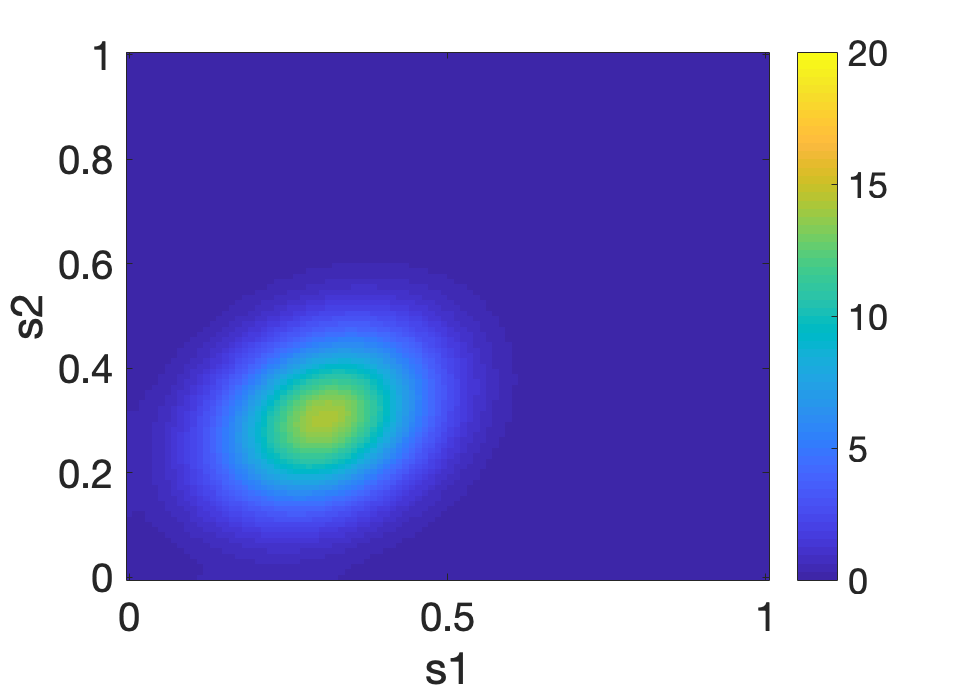}
\includegraphics[width=6cm]{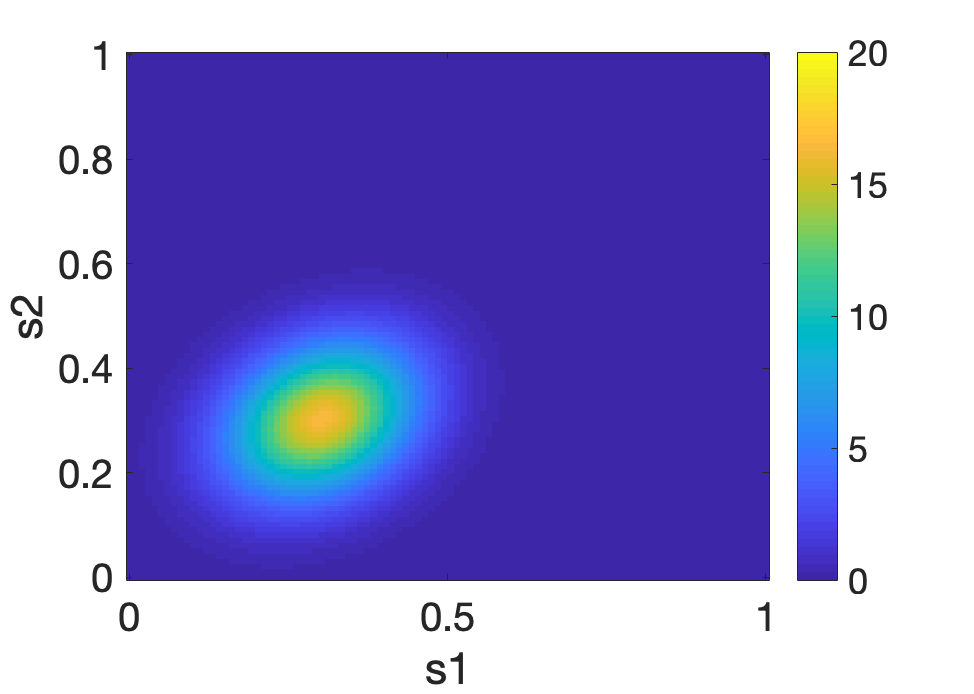}\\
\includegraphics[width=6cm]{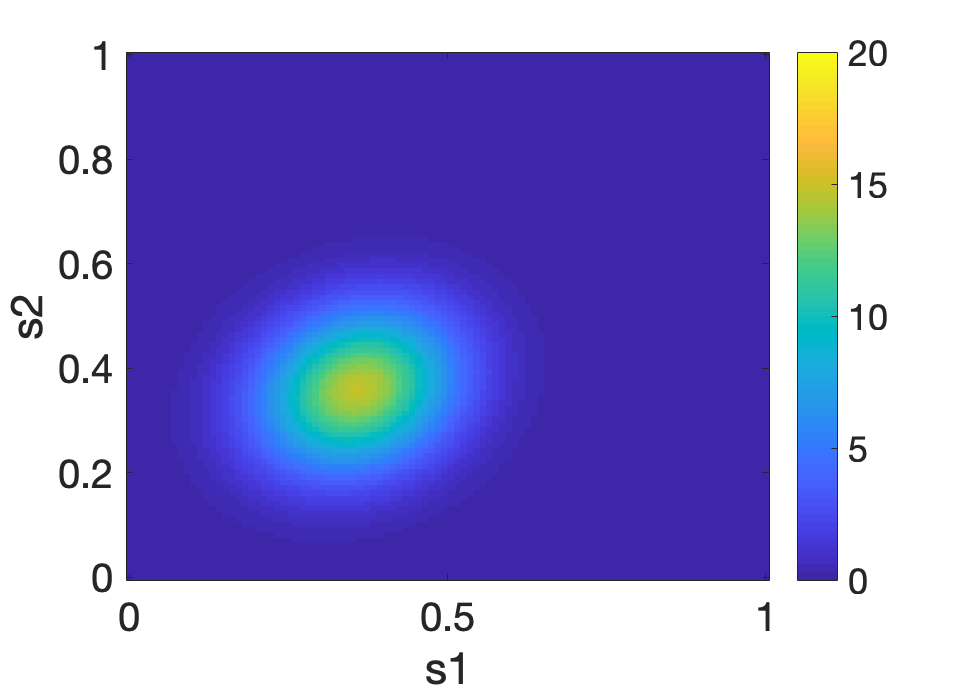}
\includegraphics[width=6cm]{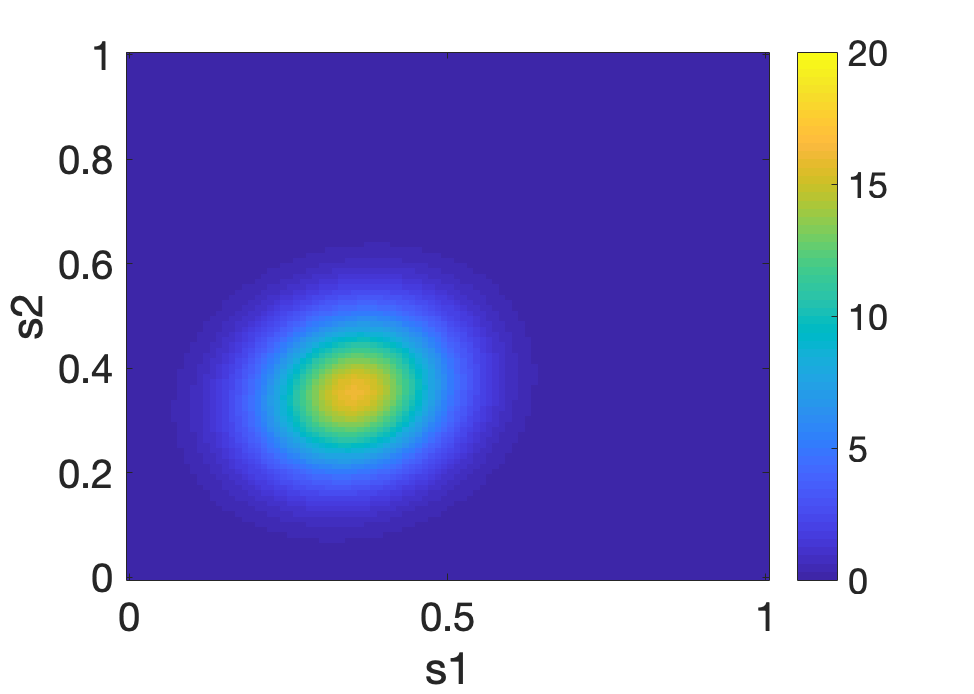}\\
\caption{Left side side extrapolations obtained  by fitting the global model (left panels) for 2-dimensional simulated data with $n=50$, where predictors are randomly sampled on $[0,1]$, and true extrapolated Wasserstein geodesic (right panels), for extrapolation levels at $x=-0.5$ (upper panels) and  at $x= -0.5$ (lower panels).} 
	\label{f3-4_2}
	\centering
	\includegraphics[width=6cm]{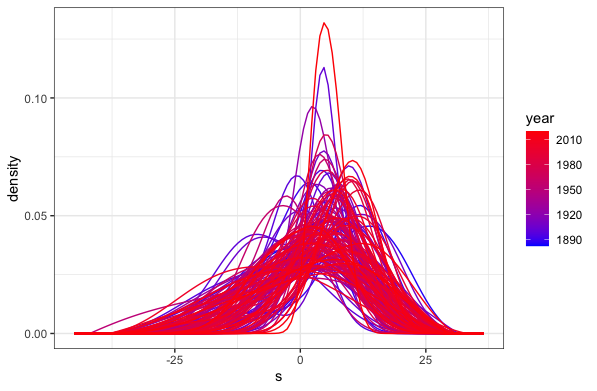}
	\includegraphics[width=6cm]{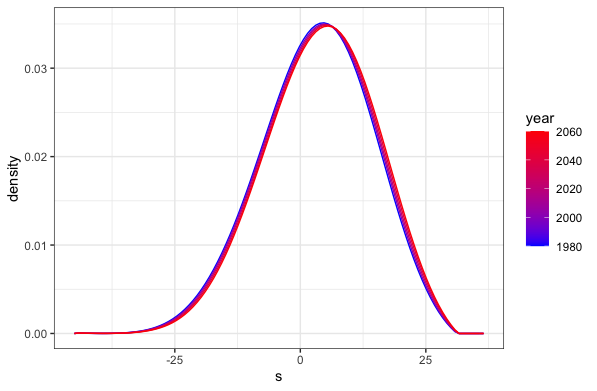}\\
	\includegraphics[width=6cm]{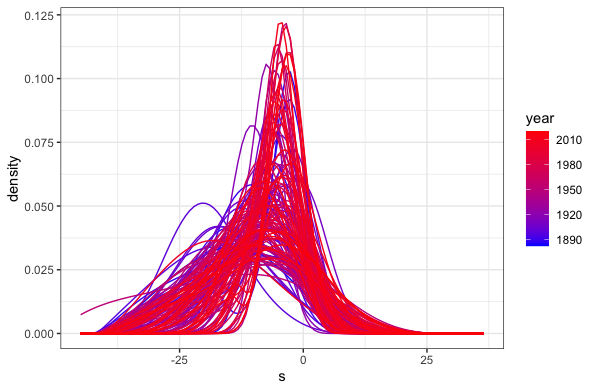}
	\includegraphics[width=6cm]{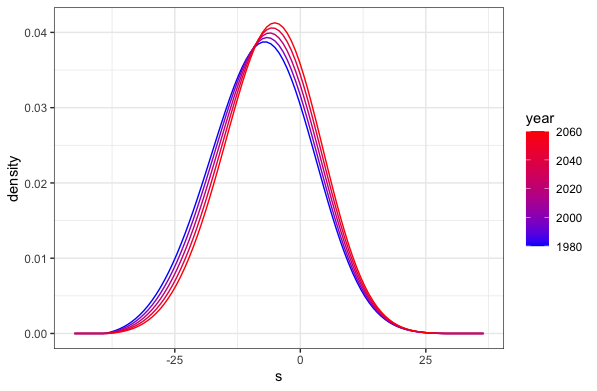}
	\caption{Marginal densities of maximum and minimum temperature in March for Calgary, with the same layout as in Figure \ref{f3-9_2}.}
	\label{f3-10_1}
\end{figure}

\begin{figure}[!htb]
\centering
	\includegraphics[width=6.5cm]{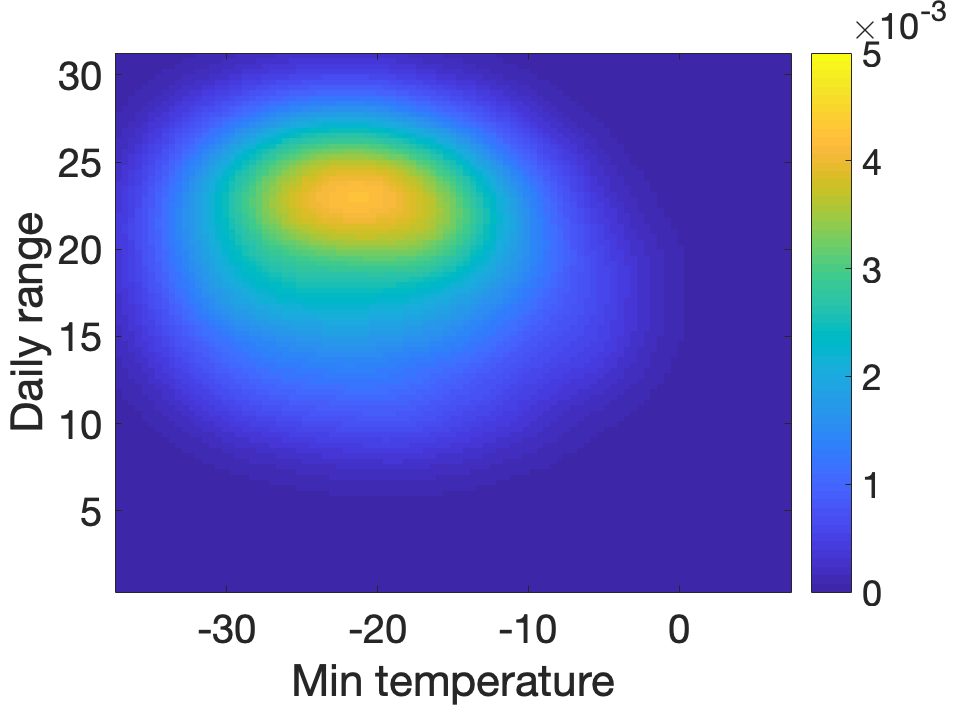}\\
	\includegraphics[width=6.5cm]{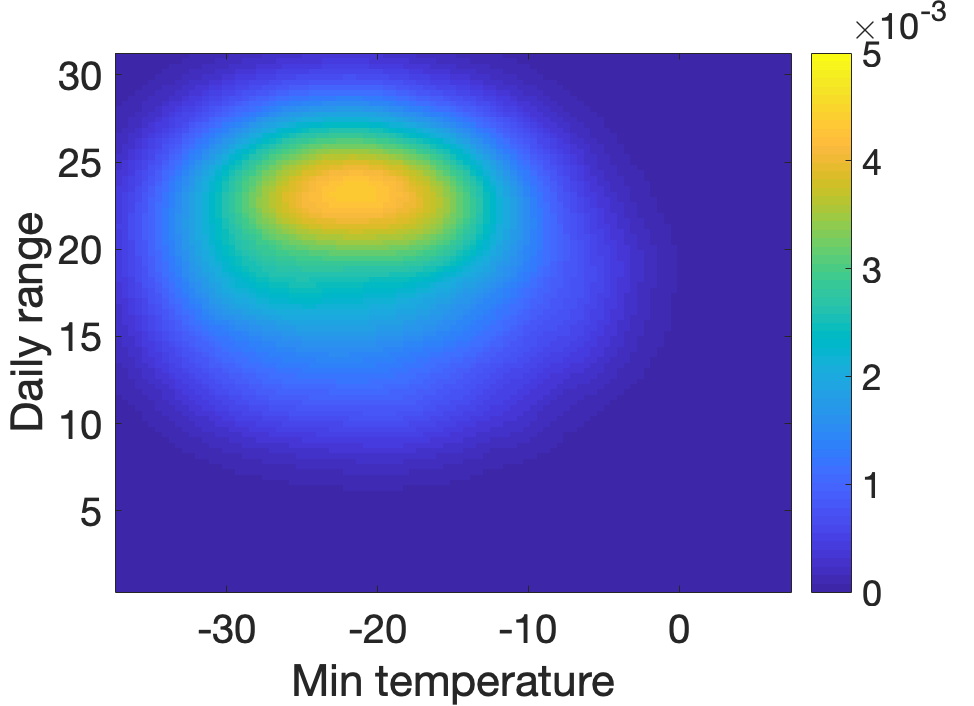}
	\includegraphics[width=6.5cm]{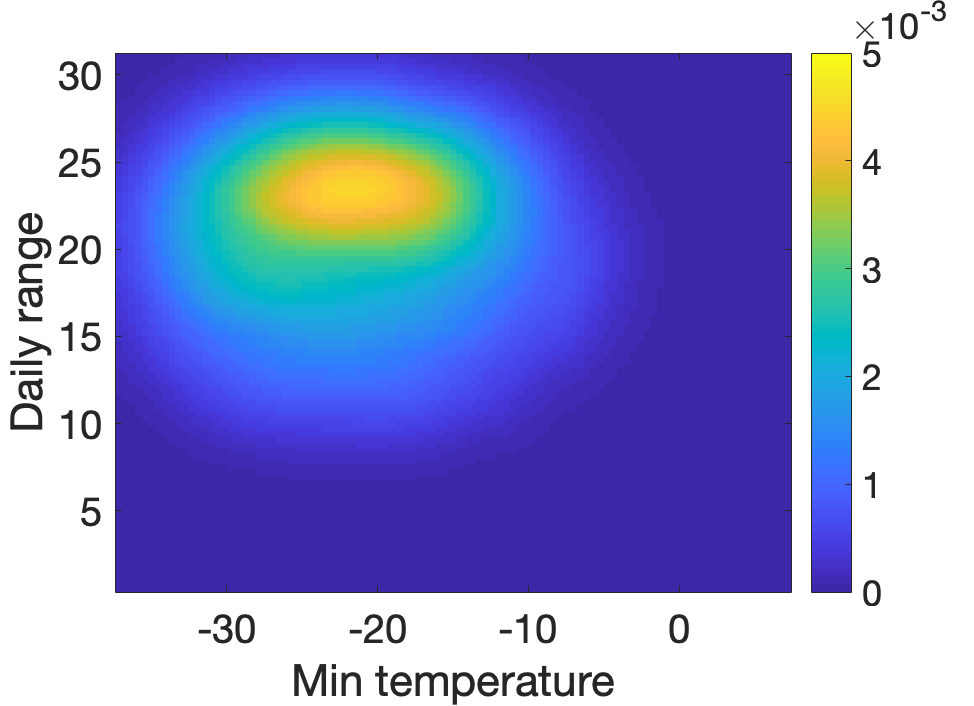}\\
	\includegraphics[width=6.5cm]{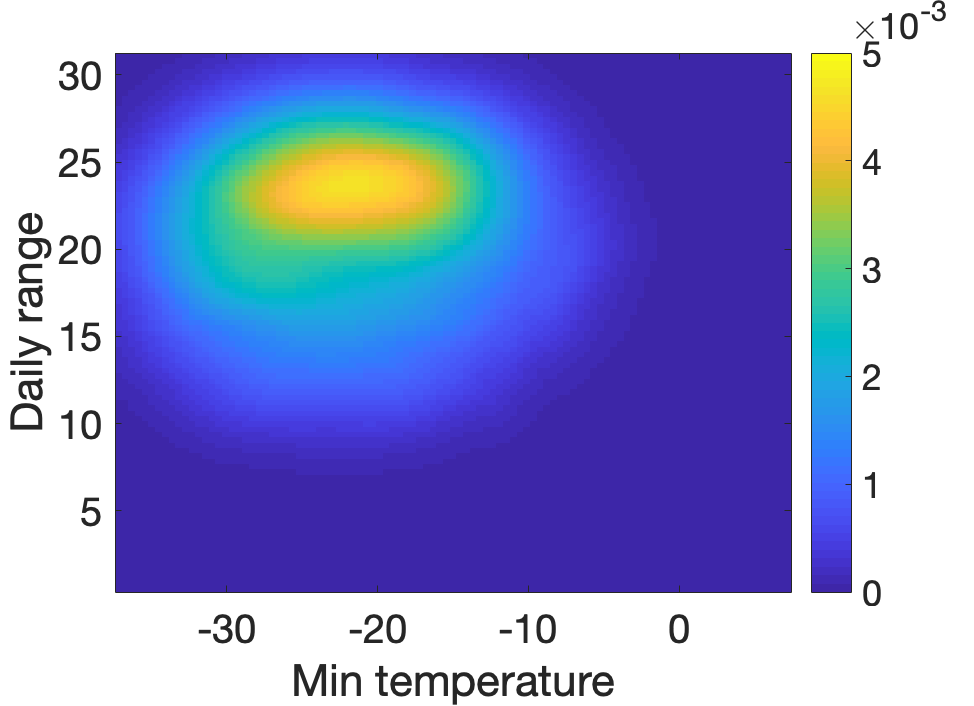}
	\includegraphics[width=6.5cm]{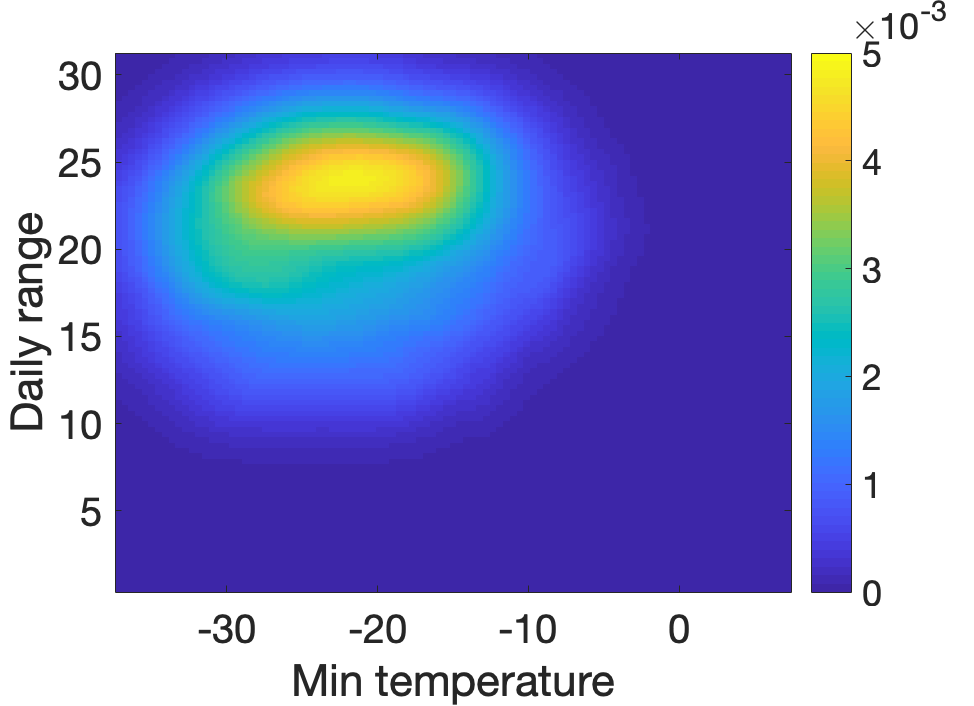}
	\caption{Globally fitted joint distributions of the maximum-minimum (temperature range) and  the minimum temperature in March. First row: Extrapolation for  the years $1980$. Second row: For the years $2000, 2020$.  Third row: For the years $2040, 2060$.}
	\label{f3-10}
\end{figure}

\begin{figure}[!htb]
	\centering
	\includegraphics[width=6.5cm]{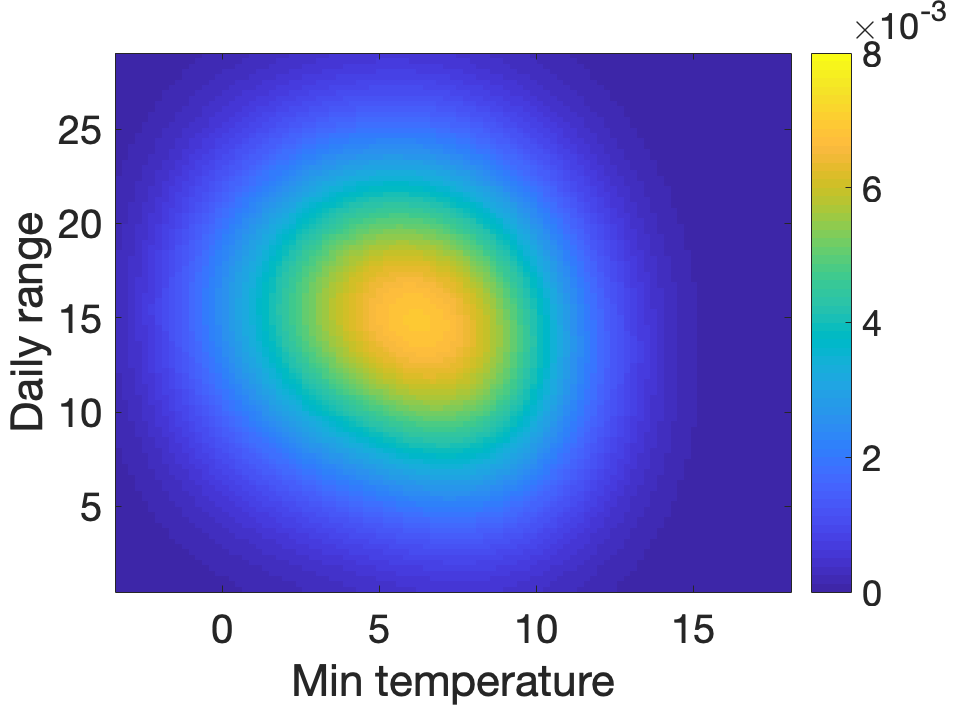}\\
	\includegraphics[width=6.5cm]{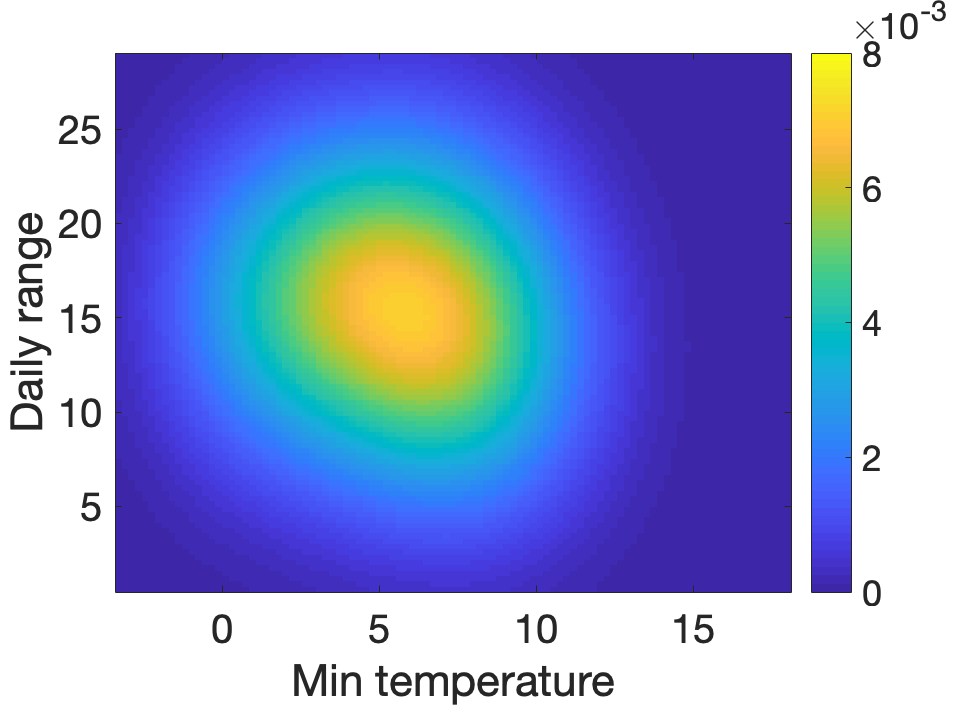}
	\includegraphics[width=6.5cm]{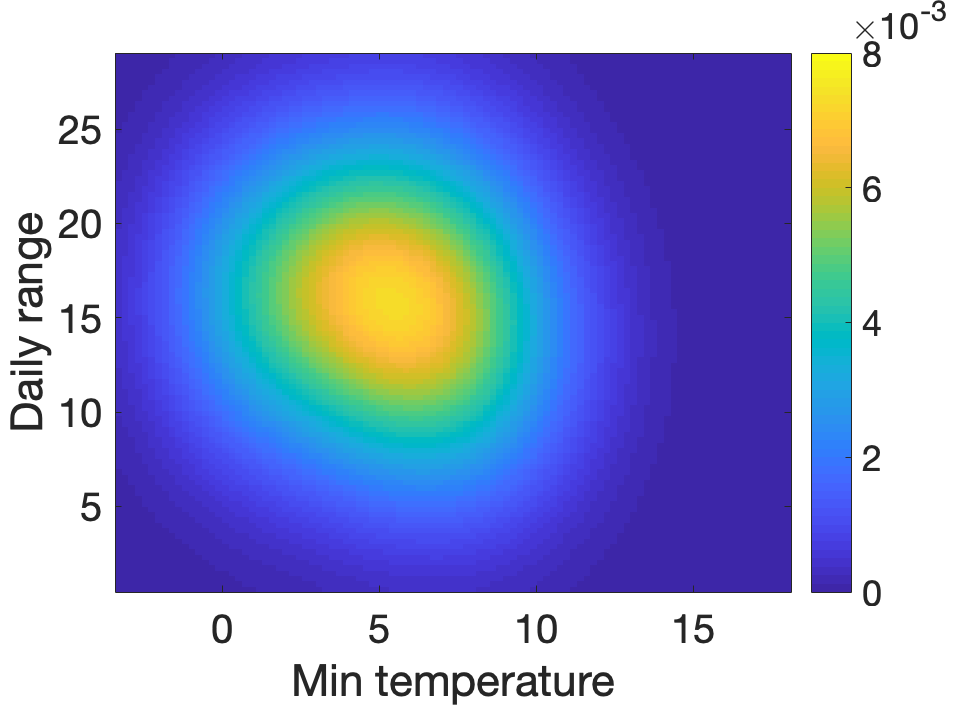}\\
	\includegraphics[width=6.5cm]{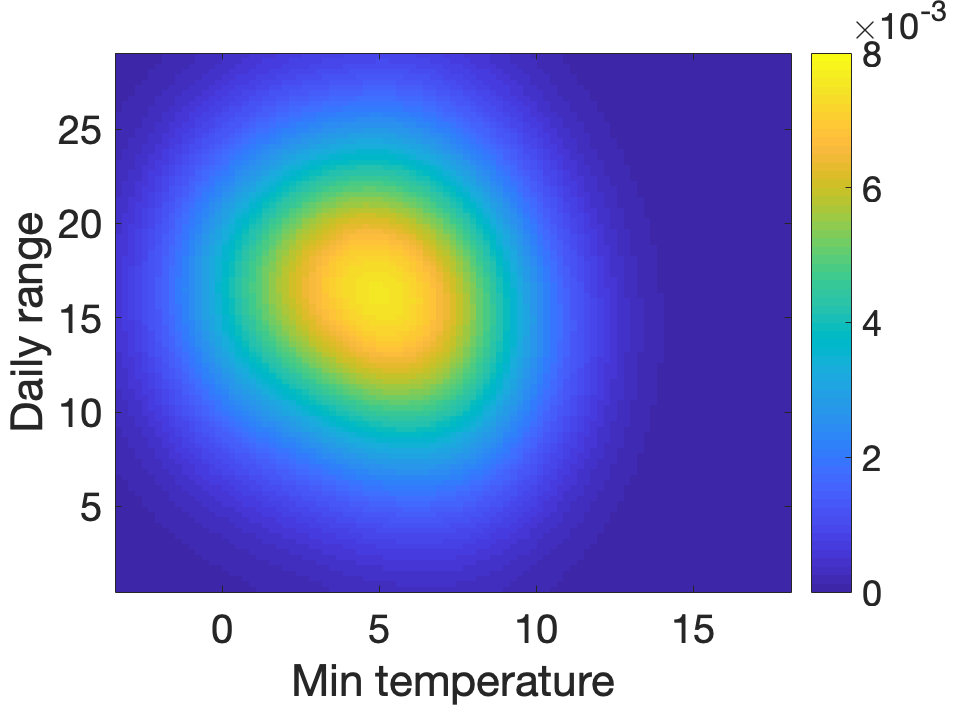}
	\includegraphics[width=6.5cm]{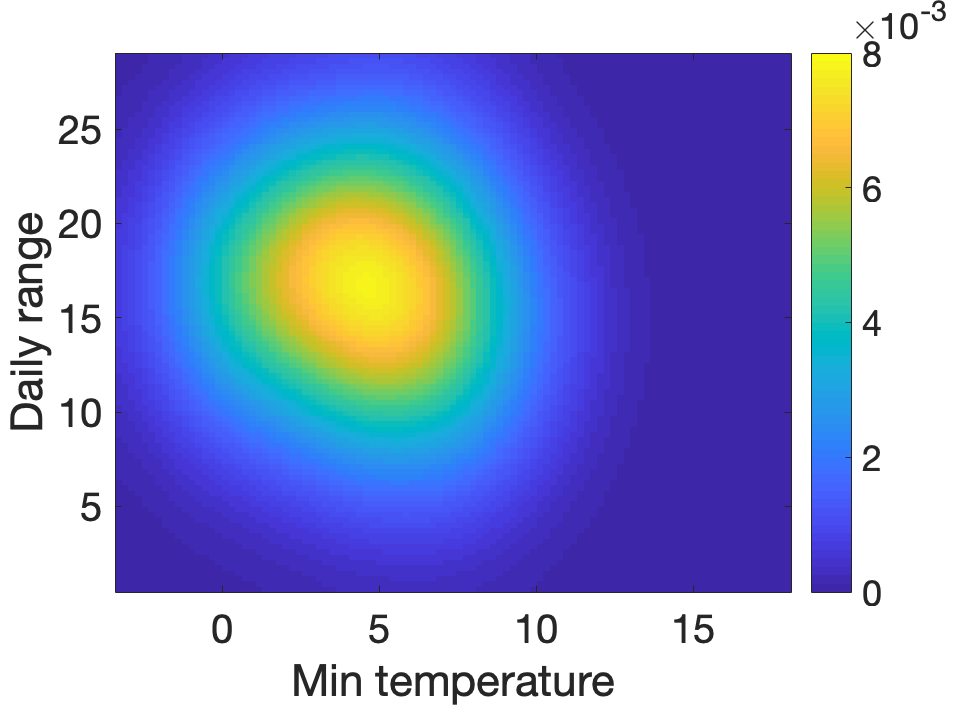}
	\caption{Same as Figure \ref{f3-10} for June.}
	\label{f3-12}
\end{figure}


\begin{thebibliography}{43}
	\expandafter\ifx\csname natexlab\endcsname\relax\def\natexlab#1{#1}\fi
	\expandafter\ifx\csname url\endcsname\relax
	\def\url#1{\texttt{#1}}\fi
	\expandafter\ifx\csname urlprefix\endcsname\relax\def\urlprefix{URL }\fi
	
	\bibitem[{Agueh and Carlier(2011)}]{agueh:2011}
	\textsc{Agueh, M.} and \textsc{Carlier, G.} (2011).
	\newblock Barycenters in the {W}asserstein space.
	\newblock \textit{SIAM Journal on Mathematical Analysis} \textbf{43} 904--924.
	
	\bibitem[{Ahidar-Coutrix et~al.(2019)Ahidar-Coutrix, Le~Gouic and
		Paris}]{ahidar:2019}
	\textsc{Ahidar-Coutrix, A.}, \textsc{Le~Gouic, T.} and \textsc{Paris, Q.}
	(2019).
	\newblock Convergence rates for empirical barycenters in metric spaces:
	curvature, convexity and extendable geodesics.
	\newblock \textit{Probability Theory and Related Fields}  1--46.
	
	\bibitem[{{\'A}lvarez-Esteban et~al.(2016){\'A}lvarez-Esteban, Del~Barrio,
		Cuesta-Albertos and Matr{\'a}n}]{alvarez:2016}
	\textsc{{\'A}lvarez-Esteban, P.~C.}, \textsc{Del~Barrio, E.},
	\textsc{Cuesta-Albertos, J.} and \textsc{Matr{\'a}n, C.} (2016).
	\newblock A fixed-point approach to barycenters in {W}asserstein space.
	\newblock \textit{Journal of Mathematical Analysis and Applications}
	\textbf{441} 744--762.
	
	\bibitem[{Ambrosio et~al.(2008)Ambrosio, Gigli and Savar{\'e}}]{ambrosio:2008}
	\textsc{Ambrosio, L.}, \textsc{Gigli, N.} and \textsc{Savar{\'e}, G.} (2008).
	\newblock \textit{Gradient Flows in Metric Spaces and in the Space of
		Probability Measures}.
	\newblock Springer Science \& Business Media.
	
	\bibitem[{Anderes et~al.(2016)Anderes, Borgwardt and Miller}]{anderes:2016}
	\textsc{Anderes, E.}, \textsc{Borgwardt, S.} and \textsc{Miller, J.} (2016).
	\newblock Discrete {W}asserstein barycenters: optimal transport for discrete
	data.
	\newblock \textit{Mathematical Methods of Operations Research} \textbf{84}
	389--409.
	
	\bibitem[{Bigot(2019)}]{bigo:19}
	\textsc{Bigot, J.} (2019).
	\newblock Statistical data analysis in the {W}asserstein space.
	\newblock \textit{arXiv:1907.08417} .
	
	\bibitem[{Bigot et~al.(2018)Bigot, Cazelles and Papadakis}]{bigo:18}
	\textsc{Bigot, J.}, \textsc{Cazelles, E.} and \textsc{Papadakis, N.} (2018).
	\newblock Data-driven regularization of {W}asserstein barycenters with an
	application to multivariate density registration.
	\newblock \textit{arXiv preprint arXiv:1804.08962} .
	
	\bibitem[{Bigot et~al.(2019)Bigot, Cazelles and Papadakis}]{bigot:2019}
	\textsc{Bigot, J.}, \textsc{Cazelles, E.} and \textsc{Papadakis, N.} (2019).
	\newblock Penalization of barycenters in the {W}asserstein space.
	\newblock \textit{SIAM Journal on Mathematical Analysis} \textbf{51}
	2261--2285.
	
	\bibitem[{Bigot et~al.(2017)Bigot, Gouet, Klein and L{\'o}pez}]{bigo:17}
	\textsc{Bigot, J.}, \textsc{Gouet, R.}, \textsc{Klein, T.} and
	\textsc{L{\'o}pez, A.} (2017).
	\newblock Geodesic {PCA} in the {W}asserstein space by convex {PCA}.
	\newblock \textit{Annales de l{'}Institut Henri Poincar{\'e} B: Probability and
		Statistics} \textbf{53} 1--26.
	
	\bibitem[{Boissard et~al.(2015)Boissard, Le~Gouic and Loubes}]{boissard:2015}
	\textsc{Boissard, E.}, \textsc{Le~Gouic, T.} and \textsc{Loubes, J.-M.} (2015).
	\newblock Distribution’s template estimate with {W}asserstein metrics.
	\newblock \textit{Bernoulli} \textbf{21} 740--759.
	
	\bibitem[{Buja et~al.(2019{\natexlab{a}})Buja, Brown, Berk, George, Pitkin,
		Traskin, Zhang and Zhao}]{buja:19}
	\textsc{Buja, A.}, \textsc{Brown, L.}, \textsc{Berk, R.}, \textsc{George, E.},
	\textsc{Pitkin, E.}, \textsc{Traskin, M.}, \textsc{Zhang, K.} and
	\textsc{Zhao, L.} (2019{\natexlab{a}}).
	\newblock Models as approximations {I}: Consequences illustrated with linear
	regression.
	\newblock \textit{Statistical Science} \textbf{34} 523--544.
	
	\bibitem[{Buja et~al.(2019{\natexlab{b}})Buja, Brown, Kuchibhotla, Berk, George
		and Zhao}]{buja:19:1}
	\textsc{Buja, A.}, \textsc{Brown, L.}, \textsc{Kuchibhotla, A.~K.},
	\textsc{Berk, R.}, \textsc{George, E.} and \textsc{Zhao, L.}
	(2019{\natexlab{b}}).
	\newblock Models as approximations {II}: A model-free theory of parametric
	regression.
	\newblock \textit{Statistical Science} \textbf{34} 545--565.
	
	\bibitem[{Burago et~al.(2001)Burago, Burago and Ivanov}]{bura:01}
	\textsc{Burago, D.}, \textsc{Burago, Y.} and \textsc{Ivanov, S.} (2001).
	\newblock \textit{A Course in Metric Geometry}.
	\newblock American Mathematical Society, Providence, RI.
	
	\bibitem[{Buttazzo et~al.(2012)Buttazzo, De~Pascale and
		Gori-Giorgi}]{buttazzo:2012}
	\textsc{Buttazzo, G.}, \textsc{De~Pascale, L.} and \textsc{Gori-Giorgi, P.}
	(2012).
	\newblock Optimal-transport formulation of electronic density-functional
	theory.
	\newblock \textit{Physical Review A} \textbf{85} 062502.
	
	\bibitem[{Carlier and Ekeland(2010)}]{carlier:2010}
	\textsc{Carlier, G.} and \textsc{Ekeland, I.} (2010).
	\newblock Matching for teams.
	\newblock \textit{Economic Theory} \textbf{42} 397--418.
	
	\bibitem[{Chen et~al.(2020)Chen, Lin and M{\"u}ller}]{mull:20:7}
	\textsc{Chen, Y.}, \textsc{Lin, Z.} and \textsc{M{\"u}ller, H.-G.} (2020).
	\newblock Wasserstein regression.
	\newblock \textit{arXiv:2006.09660} .
	
	\bibitem[{Cuturi(2013)}]{cuturi:2013}
	\textsc{Cuturi, M.} (2013).
	\newblock {S}inkhorn distances: Lightspeed computation of optimal transport.
	\newblock In \textit{Advances in neural information processing systems}.
	
	\bibitem[{Cuturi and Doucet(2014)}]{cuturi:2014}
	\textsc{Cuturi, M.} and \textsc{Doucet, A.} (2014).
	\newblock Fast computation of {W}asserstein barycenters.
	\newblock In \textit{International Conference on Machine Learning}, vol.~32.
	
	\bibitem[{Cuturi and Peyr{\'e}(2016)}]{cuturi:2016}
	\textsc{Cuturi, M.} and \textsc{Peyr{\'e}, G.} (2016).
	\newblock A smoothed dual approach for variational {W}asserstein problems.
	\newblock \textit{SIAM Journal on Imaging Sciences} \textbf{9} 320--343.
	
	\bibitem[{del Barrio and Loubes(2020)}]{del:2020}
	\textsc{del Barrio, E.} and \textsc{Loubes, J.-M.} (2020).
	\newblock The statistical effect of entropic regularization in optimal
	transportation.
	\newblock \textit{arXiv preprint arXiv:2006.05199} .
	
	\bibitem[{Dvurechenskii et~al.(2018)Dvurechenskii, Dvinskikh, Gasnikov, Uribe
		and Nedich}]{dvurechenskii:2018}
	\textsc{Dvurechenskii, P.}, \textsc{Dvinskikh, D.}, \textsc{Gasnikov, A.},
	\textsc{Uribe, C.} and \textsc{Nedich, A.} (2018).
	\newblock Decentralize and randomize: Faster algorithm for {W}asserstein
	barycenters.
	\newblock In \textit{Advances in Neural Information Processing Systems}.
	
	\bibitem[{Fr{\'e}chet(1948)}]{frechet:1948}
	\textsc{Fr{\'e}chet, M.} (1948).
	\newblock Les {\'e}l{\'e}ments al{\'e}atoires de nature quelconque dans un
	espace distanci{\'e}.
	\newblock \textit{Annales de l`Institut Henri Poincar\'e} \textbf{10} 215--310.
	
	\bibitem[{Frogner et~al.(2015)Frogner, Zhang, Mobahi, Araya and
		Poggio}]{frogner:2015}
	\textsc{Frogner, C.}, \textsc{Zhang, C.}, \textsc{Mobahi, H.}, \textsc{Araya,
		M.} and \textsc{Poggio, T.~A.} (2015).
	\newblock Learning with a {W}asserstein loss.
	\newblock In \textit{Advances in Neural Information Processing Systems}.
	
	\bibitem[{Genevay et~al.(2018)Genevay, Peyr{\'e} and Cuturi}]{genevay:2018}
	\textsc{Genevay, A.}, \textsc{Peyr{\'e}, G.} and \textsc{Cuturi, M.} (2018).
	\newblock Learning generative models with sinkhorn divergences.
	\newblock In \textit{International Conference on Artificial Intelligence and
		Statistics}.
	
	\bibitem[{Gibbs and Su(2002)}]{gibbs:2002}
	\textsc{Gibbs, A.~L.} and \textsc{Su, F.~E.} (2002).
	\newblock On choosing and bounding probability metrics.
	\newblock \textit{International Statistical Review} \textbf{70} 419--435.
	
	\bibitem[{Horvath and Kokoszka(2012)}]{horv:12}
	\textsc{Horvath, L.} and \textsc{Kokoszka, P.} (2012).
	\newblock \textit{Inference for Functional Data with Applications}.
	\newblock Springer, New York.
	
	\bibitem[{Huber(2004)}]{huber:2004}
	\textsc{Huber, P.~J.} (2004).
	\newblock \textit{Robust statistics}, vol. 523.
	\newblock John Wiley \& Sons.
	
	\bibitem[{Janati et~al.(2020)Janati, Cuturi and Gramfort}]{janati:2020}
	\textsc{Janati, H.}, \textsc{Cuturi, M.} and \textsc{Gramfort, A.} (2020).
	\newblock Debiased {S}inkhorn barycenters.
	\newblock \textit{arXiv preprint arXiv:2006.02575} .
	
	\bibitem[{Le~Gouic and Loubes(2017)}]{le:2017}
	\textsc{Le~Gouic, T.} and \textsc{Loubes, J.-M.} (2017).
	\newblock Existence and consistency of {W}asserstein barycenters.
	\newblock \textit{Probability Theory and Related Fields} \textbf{168} 901--917.
	
	\bibitem[{McCann(1997)}]{mccann:1997}
	\textsc{McCann, R.~J.} (1997).
	\newblock A convexity principle for interacting gases.
	\newblock \textit{Advances in Mathematics} \textbf{128} 153--179.
	
	\bibitem[{Monge(1781)}]{monge:1781}
	\textsc{Monge, G.} (1781).
	\newblock M{\'e}moire sur la th{\'e}orie des d{\'e}blais et des remblais.
	\newblock \textit{Histoire de l'Acad{\'e}mie Royale des Sciences de Paris} .
	
	\bibitem[{Neumayer and Steidl(2020)}]{neumayer:2020}
	\textsc{Neumayer, S.} and \textsc{Steidl, G.} (2020).
	\newblock From optimal transport to discrepancy.
	\newblock \textit{arXiv preprint arXiv:2002.01189} .
	
	\bibitem[{Panaretos and Zemel(2019)}]{pana:19}
	\textsc{Panaretos, V.~M.} and \textsc{Zemel, Y.} (2019).
	\newblock Statistical aspects of {W}asserstein distances.
	\newblock \textit{Annual Review of Statistics and its Application} \textbf{6}
	405--431.
	
	\bibitem[{Petersen and M\"uller(2016)}]{mull:16:1}
	\textsc{Petersen, A.} and \textsc{M\"uller, H.-G.} (2016).
	\newblock Functional data analysis for density functions by transformation to a
	{H}ilbert space.
	\newblock \textit{Annals of Statistics} \textbf{44} 183--218.
	
	\bibitem[{Petersen and M{\"u}ller(2019)}]{petersen:2019}
	\textsc{Petersen, A.} and \textsc{M{\"u}ller, H.-G.} (2019).
	\newblock Fr{\'e}chet regression for random objects with {E}uclidean
	predictors.
	\newblock \textit{The Annals of Statistics} \textbf{47} 691--719.
	
	\bibitem[{Peyr{\'e}(2015)}]{peyre:2015}
	\textsc{Peyr{\'e}, G.} (2015).
	\newblock Entropic approximation of {W}asserstein gradient flows.
	\newblock \textit{SIAM Journal on Imaging Sciences} \textbf{8} 2323--2351.
	
	\bibitem[{Peyr{\'e} and Cuturi(2019)}]{peyr:19}
	\textsc{Peyr{\'e}, G.} and \textsc{Cuturi, M.} (2019).
	\newblock Computational optimal transport: With applications to data science.
	\newblock \textit{Foundations and Trends in Machine Learning} \textbf{11}
	355--607.
	
	\bibitem[{Rabin et~al.(2011)Rabin, Peyr{\'e}, Delon and Bernot}]{rabin:2011}
	\textsc{Rabin, J.}, \textsc{Peyr{\'e}, G.}, \textsc{Delon, J.} and
	\textsc{Bernot, M.} (2011).
	\newblock {W}asserstein barycenter and its application to texture mixing.
	\newblock In \textit{International Conference on Scale Space and Variational
		Methods in Computer Vision}. Springer.
	
	\bibitem[{Rubner et~al.(2000)Rubner, Tomasi and Guibas}]{rubner:2000}
	\textsc{Rubner, Y.}, \textsc{Tomasi, C.} and \textsc{Guibas, L.~J.} (2000).
	\newblock The earth mover's distance as a metric for image retrieval.
	\newblock \textit{International Journal of Computer Vision} \textbf{40}
	99--121.
	
	\bibitem[{Sheather and Jones(1991)}]{sheather:1991}
	\textsc{Sheather, S.~J.} and \textsc{Jones, M.~C.} (1991).
	\newblock A reliable data-based bandwidth selection method for kernel density
	estimation.
	\newblock \textit{Journal of the Royal Statistical Society: Series B
		(Methodological)} \textbf{53} 683--690.
	
	\bibitem[{van~der Vaart and Wellner(1996)}]{van:1996}
	\textsc{van~der Vaart, A.} and \textsc{Wellner, J.} (1996).
	\newblock \textit{Weak Convergence and Empirical Processes with Applications to
		Statistics}.
	\newblock Springer.
	
	\bibitem[{Villani(2008)}]{villani:2008}
	\textsc{Villani, C.} (2008).
	\newblock \textit{Optimal transport: Old and New}, vol. 338.
	\newblock Springer Science \& Business Media.
	
	\bibitem[{Wang et~al.(2016)Wang, Chiou and M\"uller}]{mull:16:3}
	\textsc{Wang, J.-L.}, \textsc{Chiou, J.-M.} and \textsc{M\"uller, H.-G.}
	(2016).
	\newblock Functional data analysis.
	\newblock \textit{Annual Review of Statistics and its Application} \textbf{3}
	257--295.
	
\end{thebibliography}
\end{document}